\documentclass[
reprint,
superscriptaddress,
nofootinbib,
showkeys,
aps,
pra,
longbibliography
]{revtex4-1}

\usepackage[utf8]{inputenc}

\usepackage[hyperref]{xcolor}
\definecolor{goethe-blau}{cmyk}{1.0,0.2,0.0,0.4}
\definecolor{hellgrau}{cmyk}{0.04,0.04,0.05,0.02}
\definecolor{sandgrau}{cmyk}{0.12,0.09,0.13,0.0}
\definecolor{dunkelgrau}{cmyk}{0.25,0.25,0.30,0.75}
\definecolor{purple}{cmyk}{0.08,1.0,0.3,0.36}
\definecolor{emo-rot}{cmyk}{0.04,1.0,0.8,0.07}
\definecolor{senfgelb}{cmyk}{0.01,0.25,1.0,0.05}
\definecolor{gruen}{cmyk}{0.62,0.4,0.87,0.09}
\definecolor{magenta}{cmyk}{0.08,0.86,0.12,0.12}
\definecolor{orange}{cmyk}{0.0,0.7,1.0,0.04}
\definecolor{sonnengelb}{cmyk}{0.0,0.12,0.95,0.0}
\definecolor{helles-gruen}{cmyk}{0.4,0.17,0.81,0.07}
\definecolor{lichtblau}{cmyk}{0.8,0.0,0.06,0.04}

\usepackage[
colorlinks,
pdfpagelabels,
breaklinks,
pdfstartview=FitH,
bookmarksopen=true,
bookmarksnumbered=true,
bookmarksopenlevel=2,
plainpages=false,
hypertexnames=false,
citecolor=emo-rot,
linkcolor=goethe-blau,
urlcolor=purple,
pdftitle={...},		
pdfauthor={...},	
pdfkeywords={}					
]{hyperref}

\usepackage{graphicx}
\usepackage{dcolumn}
\usepackage{braket}
\usepackage[tbtags]{amsmath}
\usepackage{amssymb}
\usepackage{epstopdf}
\usepackage{color}
\usepackage{xspace}
\usepackage{cleveref}
\usepackage{longtable}

\usepackage{orcidlink}
\usepackage{upgreek}

\usepackage[acronym]{glossaries-extra}

\glssetcategoryattribute{acronym}{nohyperfirst}{true}
\setabbreviationstyle[acronym]{long-short}

\newcommand{\UNIT}[1]{\; \mathrm{#1}}

\newcommand{\MeV}{\UNIT{MeV}}

\newcommand{\fm}{\UNIT{fm}}

\newcommand{\REM}[1]{}

\definecolor{magenta}{cmyk}{0,1,0,0}

\newcommand*{\defeq}{\mathrel{\vcenter{\baselineskip0.5ex \lineskiplimit0pt
			\hbox{\scriptsize.}\hbox{\scriptsize.}}}%
	=}

\allowdisplaybreaks
\begin{document}


\title{Kadanoff-Baym approach to bound states in open quantum systems}

\author{Tim Neidig}
\email{neidig@itp.uni-frankfurt.de}
\affiliation{Institut f\"ur Theoretische Physik, Johann Wolfgang Goethe-Universit\"at, Max-von-Laue-Strasse 1, 60438 Frankfurt am Main, Germany}%
\affiliation{Helmholtz Research Academy Hessen for FAIR (HFHF), GSI Helmholtz Center, Campus Frankfurt, Max-von-Laue-Stra{\ss}e 12, 60438 Frankfurt am Main, Germany}


\author{Marcus Bleicher}
\affiliation{Institut f\"ur Theoretische Physik, Johann Wolfgang Goethe-Universit\"at, Max-von-Laue-Strasse 1, 60438 Frankfurt am Main, Germany}%
\affiliation{Helmholtz Research Academy Hessen for FAIR (HFHF), GSI Helmholtz Center, Campus Frankfurt, Max-von-Laue-Stra{\ss}e 12, 60438 Frankfurt am Main, Germany}

\author{Hendrik van Hees}
\affiliation{Institut f\"ur Theoretische Physik, Johann Wolfgang Goethe-Universit\"at, Max-von-Laue-Strasse 1, 60438 Frankfurt am Main, Germany}%
\affiliation{Helmholtz Research Academy Hessen for FAIR (HFHF), GSI Helmholtz Center, Campus Frankfurt, Max-von-Laue-Stra{\ss}e 12, 60438 Frankfurt am Main, Germany}

\author{Carsten Greiner}
\affiliation{Institut f\"ur Theoretische Physik, Johann Wolfgang Goethe-Universit\"at, Max-von-Laue-Strasse 1, 60438 Frankfurt am Main, Germany}%
\affiliation{Helmholtz Research Academy Hessen for FAIR (HFHF), GSI Helmholtz Center, Campus Frankfurt, Max-von-Laue-Stra{\ss}e 12, 60438 Frankfurt am Main, Germany}


\date{\today}

\begin{abstract}
In this paper, we extend the method of Kadanoff-Baym equations for open quantum systems to arbitrary kinds of systems and heat baths, either fermionic or bosonic. This includes three spacial dimensions and different potentials for the system-bath interaction or external traps.
We study the quantum-mechanical formation of bound states in one and also in three dimensions with the full Kadanoff-Baym equations and compare them to more simplified approaches with and without memory effects. 
An in-depth examination of the thermodynamics of open systems is performed, showing perfect equilibration of the system's degrees of freedom along with a comprehensive investigation of the influence of the heat bath on the system's wave functions. The formation time, decay time and regeneration of bound states and their dependence on the temperature and coupling strength is explored

We evaluate the non-equilibrium Kadanoff-Baym equations for the system particles, assuming that interactions are elastic two-particle collisions with the heat-bath particles. Finally, we describe in detail the method used to numerically solve the corresponding spatially heterogeneous integro-differential equations for the set of one-particle Green's functions. 



\end{abstract}

\maketitle

\section{Introduction} 

Open quantum systems, i.e. quantum systems that interact with external environments, represent a fundamental and very often encountered class of physical systems. Unlike their idealized closed counterparts, real quantum systems are never perfectly isolated; they are always coupled to reservoirs, fluctuating fields, or measurement devices. This coupling leads to phenomena such as decoherence, dissipation, and entanglement with the environment, all of which are crucial in modern quantum technologies including quantum information processing, quantum thermodynamics, and mesoscopic transport, but of course also for fundamental aspects of particle physics.

In recent years, the Keldysh/Kadanoff–Baym framework has been increasingly applied to open quantum systems \cite{Sieberer_2016,handle:20.500.11811/8961}, offering a unified treatment of equilibrium and non-equilibrium dynamics, as well as the ability to incorporate structured environments, time-dependent driving, and thermalization processes. Nevertheless, solving the Kadanoff–Baym equations is computationally demanding, and approximations, such as conserving self energies, must often be employed to make the problem tractable.

Our aim is to demonstrate how the Kadanoff–Baym equation framework can serve as a versatile and systematically improvable tool for modeling open quantum systems across a wide range of physical settings.

In a previous letter \cite{2024PhLB..85138589N}, we studied a one-dimensional fermionic system in a bosonic bath, wherein we had explicitly verified the fundamental conditions of decoherence and thermalization of the occupied quantum states. In addition, the spectral characteristics of the system have been shown.

This approach represents a novel contribution to the field of open quantum systems, offering a more expansive range of potential applications than established Markovian master equations, such as Lindblad master equations \cite{Gorini:1975nb, 1976CMaPh..48..119L,DIOSI1993517,PhysRevE.110.054116} specifically for the Caldeira-Leggett model \cite{CALDEIRA1983587, gardiner00, BRE02}. One should note that such approaches are applicable only in the high-temperature limit with weak coupling, because otherwise it cannot be guaranteed that the time evolution will always result in the system reaching thermal equilibrium with the bath \cite{PhysRevLett.80.5702,Rais:2025fps}. 

The underlying Hamiltonian can be decomposed into a pure system component $S$, a pure bath component $B$, and the interaction between the system and the bath $SB$,

\begin{equation}
    \hat{H} = \hat{H}_S + \hat{H}_B + \hat{H}_{SB}.
\end{equation}

The subsequent discussion employs natural units ($c=k_B=\hbar=1$).
The common master equation setup is stated, where the evolution of the density matrix $\hat{\rho}_S$ is governed by a Lindblad equation \cite{Gorini:1975nb, 1976CMaPh..48..119L, DIOSI1993517}, 

\begin{equation}
\begin{split}
\label{Lindblad}
    \frac{\partial \hat{\rho}_S}{\partial t} = \text{i} \left[ \hat{\rho}_S, \hat{H}_S\right] + \hat{L}(\hat{\rho}_S, \hat{L}_\lambda),
\end{split}
\end{equation}

but with system-bath interactions included too. 
In the given context, the Lindblad superoperator, represented by the symbol $\hat{L}$, comprises the Lindblad operators, denoted here by $\hat{L}_\lambda$. The determination of these operators is a crucial aspect of the analysis, as they represent the fundamental components necessary to comprehend the physical dynamics of the system under investigation. The identification of these operators often involves intuitive considerations, as discussed e.g. in \cite{May:1416853}. This contrasts the strict formalism of the applied Kadanoff-Baym approach, where the system-bath interaction is incorporated through self energies, that are based on resummed perturbation theory, which enables the development of expansions in the coupling constants.

The approach of considering non-equilibrium Green's functions to describe systems out of equilibrium was initially developed by Schwinger \cite{10.1063/1.1703727} and further specified by Kadanoff, Baym and Keldysh \cite{KBBook,PhysRev.124.287,PhysRev.127.1391,Keldysh:1964ud}. 

It is evident that the Kadanoff-Baym equations are constructed in such a manner as to ensure the preservation of local conservation laws, as well as the norm and complete positivity of the occupation numbers. In the context of an open quantum system, it is guaranteed that thermalization is attained for the dressed quantum degrees of freedom. Furthermore, it has been demonstrated that quantum correlations decohere as a consequence of interactions between the system and the bath degrees of freedom \cite{Schlosshauer:2007:un,2024PhLB..85138589N,Rais:2025fps}.

In this study, we consider the following class of particles: fermionic or bosonic particles residing in a finite volume in one or three dimensions. These particles are confined through a potential, e.g. a harmonic potential or a double square-well potential to mimic a confining box. These conditions are necessary for the numerical method because they yield a discrete spectrum of the corresponding single-particle Hamiltonian.
In particular, we focus on bound-state formation in one dimension \cite{Rais:2022} and three dimensions. It is important to elucidate the impact of memory effects and decoherence on the thermalization process. To this aim, an extensive analysis of the Kadanoff-Baym equations is needed. This analysis should encompass a range of approximations, including the homogeneous \cite{KOHLER1999123,Juchem_2004,PhysRevD.102.016012} or diagonal approximation, which preserves the non-Markovian nature but entirely neglects the phenomenon of decoherence. Another approximation to consider is the quantum kinetic master equation, which, akin to the Lindblad or Boltzmann equation, exhibits a Markovian character.

In accordance with the approach outlined in \cite{2024PhLB..85138589N}, system particles are linked to a heat bath of free, thermalized particles through the process of elastic scattering in a conventional many-body quantum framework \cite{10.1063/1.1703727,KBBook, Keldysh:1964ud, Danielewicz:1982kk}. This configuration represents an open quantum system, wherein the energy exchange between the system and the bath degrees of freedom drives the system towards the thermal equilibrium state, dictated by the temperature of the bath, $T_{\rm{bath}}$, chemical potential, $\mu_{\rm{bath}}$, and the interaction strength between system and bath. The underlying philosophy of the Kadanoff-Baym equations \cite{Greiner:1998vd} is analogous to the Influence Functional of Feynman and Vernon \cite{FEYNMAN1963118} in the well-known Caldeira-Leggett framework \cite{CALDEIRA1983587}.

In the next \cref{sec:2,sec:3}, the framework for the resulting Kadanoff-Baym equations in an open quantum system for arbitrary particle species in the system/bath is established. Subsequently, the approach to obtain numerical solutions is explained and the fully decomposed matrix-valued equations for the Green's functions are shown. Furthermore, we shortly summarize the advantages and possible further investigations obtained by using the Kadanoff-Baym approach. In subchapters, we will elaborate on the direct relation between Green's functions and density matrices and determine some interesting thermodynamic properties, such as entropy and energy functionals \cite{galitskii1958application} and explain the derivation and short-comings of two common approximations of the Kadanoff-Baym equations, the diagonal Kadanoff-Baym equations and the quantum kinetic master equation, in more detail. 

In \cref{sec:results}, we then present extensive numerical results, and also put a focus on the \textit{quantum mechanical} formation of bound states, e.g. deuteron \cite{KO2010253c,PhysRevC.104.034908,Glassel:2021rod,Sun:2021dlz,PhysRevC.105.044909,Neidig:2021bal} or $J/\Psi$ \cite{Hammou:2024dtj,Oei:2024zyx,Oei:2024fbg}, which are currently of high interest in the heavy ion physics community \cite{2024PhLB..85138589N,Rais:2022,Rais:2025fps}. We demonstrate the thermalization of the system within a hot medium and contrast it to classical transport approaches or quantum master equations. We showcase the effects of different approximations of the Kadanoff-Baym equations and their impact on the evolution of the system particles and thermalization times. Furthermore, we will demonstrate the possibility to capture potential changes in wave functions and energy eigenvalues by $H_{\mathrm{SB}}$. In the following subsection, we provide an insight to the 3-dimensional extension of the present study and point out differences to the one-dimensional case in the early-time dynamics and in the spectral functions. Finally, the application to an interacting Bose gas in one dimension will be illustrated, where we point out differences in the spectral functions compared to fermionic systems and show how the ground state will be populated during the evolution of the system.

\section{Setup for the model} \label{sec:2}

\begin{figure}
     \centering
     \begin{tikzpicture}
     \draw [black, very thick]  (-0.5,0) -- (4.94,0);
     \draw [black, thick] (4.9,-0.2) -- (4.98,0.2);
     \draw [black, thick] (5.02,-0.2) -- (5.1,0.2);
     \draw [black, very thick,-stealth]  (5.06,0) -- (7.5,0);
     \draw [red, very thick]  (0,0.5) -- (4.5,0.5);
     \draw [red, very thick,dotted]  (4.5,0.5) -- (5.5,0.5);
     \draw [red, very thick]  (5.5,0.5) -- (7.0,0.5);
     \draw [red, very thick, -latex]  (5.5,0.5) -- (6,0.5);
     \draw [red, very thick, latex-]  (5.5,-0.5) -- (6,-0.5);
     \draw [red, very thick]  (0,-0.5) -- (4.5,-0.5);
     \draw [red, very thick,dotted]  (4.5,-0.5) -- (5.5,-0.5);
     \draw [red, very thick]  (5.5,-0.5) -- (7.0,-0.5);
     \draw [red, very thick] (7.0,0.5) arc[radius =0.5, start angle= 90, 
end angle= -90];
     \draw[fill=blue!50, draw=black, thick] (0,0.5) circle (2.5pt);
     \draw[fill=blue!50, draw=black, thick] (0,-0.5) circle (2.5pt);
     \node (a) at (7.8,0) {$\infty$};
     \fill (0,0) circle (0.05) node[above]{$t_0$};
     \fill (4,0.5) circle (0.05) node[above]{$t$};
     \fill (3,-0.5) circle (0.05) node[above]{$t'$};
     \fill (7.2,0) circle (0.0) node[below]{$t$};
     \end{tikzpicture}
     \caption{The closed-time path $C$ with the times ordered as it is the case for $S^{<}$.}
     \label{fig:CTP}
\end{figure}
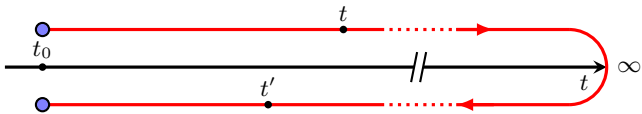

The non-equilibrium Green's function $S(1,1')$ is defined by (we follow the conventions of \cite{10.1063/1.1703727,KBBook, Keldysh:1964ud, Danielewicz:1982kk})

\begin{equation}
\begin{split}
S(1,1') &= -i \bigl\langle T_{c} \bigl[ \hat{\psi}(r,t) \hat{\psi}(r',t')^{\dagger} \bigr] \bigr\rangle \\
&= \Theta_c(t,t') S^{>}(1,1') \pm \Theta_c(t', t) S^{<}(1,1'), 
\label{1}
\end{split}
\end{equation}

where the $\pm$ is connected to (upper) bosons and (lower sign) fermions, which takes the permutation into account. In this notation $1=(r,t)$. $T_c$ is the contour-time ordering operator along the ``Schwinger-Keldysh" contour \cref{fig:CTP},

\begin{equation}
\begin{split}
T_{c} \bigl[ \hat{\psi}(r,t) \hat{\psi}(r',t')^{\dagger} \bigr] \defeq \begin{cases}
\hat{\psi}(r,t) \hat{\psi}(r',t')^{\dagger} &\text{if $t > t'$,}\\
\pm \hat{\psi}(r',t')^{\dagger} \hat{\psi}(r,t) &\text{if $t \le t'$}.
\end{cases} 
\label{2}
\end{split}
\end{equation}

Furthermore, it is possible to substitute the contour time ordering operator by a contour Heaviside function, which is defined as \cite{DANIELEWICZ1984305}

\begin{equation}
\begin{split}
\Theta_{c}(t,t') \defeq \begin{cases}
1, &\text{if $t$ is later on the contour than $t'$,}\\
0, &\text{else}.
\end{cases} 
\label{1.1}
\end{split}
\end{equation}

Here $t < t'$ means, $t$ is earlier on the ``Schwinger-Keldysh" contour than $t'$. We assume a spin-saturated system, such that there are no spin-dependent interactions, which could make the Green's function non-diagonal in spin-space. 

The system particles are described within an external, attractive, static, and spatially extended potential coupled to an environment (heat bath) by the many-body Hamiltonian
\begin{equation}
\begin{split}
 &\hat{H}(t) = \underbrace{\int dr \, \hat{\psi}(r,t)^{\dagger} \Bigl( \underbrace{-\frac{\Delta}{2\mu_S} + V_S(r)}_{\defeq h_{0}} \Bigr) \hat{\psi}(r,t)}_{\defeq \hat{H}_{S}(t)} \\
 &+ \underbrace{ \int dr \int dr' \, \hat{\psi}(r,t)^{\dagger} \hat{\psi}(r,t) V_{\rm{int}}(|r-r'|) \hat{\phi}(r',t)^{\dagger} \hat{\phi}(r',t)}_{\defeq\hat{H}_{\mathrm{SB}}(t)} 
 \label{Hamiltonian}
\end{split}
\end{equation}
with (reduced) mass $\mu_S$.
The degrees of freedom of the bath are subject to an uncoupled standard Hamiltonian, although they may also be influenced by an external potential. The external potential $V_{S}(r)$ is selected to emulate, for instance, a gas of bosons confined within a harmonic potential or a gas of nucleons influenced by other nucleons, thereby facilitating the formation of bound states. $\hat{\psi}(r,t)$ represents the system-operator and $\hat{\phi}(r,t)$ the bath-operator. $\hat{H}_{\mathrm{SB}}$ represents the interaction between system and heat-bath particles, which can be arbitrarily chosen up to now. In many cases one can think of a screened Coulomb potential, a smeared gaussian potential or a local s-wave interaction as a choice of $V_{\rm{int}}(|r-r'|)$. 
The explicit consideration of self-interactions of the system particles or the bath particles is omitted in this study. 

The heat-bath particles are assumed to be at constant temperature, $T_{\mathrm{bath}}$, in a state of thermal equilibrium. The number of bath particles is controlled by a separate chemical potential. 

This set-up allows for a sufficiently comprehensive analysis of the open quantum system. The next step is now to derive the Green's functions relevant for the Kadanoff-Baym formalism. These are obtained from the Hamiltonian described in \cref{Hamiltonian} and the field operators contained therein. The final step in this process is to calculate the self-energies composed of these Green's functions.

The field operators can be expanded into any complete basis without restriction of generality. According to the formalism of the second quantization, the expansion coefficients occurring there are the corresponding creation or annihilation operators,

\begin{equation}
\begin{split}
\hat{\psi}(r,t) &\defeq \sum_{n=0}^{S} \hat{c}_{n}(t) \phi_{n}(r).
\label{4}
\end{split}
\end{equation}

It is customary to adapt the basis in accordance with the specific problem at hand, mostly to include symmetries of the system.

In the example of an open quantum system presented here, the one-particle eigenstates of $h_0$ in \cref{Hamiltonian} \cite{Keldysh:1964ud, dahlen2006propagating, dahlen2007solving, Stan_2009}, naturally lend themselves as the preferred basis,

\begin{equation}
\begin{split}
h_{0} \, \phi_{n}(r) &= E_{n} \, \phi_{n}(r), \\
\int dr \, \phi_{m}(r)^{*} \, \phi_{n}(r) &= \delta_{m,n}. 
\label{eigenfunction}
\end{split}
\end{equation}

For the one-dimensional bound-state problem, the eigenfunctions can be found in \cite{Rais:2022} for a special choice of $V_S(r)$, which was already elaborated in \cite{2024PhLB..85138589N}. 
For the numerical solution we truncate the Hilbert space of the modes at a cutoff mode $S$.
Inserting (\ref{4}) in (\ref{1}) leads to the energy-basis representation of the inhomogenous Green's function as

\begin{equation}
\begin{split}
S^{>}(1,1') &= -i \sum_{n,m=0}^{S} \langle \hat{c}_{n}(t) \hat{c}_{m}(t')^{\dagger} \rangle \phi_{n}(r) \phi^{*}_{m}(r'), \\
S^{<}(1,1') &= \mp i \sum_{n,m=0}^{S} \langle \hat{c}_{m}(t')^{\dagger} \hat{c}_{n}(t) \rangle \phi_{n}(r) \phi^{*}_{m}(r').
\label{5}
\end{split}
\end{equation}

The temporal evolution of these Green's functions is governed by the Kadanoff-Baym equations \cite{KBBook,Danielewicz:1982kk}

\begin{equation}
\begin{split}
\Bigl( i \frac{\partial}{\partial t} + \frac{\Delta_1}{2\mu_S} -
  V_{\mathrm{eff}}(1) \Bigr) S^{\gtrless}(1,1') &= I_{\mathrm{coll}_{1}}^{\gtrless}(t,t'), 
\label{6.1}
\end{split}
\end{equation}

\begin{equation}
\begin{split}
\Bigl( -i \frac{\partial}{\partial t'} + \frac{\Delta_{1'}}{2\mu_S} - V_{\mathrm{eff}}(1') \Bigr) S^{\gtrless}(1,1') &= I_{\mathrm{coll}_{2}}^{\gtrless}(t,t'),
\label{6.2}
\end{split}
\end{equation}

where we introduced an effective potential as the sum of the external potential and the Hartree self-energy
\begin{equation}
\begin{split}
V_{\mathrm{eff}}(1) &= V_S(1) + \Sigma_{H}(1). 
\label{effpot}
\end{split}
\end{equation}

The collision terms on the right-hand side are

\begin{equation}
\begin{split}
I_{\mathrm{coll}_{1}}^{\gtrless}(t,t') &= \int_{t_0}^{t} d\bar{1}\biggl[ \Sigma^{>}(1, \bar{1}) - \Sigma^{<}(1, \bar{1}) \biggr] S^{\gtrless}(\bar{1},1') \\
&- \int_{t_0}^{t'} d\bar{1} \Sigma^{\gtrless}(1, \bar{1}) \biggl[S^{>}(\bar{1},1') - S^{<}(\bar{1},1') \biggr], \\
I_{\mathrm{coll}_{2}}^{\gtrless}(t,t') &= \int_{t_0}^{t} d\bar{1} \biggl[S^{>}(1,\bar{1}) - S^{<}(1,\bar{1}) \biggr] \Sigma^{\gtrless}(\bar{1},1') \\
&- \int_{t_0}^{t'} d\bar{1} S^{\gtrless}(1,\bar{1}) \biggl[ \Sigma^{>}(\bar{1},1') - \Sigma^{<}(\bar{1},1') \biggr]. \label{7}
\end{split}
\end{equation}

A second-order (direct Born) sunset self energy is incorporated, which itself contains the system Green's function $S$. This is known as ``resummation", in fact the Kadanoff-Baym equations are resumming the Green's function as they are just a rewritten Dyson equation,

\begin{equation}
\begin{split}
S(1,1') &= S_0(1,1') + \int_{c} d2 \int_{c} d3 \, S_0(1,2) \Sigma(2,3) S(3,1').
\label{Dyson}
\end{split}
\end{equation}

 However, the Kadanoff-Baym equations \cref{6.1,6.2} are obtained from \cref{Dyson} by multiplying with the free inverse Green's function $S_0^{-1}$, which is just the Schrödinger operator on the left hand side of \cref{6.1,6.2} in front of the Green's function.

To make the system open, it is first necessary to establish the thermal equilibrium Green's function for the bath particle prior to the explicit definition of the self energies,

\begin{equation}
\begin{split}
\label{8}
D^{>}_0(1, 1') &= -i \sum_{n}^{B} \int \frac{d\mathrm{\omega}}{2\pi} \tilde{a}_{n}(\omega) e^{-i \omega (t-t')} \\
& \,\,\,\,\,\,\,\, (1 \pm n_{B/F}(\omega)) \tilde{\phi}_{n}(r) \tilde{\phi}^{*}_{n}(r'), \\
D^{<}_0(1, 1') &= \mp i \sum_{n}^{B} \int \frac{d\mathrm{\omega}}{2\pi} \tilde{a}_{n}(\omega) e^{-i \omega (t-t')} \\
& \,\,\,\,\,\,\,\, n_{B/F}(\omega) \tilde{\phi}_{n}(r) \tilde{\phi}^{*}_{n}(r'),
\end{split}
\end{equation}

where $n_{B/F}(\epsilon_{n})=\frac{1}{\mathrm{exp}((\epsilon_{n}-\mu_{\mathrm{bath}})/T) \mp 1}$, with $\mu_{\mathrm{bath}}$ the chemical potential of the bath and $\tilde{\phi}_{n}(r)$ are the eigenfunctions of the bath, which in general can be also affected by an external potential. The chemical potential of the bath fixes the number of particles in the bath and $B$ is a cutoff mode similar to the system case.
A general, dissipative spectral function $\tilde{a}_{n}(\omega)$ can be straightforwardly incorporated to mimic the effect of scattering on the bath degrees of freedom. For simplicity we use here the on-shell approximation $\tilde{a}_{n}(\omega) = \delta(\omega - \epsilon_n)$ with $\epsilon_n$ the energy of the n-th eigen state of the bath Hamiltonian. These details are, however, not relevant for our later discussions.

\begin{figure}[h] 
\begin{center}
\includegraphics[width=0.5\textwidth]{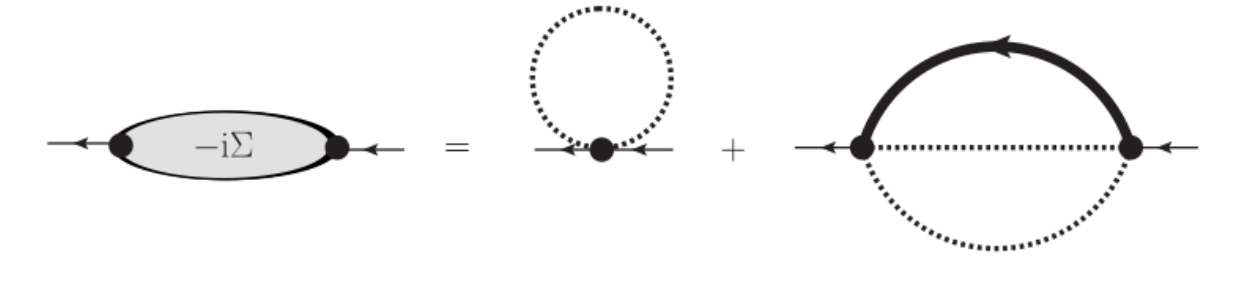}
\end{center}
\caption{Self energy diagrams for the scattering process in the open quantum system. The dotted lines are bath propagators, fixed to the free thermal real-time propgators, and full lines depict self-consistently evaluated system-particle propagators \cite{BINOSI200476}. }
\label{fig:Sunset}
\end{figure}

\begin{figure}[h] 
\begin{center}
\includegraphics[width=0.35 \textwidth]{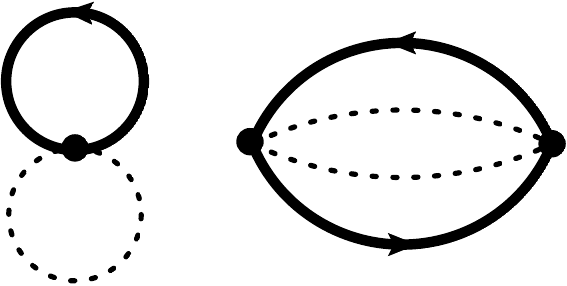}
\end{center}
\caption{The basketball-diagram (right) and the diagram, that creates the tadpole (left), as a part of the $\Phi$-functional of the 2-PI effective action. The dotted lines denote bath propagators, affixed to the free thermal real-time propagators. The full lines depict system-particle propagators, which have been self-consistently evaluated \cite{BINOSI200476}. }
\label{fig:Basketball}
\end{figure}

The self energies that are guided by an expansion in the interacting potential, denoted by $V_{\rm{int}}$, are given by
\begin{equation}
\begin{split}
\Sigma_{H}(1) &= \underbrace{\pm}_{\rm{bath-type}}i \int dr' \, V_{\rm{int}}(|r-r'|) \, D^{<}_0(r',t,r',t^{+}), \\
\Sigma^{\gtrless}(1, 1') &= \rm{sign(\mathrm{\cref{table:sign}})} \, \int dr_{2} \int dr_{2'} \, S^{\gtrless}(1,1') \\ & V_{\rm{int}}(|r-r_{2}|) \, V_{\rm{int}}(|r_{2'}-r'|) \, D^{\gtrless}_0(r_{2},t,r_{2'},t') \\
& D^{\lessgtr}_0(r_{2'},t',r_{2},t).
\label{9}
\end{split}
\end{equation}
\begin{table}
\begin{center}
\begin{tabular}{|c|c|c|}
\hline
S\textbackslash B & Boson & Fermion \\
\hline
Boson & - & + \\
\hline
Fermion & - & + \\
\hline
\end{tabular}
\end{center}
\caption{The sign in \cref{9} as a function of the system- and bath-particle species. }
\label{table:sign}
\end{table}
The two diagrams under consideration correspond to the non-dissipative (Hartree) tadpole and dissipative (direct Born) sunset diagram. The derivation of these is possible through second quantization, with the one-particle irreducible (1PI) diagrams being identified from the two-part Green's function \cite{KBBook,Danielewicz:1982kk}. Alternatively, this can be achieved through the more elegant process of deriving the functional derivative of the $\Phi$ functional appearing in the path integral formalism of the two-particle irreducible (2PI) effective action \cite{PhysRevD.10.2428,PhysRevA.72.063604,Juchem_2004}. In this case, the self energies can be obtained by functional derivation of the $\Phi$ functional. For the important contributions up to second order in the (local) system-bath interaction, the (Hartree) tadpole and the sunset diagram \cref{fig:Sunset}, the corresponding parts in the $\Phi$ functional are presented in \cref{fig:Basketball}.
The sunset diagram is of greater significance as it results in a dissipative energy exchange in the (two-body) elastic scatterings between the bath particles and the system particles. In the absence of a sunset contribution, thermal equilibrium can not be achieved as only true scattering processes can guarantee thermalization. The sign of the Hartree term is determined by the type of bath, however, for the sunset diagram, it is also influenced by the type of system particles.
\section{Numerical solution of the Kadanoff-Baym equations}\label{sec:3}

In order to numerically solve the Kadanoff-Baym equations, it is necessary to substitute the expansions for the Green's functions, as specified in \cref{5} and \cref{8}, into \cref{6.1,6.2}. It can be shown that, with the help of \cref{4}, the set of partial integro-differential equations \cref{6.1,6.2} can be reduced to a set of ordinary integro-differential equations for the matrix-valued expectation values $c^{>}_{n,m}(t,t') = \langle \hat{c}_{n}(t) \hat{c}_{m}(t')^{\dagger} \rangle $ and $c^{<}_{n,m}(t,t') = \langle \hat{c}_{m}(t')^{\dagger} \hat{c}_{n}(t) \rangle $ in the two-time plane. The self energies can be expanded in the same basis as the Green's functions

\begin{equation}
\begin{split}
 \Sigma^{>}(1, 1') &= -i \sum_{b,a}^{S} \Sigma^{>}_{b,a}(t,t') \phi_{b}(r) \phi^{*}_{a}(r'), \\
 \Sigma^{<}(1, 1') &= \mp i \sum_{b,a}^{S} \Sigma^{<}_{b,a}(t,t') \phi_{b}(r) \phi^{*}_{a}(r'), \\
 \Sigma^{>}_{b,a}(t,t') &= \sum_{n,m}^{S} \Biggl[ \sum_{j,k}^{B} e^{- i (\epsilon_{j}-\epsilon_{k})(t-t')} \, (1 \pm n_{B/F}(\epsilon_{j}))  \\
 & n_{B/F}(\epsilon_{k}) \, V_{b,n,j,k} \, V_{m,a,k,j}  \Biggr ] \, c^{>}_{n,m}(t,t'), \\
\Sigma^{<}_{b,a}(t,t') &= \sum_{n,m}^{S} \Biggl[ \sum_{j,k}^{B} e^{ i (\epsilon_{j}-\epsilon_{k})(t-t')} \, (1 \pm n_{B/F}(\epsilon_{j}))  \\
& n_{B/F}(\epsilon_{k}) \, V_{b,n,j,k} \, V_{m,a,k,j}  \Biggr ] \, c^{<}_{n,m}(t,t'), \\
\Sigma_{H_{b,a}} &= \sum_{j}^{B} e^{-i \epsilon_{j}0^{+}} n_{B}(\epsilon_{j}) V_{b,a,j,j},
\label{10}
\end{split}
\end{equation}

with the transition amplitudes defined as

\begin{equation}
\begin{split}
V_{b,n,j,k} = \int dr \int dr' \phi^{*}_{b}(r) \phi_{n}(r) V_{\rm{int}}(|r-r'|) \tilde{\phi}_{j}(r')\tilde{\phi}^{*}_{k}(r').
\label{10.1}
\end{split}
\end{equation}
The coefficients $c^{\gtrless}_{n,m}$ are thus propagated in the discretized two-time plane using \cref{6.1,6.2}. However, it should be noted that not all four equations are required, given that the Green's functions are skew-hermitian,

\begin{equation}
S^{\gtrless}(1,1') = -S^{\gtrless}(1',1)^{\dagger} \rightarrow
  c^{\gtrless}_{n,m}(t,t') = c^{\gtrless *}_{m,n}(t',t).
\label{11}
\end{equation}

Therefore, it is sufficient to compute $c^{\gtrless}_{n,m}$ within the corresponding lower/upper triangle of the two-time plane \cite{DANIELEWICZ1984305, KOHLER1999123, Juchem_2004, dahlen2006propagating, dahlen2007solving, Stan_2009, PhysRevB.82.155108, Schenke:2005ry, Meirinhos_2022}. 
As a consequence, solely \cref{6.1} is employed to derive $c^{>}_{n,m}$ for $t$-direction, and \cref{6.2} is utilised to derive $c^{<}_{n,m}$ for $t'$-direction. Along the time-diagonal, instead, only $c^{<}_{n,m} (t,t)$ is evolved through a combination of both \cref{6.1,6.2}, with $c^{>}_{n,m}(t,t)$ being fixed by the equal-time (anti-)commutation relation for fermions and bosons, $c^{>}_{n,m}(t,t) \mp c^{<}_{n,m}(t,t) = \delta_{n,m} $. The resulting equations are

\begin{equation}
\begin{split}
\frac{\partial}{\partial t} c^{>}_{n,m}(t,t') + i \sum_{i}^{S}
  V_{\mathrm{eff}n,i}(t) c^{>}_{i,m}(t,t') &= I^{>}_{n,m 1}(t,t'), \\
\mp \frac{\partial}{\partial t'} c^{<}_{n,m}(t,t') \pm i \sum_{i}^{S}
  c^{<}_{n,i}(t,t') V_{\mathrm{eff}i,m}(t') &= I^{<}_{n,m 2}(t,t'), \\
\pm \frac{\partial}{\partial t} c^{<}_{n,m}(t,t) \mp i [c^{<}, V_{\mathrm{eff}}
   ]_{n,m}(t) = I^{<}_{n,m 1}(t,t)& - I^{<}_{n,m 2}(t,t) 
\label{12}
\end{split}
\end{equation}

with 

\begin{equation}
\begin{split}
V_{\mathrm{eff}n,m}(t) &= E_n \delta_{n,m} + \Sigma_{Hn,m},  \\
[c^{<}, V_{\mathrm{eff}} ]_{n,m}(t) &= \sum_{i}^{S} c^{<}_{n,i}(t,t) V_{\mathrm{eff}i,m}(t) - V_{\mathrm{eff}n,i}(t) c^{<}_{i,m}(t,t) , 
\end{split}
\label{12.1}
\end{equation}

and

\begin{equation}
\begin{split}
I^{>}_{n,m 1}(t,t') &= -\int_{t_0}^{t} d\bar{t} \sum_{i}^{S} \biggl[ \Sigma^{>}_{n,i}(t,\bar{t}) \mp \Sigma^{<}_{n,i}(t,\bar{t}) \biggr] c^{>}_{i,m}(\bar{t},t')  \\
&+ \int_{t_0}^{t'} d\bar{t} \sum_{i}^{S} \Sigma^{>}_{n,i}(t,\bar{t}) \biggl[c^{>}_{i,m}(\bar{t},t') \mp c^{<}_{i,m}(\bar{t},t') \biggr] , \\
I^{<}_{n,m 1}(t,t') &= \mp \int_{t_0}^{t} d\bar{t} \sum_{i}^{S} \biggl[ \Sigma^{>}_{n,i}(t,\bar{t}) \mp \Sigma^{<}_{n,i}(t,\bar{t}) \biggr] c^{<}_{i,m}(\bar{t},t')  \\
&\pm \int_{t_0}^{t'} d\bar{t} \sum_{i}^{S} \Sigma^{<}_{n,i}(t,\bar{t}) \biggl[c^{>}_{i,m}(\bar{t},t') \mp c^{<}_{i,m}(\bar{t},t') \biggr] , \\
I^{<}_{n,m 2}(t,t') &= \mp \int_{t_0}^{t} d\bar{t} \sum_{i}^{S} \biggl[c^{>}_{n,i}(t,\bar{t}) \mp c^{<}_{n,i}(t,\bar{t}) \biggr] \Sigma^{<}_{i,m}(\bar{t},t')  \\
& \pm \int_{t_0}^{t'} d\bar{t} \sum_{i}^{S} c^{<}_{n,i}(t,\bar{t}) \biggl[ \Sigma^{>}_{i,m}(\bar{t},t') \mp \Sigma^{<}_{i,m}(\bar{t},t') \biggr] .
\label{13}
\end{split}
\end{equation}

The solution to this highly coupled systems of ordinary integro-differential equations is obtained through numerical means. The predictor-corrector method known as the ``Heun method'' is used, a technique that has already been used successfully in similar studies \cite{DANIELEWICZ1984305, KOHLER1999123, Juchem_2004, dahlen2006propagating, dahlen2007solving, Stan_2009, PhysRevB.82.155108, Meirinhos_2022}. 
For a comprehensive overview of the numerical treatment, readers are directed to \cref{sec:Numerics}. 

\subsection{Features of the Kadanoff-Baym equations}\label{subsec:Advantages_of_the_KBE}

\textit{Spectral properties.}
It is well known, that the Green's function provides spectral information of the
particular quantum states of the system during the
full time evolution and not only in
equilibrium. The spectral coefficients are defined as $a_{n,m}(t,t') \defeq c^{>}_{n,m}(t,t') \mp c^{<}_{n,m}(t,t')$ and a central time, $\bar{t} = \frac{t+t'}{2}$, and a relative time, $\Delta{t} = t-t'$ \cite{KBBook, Danielewicz:1982kk, Juchem_2004} are introduced. A Wigner transform of the spectral coefficients in the relative time yields then 
\begin{equation}
\tilde{a}_{n,m}(\omega,\bar{t}) = \int d\Delta{t} \, e^{i \omega \Delta{t}} a_{n,m}\Bigl(\bar{t}+\frac{\Delta{t}}{2},\bar{t}-\frac{\Delta{t}}{2} \Bigr).
\label{14}
\end{equation}

The diagonal elements $\tilde{a}_{n,n}(\omega,\bar{t})$ are the corresponding spectral densities of the quantum state $n$. The interaction with the heat bath results in the emergence of finite self energies, which give rise to a shift of the peak (related to the real part of the retarded self energy) and a broadening of the spectral function (due to the imaginary part of the retarded self energy). The width, $\Gamma_{n,n}(\bar{t},\omega)$, reflects the inverse of the n-th state's lifetime \cite{KBBook, Danielewicz:1982kk, Juchem_2004}:
\begin{equation}
\begin{split}
& \Gamma_{n,m}(\bar{t},\omega) = -2 \, \rm{Im}(\Sigma^{\mathrm{ret}}_{n,m}(\bar{t},\omega)) = \int d\Delta{t} \, e^{i \omega \Delta{t}} \\
&\Bigl[  \Sigma^{>}_{n,m}\Bigl(\bar{t}+\frac{\Delta{t}}{2},\bar{t}-\frac{\Delta{t}}{2}\Bigr) 
\mp \Sigma^{<}_{n,m}\Bigl(\bar{t}+\frac{\Delta{t}}{2},\bar{t}-\frac{\Delta{t}}{2}\Bigr)  \Bigr],
\label{15.1}
\end{split}
\end{equation}
\begin{equation}
\begin{split}
& \rm{Re}(\Sigma^{\mathrm{ret}}_{n,m}(\bar{t},\omega)) = \Sigma_{H_{b,a}} -\frac{i}{2} \int d\Delta{t} \, e^{i \omega \Delta{t}} \Bigl[ sign(\Delta{t}) \\ &\Bigl( \Sigma^{>}_{n,m}\Bigl(\bar{t}+\frac{\Delta{t}}{2},\bar{t}-\frac{\Delta{t}}{2}\Bigr)
\mp \Sigma^{<}_{n,m}\Bigl(\bar{t}+\frac{\Delta{t}}{2},\bar{t}-\frac{\Delta{t}}{2}\Bigr) \Bigr) \Bigr].
\label{15.2}
\end{split}
\end{equation}
It is important to mention, that $\Gamma$ can become negative for bosonic particles, when $\Sigma^{<}_{n,m} > \Sigma^{>}_{n,m}$. This leads to exponential increase in the occupation number of the corresponding quantum state and is known as ``laser effect" \cite{Danielewicz:1982kk}.

\textit{Equilibration and Thermalization.}
We may now proceed to examine the process of equilibration and thermalization of the system under examination. In the limit of long times, the system will approach a fixed point of thermal equilibrium defined by the temperature of the surrounding environment, $T_{\text{bath}}$. 
Thereby the occupation numbers of the states, $c^{<}_{n,n}(t,t)$, should approach the Bose-Einstein or Fermi-Dirac distribution
\begin{equation}
\begin{split}
\lim_{t \rightarrow \infty} c^{<}_{n,n}(t,t) &= \int \frac{d\mathrm{\omega}}{2\pi} \, n_{\mathrm{B/F}}(T_{\mathrm{syst}}, \mu_{\mathrm{syst}}, \omega) \, \tilde{a}_{n,n}(\omega,\bar{t} ).
\label{16}
\end{split}
\end{equation}
It is crucial to utilize the complete spectral functions of the energy states, because an incomplete approach will inevitably fail to align with the actual bath temperature. The temperature, $T_{\mathrm{syst}}=T_{\mathrm{bath}}$, should be achieved in a manner that is independent of the interaction strength or initial values. However, it is important to note that the chemical potential in thermal equilibrium is contingent upon the total number of bosons or fermions. 
This is due to the fact, that the presented Kadanoff-Baym equations are conserving particle number (and energy/momentum for closed systems), if and only if the self energies are $\Phi$-derivable \cite{PhysRev.127.1391}. 

\textit{Decoherence.}
The results of numerical simulations consistently demonstrated \cite{2024PhLB..85138589N} that off-diagonal elements, if initially nonzero, approach zero over extended time periods. Consequently, it should be stressed that the phenomenon of quantum decoherence is inherently embedded within the aforementioned equations.

In our recent studies, it is observed that the off-diagonal elements initially exhibited a decaying trend, followed by the system's eventual attainment of thermal equilibrium. The equilibration timescales for all occupation numbers were found to be longer than in the case of initially non-zero coherences, occurring on a timescale that was determined by the coupling, temperature, and density. Decoherence provides the theoretical basis for the utilization of approximations such as assuming of homogeneous/purely diagonal Kadanoff-Baym equations \cite{KOHLER1999123,Juchem_2004,PhysRevA.72.063604,PhysRevD.102.016012} and quantum-kinetic Master equations \cite{PhysRevC.49.1693,PhysRevC.51.3232,Juchem_2004} that are less complex. \cref{subsec:Approximations_to_the_KBE} will provide a more detailed examination of these approximations.

\subsection{Eigenstates and eigenvalues of the density matrix at large times, Entropy and Heat}\label{subsection:Eigenstates_and_eigenvalues}

Having a direct access to the density matrix, defined as $\rho(r,r',t) = \pm i \, S^{<}(r,t,r',t)$ \cite{Danielewicz:1982kk,PhysRevA.81.022510}, over the entire time evolution range, one can examine specific thermodynamic properties. It may further be of interest to observe the potential change in the wave functions when the interaction is introduced. To do so, one must consider the eigenstates and eigenvalues of the density matrix,
\begin{equation}
\begin{split}
\int dr' \, \rho(r,r',t) \, \psi_n(r',t) = \xi_n(t) \, \psi_n(r,t).
\label{Entropy_Heat_0}
\end{split}
\end{equation}
It is important to note that due to the conservation of particle number, the trace of $\rho$ is constant over time. This leads to the conclusion that the sum of all eigenvalues at any given point in time is equal to the constant particle number, $\sum_n \xi_n(t) = N_{\rm{system}}$. Note, that the eigenstates of the single-particle Hamiltonian, $h_0$, and the initial density matrix, $\rho(t=0)$, do not have to be identical, 
\begin{equation}
\begin{split}
\phi_n \neq \psi_n(t=0),
\label{Entropy_Heat_1}
\end{split}
\end{equation}
but depend on the initial condition. This also directly follows from \cref{5}.
It is useful to write $\rho$ in its eigenbasis, where it is diagonal,
\begin{equation}
\begin{split}
\rho(r,r',t) = \sum^{S}_n \, \xi_n(t) \, \psi^{*}_n(r,t) \, \psi_n(r',t).
\label{Entropy_Heat_4}
\end{split}
\end{equation}
From a numerical point of view, it is obvious that the integral eigenvalue equation (\ref{Entropy_Heat_0}) does not require a solution within the boundaries of the position space. Instead, the already existing basis of the one-particle Hamiltonian $h_0$ is used to formulate the eigenvalue problem for the discrete matrix-valued $c^{<}_{n,m}(t,t)$ according to
\begin{equation}
\begin{split}
\sum^{S}_{m} c^{<}_{n,m}(t,t) \, \nu_{a, m}(t) = \xi_a (t) \, \nu_{a, n}(t) .
\label{Entropy_Heat_0.1}
\end{split}
\end{equation}
In light of the spectral theorem, the existence of a unitary matrix $U_{n,a}(t) = \nu_{a, n}(t)$ that diagonalizes $c^{<}_{n,m}$ is thus postulated. 
\begin{equation}
\begin{split}
c^{<}_{n,m}(t,t) = \sum^{S}_{a} \, U_{n,a}(t) \, \xi_a (t) \, U^{\dagger}_{a,m}(t).
\label{Entropy_Heat_0.2}
\end{split}
\end{equation}
Inserting (\ref{Entropy_Heat_0.2}) in (\ref{5}) yields then 
\begin{equation}
\begin{split}
\rho(r,r',t) &= \sum_{n,m,a}^{S} \, U_{n,a}(t) \, \xi_a (t) \, U^{\dagger}_{a,m}(t) \phi_{n}(r) \phi^{*}_{m}(r') \\
&= \sum_{a}^{S} \, \xi_a (t) \, \sum_{n}^{S} \, U_{n,a}(t) \, \phi_{n}(r) \, \sum_{m}^{S} \, U^{\dagger}_{a,m}(t) \, \phi^{*}_{m}(r').
\label{Entropy_Heat_0.3}
\end{split}
\end{equation}
This allows to ascertain the transformation of the initial free energy eigenbasis into the final (interacting) eigenbasis, when one compares \cref{Entropy_Heat_4} with \cref{Entropy_Heat_0.3},
\begin{equation}
\begin{split}
\sum_{n}^{S} \, U_{n,a}(t) \, \phi_{n}(r) = \psi_{a}(r,t).
\label{Entropy_Heat_0.4}
\end{split}
\end{equation}
It is also essential to consider the transformation of the associated spectral functions, as well as the widths and self energies, in order to maintain consistency with the new basis. The spectral functions in the interacting basis are obtained via
\begin{equation}
\begin{split}
\tilde{a}^{\rm{int}}_{n,m}(\omega,\bar{t}) = \sum^{S}_{a,b} \, U^{\dagger}_{n,a}(\bar{t}) \, \tilde{a}_{a,b}(\omega,\bar{t}) \, U_{b,m}(\bar{t}).
\label{Entropy_Heat_0.7}
\end{split}
\end{equation}
It is interesting to consider the thermalization process in this context. 
In \cref{16}, we have seen the relation that is obtained for the equilibrium occupation numbers in the (computational) eigenbasis of the Hamiltonian $h_0$. To connect them to the interacting basis, we can use \cref{Entropy_Heat_0.2,Entropy_Heat_0.7} resulting in
\begin{equation}
\begin{split}
\lim_{t \rightarrow \infty} \xi_n (t) &= \int \frac{d\mathrm{\omega}}{2\pi} \, n_{\mathrm{B/F}}(T_{\mathrm{syst}}, \mu_{\mathrm{syst}}, \omega) \, \tilde{a}^{\rm{int}}_{n,n}(\omega,\bar{t} ).
\label{Entropy_Heat_0.8}
\end{split}
\end{equation}
With this at hand, we can proceed by calculating the von Neumann entropy during the evolution of the system,
\begin{equation}
\begin{split}
S(t) \defeq - \int dr \, \int dr' \, \rho(r,r',t) \, \rm{ln}\bigl( \rho(r',r,t) \bigr ).
\label{Entropy_Heat_3}
\end{split}
\end{equation}
To efficiently calculate the trace, we use that the matrix logarithm of a diagonalizable matrix is given by 
$\rm{ln}(\rho) = U \, \rm{diag} \bigl( \rm{ln}( \xi_n) \bigr ) \, U^{-1}$,
where U is the matrix of eigenvectors of the matrix $\rho$.
It can be shown that the entropy can be calculated by using only the eigenvalues of $\rho$,
\begin{equation}
\begin{split}
S(t) &= - \int dr \, \int dr' \, \sum^{S}_n \xi_n(t) \, \psi^{*}_n(r,t) \, \psi_n(r',t) \\
& \, \quad \quad \sum^{S}_{m} \mathrm{ln}\bigl(\xi_m(t) \bigr) \, \psi^{*}_m(r',t) \, \psi_m(r,t) \\
&= - \sum^{S}_n \xi_n(t) \, \mathrm{ln}\bigl(\xi_n(t) \bigr).
\label{Entropy_Heat_5}
\end{split}
\end{equation}
In order to calculate the internal energy and the heat, it is preferable to use the eigenbasis of $h_0$ once more,
\begin{equation}
\begin{split}
U(t) &\defeq \int dr' \, \int dr \, \rho(r,r',t) \, h_0(r,r') \\
&= \int dr' \, \int dr \, \sum_{n,m,i}^{S} c^{<}_{n,m}(t,t) \phi_{n}(r) \phi^{*}_{m}(r') E_i \, \phi_{i}(r') \phi^{*}_{i}(r) \\ 
&= \sum^{S}_{i} c^{<}_{i,i}(t,t) \, E_i,
\label{Entropy_Heat_6}
\end{split}
\end{equation}
where we used the spectral representation of $h_0$. Similarly for the heat transferred to the system, we get
\begin{equation}
\begin{split}
\frac{d}{dt} U(t) &= \int dr' \, \int dr \, \frac{\partial \rho(r,r',t)}{\partial t} \, h_0(r,r') \\
&= \sum^{S}_{n} \frac{\partial c^{<}_{n,n}(t,t)}{\partial t} E_n \\
&= \sum^{S}_{n} E_n \biggl[ i [c^{<}, V_{\mathrm{eff}}]_{n,n}(t) \pm ( I^{<}_{n,n 1}(t,t) - I^{<}_{n,n 2}(t,t)) \biggr].
\label{Entropy_Heat_7}
\end{split}
\end{equation}
As one can see nicely, the transferred heat is depending on the collision integrals, which are of course the way to exchange energy/heat in this approach.

The Kadanoff-Baym approach allows for a more comprehensive understanding of the kinetic and thermodynamic properties of open quantum systems. It enables the definition of a total energy for the system, which encompasses both the interacting component and the correlations in accordance with the Galitskii-Migdal functional \cite{galitskii1958application}. For closed systems such investigations have been done in \cite{KOHLER1999123,Juchem_2004,PhysRevB.82.155108}, where the conservation of total energy was demonstrated explicitly through the application of the Kadanoff-Baym equations. The Galitskii-Migdal functional \cite{galitskii1958application} reads 
\begin{equation}
\begin{split}
E(t)\bigl[S \bigr] &= \pm i \, \int dr \, \Biggl[ \int dr' \, h_0(r,r') \, S^{<}(r',r,t,t^{+}) \\
&+ \frac{1}{2} \, \Sigma_{H}(r,r') \, S^{<}(r',r,t,t^{+}) \Biggr] + I_{coll_{1}}^{<}(r,r,t,t^{+}) \\
&= \sum^{S}_{i} \Biggl[ E_i \cdot c^{<}_{i,i}(t,t^{+})+ \biggl( \sum^{S}_{j} \frac{1}{2} \Sigma_{Hi,j} \cdot c^{<}_{j,i}(t,t^{+}) \biggr) \\
& \,\pm i \, \frac{1}{2} \, I^{<}_{i,i 1}(t,t^{+}) \Biggr] .
\label{Entropy_Heat_8}
\end{split}
\end{equation}
Again, because we have not varied the volume, e.g. the size of the box, or changed any parameter of the interacting Hamiltonian, e.g. the interacting potential $V_{\rm{int}}$, with time, the complete energy exchange between the system and the bath through scattering can be interpreted as heat flow.

\subsection{Approximations to the Kadanoff-Baym equations}\label{subsec:Approximations_to_the_KBE}

In order to gain a deeper understanding of the impact of decoherence on processes such as equilibration and thermalization, we will undertake a comparative analysis of the complete Kadanoff-Baym solution with various approximations. 

One approximation is to eliminate all possible correlations from the start, by forcing the matrix-valued coefficients $c^{\gtrless}$ to be diagonal in energy space and therefore decoherence is completely neglected,
\begin{equation}
\begin{split}
c^{\gtrless}_{n,m}(t,t') \rightarrow  \delta_{n,m} \, c^{\gtrless}_{n,m}(t,t') = c^{\gtrless}_{n}(t,t').
\label{Comparison_0}
\end{split}
\end{equation}
This approach gives rise to equations that bear a striking resemblance to the homogeneous Kadanoff-Baym equations in momentum space \cite{KOHLER1999123,Juchem_2004}. In accordance with this significant limitation, the self energies are also transformed into vector-like entities. Nevertheless, the two-time structure of the equations remains intact. The \cref{12,12.1,13} can be easily reduced by the use of \cref{Comparison_0} resulting in 
\begin{equation}
\begin{split}
\frac{\partial}{\partial t} c^{>}_{n}(t,t') + i 
  V_{\mathrm{eff}n}(t) c^{>}_{n}(t,t') &= I^{>}_{n 1}(t,t'), \\
\mp \frac{\partial}{\partial t'} c^{<}_{n}(t,t') \pm i 
  c^{<}_{n}(t,t') V_{\mathrm{eff}n}(t') &= I^{<}_{n 2}(t,t'), \\
\pm \frac{\partial}{\partial t} c^{<}_{n}(t,t) = I^{<}_{n 1}(t,t)& - I^{<}_{n 2}(t,t), 
\label{Comparison_1}
\end{split}
\end{equation}
with 
\begin{equation}
\begin{split}
V_{\mathrm{eff} b} &= E_b + \Sigma_{H b}, \\
\Sigma_{H_{b}} &= \sum_{j}^{B} e^{-i \epsilon_{j}0^{+}} n_{B}(\epsilon_{j}) V_{b,b,j,j}, \\
\Sigma^{<}_{b}(t,t') &= \sum_{n}^{S} \Biggl[ \sum_{j,k}^{B} e^{ i (\epsilon_{j}-\epsilon_{k})(t-t')} \, (1 \pm n_{B/F}(\epsilon_{j})) \, n_{B/F}(\epsilon_{k}) \\
& |V_{b,n,j,k}|^2 \, c^{<}_{n}(t,t')  \Biggr ], \\
\Sigma^{>}_{b}(t,t') &= \sum_{n}^{S} \Biggl[ \sum_{j,k}^{B} e^{- i (\epsilon_{j}-\epsilon_{k})(t-t')} \, (1 \pm n_{B/F}(\epsilon_{j})) \, n_{B/F}(\epsilon_{k}) \\
 & |V_{b,n,j,k}|^2 \, c^{>}_{n}(t,t') \Biggr ], \\
\label{Comparison_2}
\end{split}
\end{equation}
and
\begin{equation}
\begin{split}
I^{>}_{n 1}(t,t') &= -\int_{t_0}^{t} d\bar{t} \biggl[ \Sigma^{>}_{n}(t,\bar{t}) \mp \Sigma^{<}_{n}(t,\bar{t}) \biggr] c^{>}_{n}(\bar{t},t')  \\
&+ \int_{t_0}^{t'} d\bar{t} \Sigma^{>}_{n}(t,\bar{t}) \biggl[c^{>}_{n}(\bar{t},t') \mp c^{<}_{n}(\bar{t},t') \biggr] , \\
I^{<}_{n 1}(t,t') &= \mp \int_{t_0}^{t} d\bar{t} \biggl[ \Sigma^{>}_{n}(t,\bar{t}) \mp \Sigma^{<}_{n}(t,\bar{t}) \biggr] c^{<}_{n}(\bar{t},t')  \\
&\pm \int_{t_0}^{t'} d\bar{t} \Sigma^{<}_{n}(t,\bar{t}) \biggl[c^{>}_{n}(\bar{t},t') \mp c^{<}_{n}(\bar{t},t') \biggr] , \\
I^{<}_{n 2}(t,t') &= \mp \int_{t_0}^{t} d\bar{t} \biggl[c^{>}_{n}(t,\bar{t}) \mp c^{<}_{n}(t,\bar{t}) \biggr] \Sigma^{<}_{n}(\bar{t},t')  \\
& \pm \int_{t_0}^{t'} d\bar{t} c^{<}_{n}(t,\bar{t}) \biggl[ \Sigma^{>}_{n}(\bar{t},t') \mp \Sigma^{<}_{n}(\bar{t},t') \biggr] .
\label{Comparison_3}
\end{split}
\end{equation}
The most significant benefit of this approximation is the reduction in computational effort while maintaining the intrinsic non-Markovianity and two-time evolution, which enables the calculation of spectral functions.

The second, more substantial approximation is the utilization of a quantum-kinetic master equation. This also removes the two-time evolution and is entirely Markovian, akin to the Lindblad equation or the Boltzmann equation, as outlined in \cite{Juchem_2004}. A derivation similar to earlier ones \cite{PhysRevC.49.1693,PhysRevC.51.3232,Juchem_2004} is given in \cref{sec:Master}. We will only use the result, 
\begin{equation}
\begin{split}
&\frac{\partial}{\partial t} c^{<}_{b}(t) = \sum_{n}^{S} \sum_{j,k}^B \Bigl( (1 \pm n_{B/F}(\epsilon_{j})) \, n_{B/F}(\epsilon_{k}) \, c^{>}_{b}(t) \, c^{<}_{n}(t) - \\
&(1 \pm n_{B/F}(\epsilon_{k})) \, n_{B/F}(\epsilon_{j}) c^{>}_{n}(t) \, c^{<}_{b}(t) \Bigr) |V_{b,n,k,j}|^{2} \\ 
&  \frac{2}{(\epsilon_{j} -E_n -\epsilon_{k} +E_b)} \rm{sin}\biggl((\tilde{t}-t_0) (\epsilon_{j} -E_n -\epsilon_{k} +E_b)\biggr), \\
\label{Comparison_4}
\end{split}
\end{equation}
where 
\begin{equation}
\begin{split}
\tilde{t} \defeq \begin{cases} t_{\rm{max}} &\text{if $t > t_{\rm{max}}$,}\\ t &\text{if $t \le t_{\rm{max}}$} \end{cases}
\label{Comparison_4.1}
\end{split}
\end{equation}
needs to be defined by hand because otherwise the sinc function would diverge for $t \rightarrow \infty$, if $\epsilon_{j} -E_n -\epsilon_{k} +E_b=0$!
The optimal choice for $t_{\rm{max}}$ is one that yields a final width in the order of the temperature, as evidenced by the findings in \cite{PhysRevA.55.2902,PhysRevA.56.575,PhysRevC.49.1693,PhysRevC.51.3232}. Additionally, the necessity arises to ascertain that the sinc function possesses adequate width, thereby enabling sufficient scattering processes. This is imperative to ensure that the system can attain thermal equilibrium.

The following section will present a comparative analysis of the aforementioned approaches, elucidating the distinctions in transition dynamics towards a steady state or thermal equilibrium and the influence of decoherence. Additionally, a comparison will be made between the spectral functions derived from the diagonally forced solution and those obtained from the full Kadanoff-Baym solution.
 
\section{Results} \label{sec:results}

Our primary focus is to formulate a correct quantum-mechanical description of the formation of bound states \cite{Rais:2022,2024PhLB..85138589N,Rais:2025fps}.
 The present study investigates in detail the dynamics of bound states in an open quantum system coupled to a heat bath. 
 

In many phenomenological transport approaches one aims to describe the yields of light nuclei in heavy-ion collisions, clustering methods \cite{PURI2000245} and the coalescence model \cite{PhysRevC.59.1585} are applied, which allow, e.g. nucleons which are close in momentum (and position space) to form a light nucleus, e.g. a deuteron. This has been used in a wide range of experimental studies to extract so called ``coalescence factors" \cite{ALICE:2022veq} from the data or to employ coarse graining in transport approaches \cite{PhysRevC.105.044909,PhysRevC.99.014901}. 
A microscopic formation of bound states like light nuclei and the strict energy conservation are not touched in these approaches. In principle two free nucleons can never form a bound state like a deuteron. 
Microscopic deuteron formation can only be achieved by three-body reactions as elaborated a long time ago in \cite{DANIELEWICZ1991712}. In our picture, the particle undergoes continuous collisions, resulting in the perpetual creation and destruction of bound states.

In recent years, more elaborated transport codes have made significant strides in this area, as evidenced by the propagation of light nuclei as explicit degrees of freedom during the hadronic phase, rather than merely coalescing with the remaining nucleons in the final state after kinetic freeze-out \cite{KO2010253c,PhysRevC.104.034908,Glassel:2021rod,Coci:2023daq,Sun:2021dlz,PhysRevC.105.044909}.

The argument that weakly bound nuclei could not exist in such a hot environment \cite{VOVCHENKO2020135131} is replaced by the fact that they are, of course, constantly destroyed by collisions with other particles, but are also regenerated again \cite{Neidig:2021bal,Knoll:2024gaf}. In transport theory, this is referred to as single-particle collisional broadening \cite{KBBook,Danielewicz:1982kk}, which is named due to the fact that the occurrence of collisions results in a broadening of the spectral function. 
For the spectral function in \cref{14}, it follows from the Dyson equation for the retarded Green's function \cite{KBBook,Danielewicz:1982kk} and by using $\tilde{a}_{n,n}(\bar{t},\omega)) = -2 \, \rm{Im}(S^{\mathrm{ret}}_{n,n}(\bar{t},\omega))$, and \cref{15.1,15.2}, 
\begin{equation}
\begin{split}
\tilde{a}_{n,n}(\omega,\bar{t}) = \frac{\Gamma_{n,n}(\omega,\bar{t})}{\Bigl[\omega -E_n - \mathrm{Re}(\Sigma^{\mathrm{ret}}_{n,n}(\omega,\bar{t})) \Bigr]^{2} + \Bigl[\frac{\Gamma_{n,n}(\omega,\bar{t})}{2}\Bigr]^{2}}.
\end{split}
\label{Formation_of_bound_states_0}
\end{equation}
In order to see how fast a bound state may be created or destroyed in a hot medium and how long the bound state will live in the medium, one has to extract this information from the Green's functions using spectral functions.  

\subsection{Formation of bound states in the Kadanoff-Baym approach} \label{Formation of bound states in the Kadanoff-Baym approach}

For the first simulation, we consider fermions (more precisely nucleons) with the mass $\mu_{S}=m_N=938 \MeV$ in a potential $V_S(r)$, 
\begin{equation}
\begin{split}
V_S(r) = \begin{cases}
-V_{0} &\text{if $|r| \leq \frac{a}{2}$},\\
\,\,\,\,\, 0 &\text{if $|r| > \frac{a}{2}$,}\\
\,\,\,\,\, \infty &\text{if $|r| > \frac{L}{2}$,} 
\end{cases} 
\end{split}
\end{equation}
where $V_{0}=12.8 \MeV $ and  $a= 1.2 \fm$. The particles exist in a system of $L=20 \fm$ and the first 25 eigenstates are considered as the basis in the following, c.f $S=25$.
The central potential allows for one bound state ($n=0$) with a binding energy of $E_{\mathrm{binding}}=-2.23 \MeV$ (``deuteron" binding energy). The other 24 eigenstates are unbound with the energy cut-off at $E_{24}=319.23 \MeV$.

For the bath particles we assume bosons of the mass of $m_{B}=138 \MeV$ (pion mass), which are situated in a box of size $L_{\mathrm{bath}}=50 \fm$. This bath of bosons contains in its center the full system box. The bosons move freely in the box and we allow up to the first 30 quantum states, $B=30$.
The bath is at a temperature of $T_{\mathrm{bath}}= 100 \MeV$ and chemical potential of $\mu_{\mathrm{bath}} = 4 \MeV$. These parameters are typical for the kinetic freez-out in high energy heavy ion collisions.

The interaction, c.f \cref{Hamiltonian}, between the system and the bath particles is assumed to be
\begin{equation}
\begin{split}
V_{\rm{int}}(|r-r'|) = \lambda \, \delta(r-r').
\end{split}
\end{equation}
with $\lambda=0.42$. 

With this at hand, we solve now the set of equations in \cref{12} up to a tolerance of $\epsilon_{\rm{abs}}=10^{-5}$, defined in \cref{subsec:Adaptivetimestepping}.

\begin{figure}[]
    \hspace*{\fill}%
    \includegraphics[width=1.1\columnwidth,clip=true]{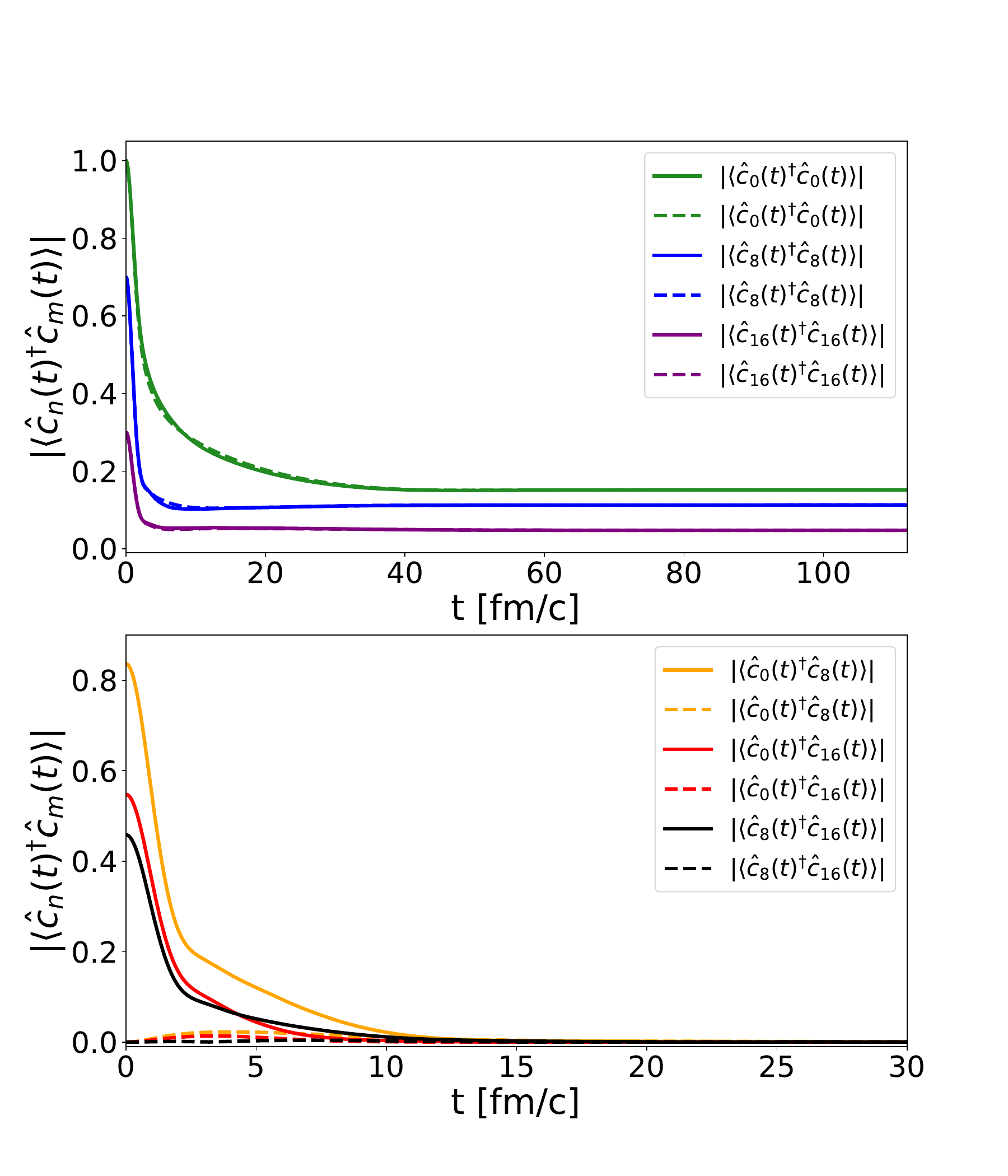}
    \hspace*{\fill}%

    \caption{
     The time evolution of the pure initial state \cref{Formation_of_bound_states_2} (full line) vs mixed state \cref{Formation_of_bound_states_1} (dashed line) plotted for the relevant matrix elements. In the top for the diagonal entries and in the bottom for the off-diagonal ones.
    }
    \label{fig:Formation_of_bound_states_0}
\end{figure}

\begin{figure}[]
    \hspace*{\fill}%
    \includegraphics[width=1.125\columnwidth,clip=true]{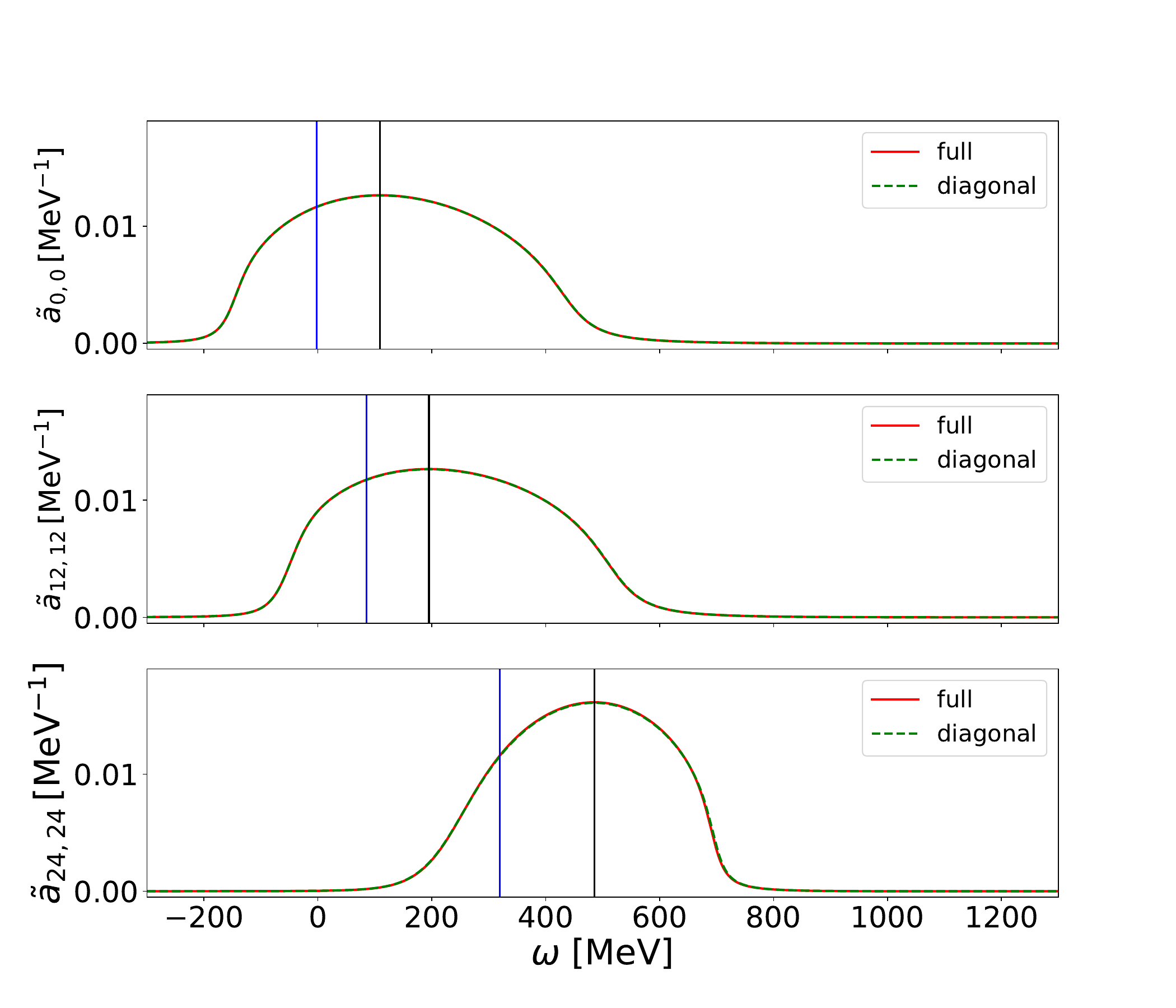}
    \hspace*{\fill}%

    \caption{
     The spectral functions of the states 0 (top), 12 (middle), and 24 (bottom) at $\bar{t}= 63 \fm$ of the full Kadanoff-Baym (full line) vs the diagonal approximation (dashed line) plotted for different states. The blue vertical line marks the bare on-shell energies of $h_0$, cf. \cref{eigenfunction}, and the black vertical line denotes the shifted peak energy for the diagonal/full Kadanoff-Baym equations. 
    }
    \label{fig:Formation_of_bound_states_1}
\end{figure}

To set the stage, we assume a mixed state as the initial state, where the bound state and in addition two higher states, namely the 8th and 16th states, 
\begin{equation}
\begin{split}
c^{<}_{0,0}(0,0)=1.0, \, c^{<}_{8,8}(0,0)=0.7, \, c^{<}_{16,16}(0,0)=0.3,
\end{split}
\label{Formation_of_bound_states_1}
\end{equation}
are initially populated.
As another choice, we initialize, in contrast to the mixed state before, a pure state 
\begin{equation}
\begin{split}
\ket{\phi}_{\text{pure}}&= \ket{\phi_{0}} + \sqrt{0.7} \ket{\phi_{8}} + \sqrt{0.3} \ket{\phi_{16}} \\[2mm]    &\rightarrow c^{<}(0,0) = \ket{\phi}_{\text{pure}}\bra{\phi}_{\text{pure}}.
\end{split}
\label{Formation_of_bound_states_2}
\end{equation}
A comparison of the time evolution of these two initial conditions are depicted in the upper part of \cref{fig:Formation_of_bound_states_0}. First we notice that both relax to the same equilibrium state in the long-time limit as it should be for the same temperature, particle number, and coupling constant. However, the time scales are also quite similar. For the pure state, we nicely see the evolving decoherence into a mixed state \cite{BRE02,Schlosshauer:2007:un}, before it fully thermalizes, as expected for open quantum system. We can see in addition that for the mixed state coherences build up first, as shown in the lower part of \cref{fig:Formation_of_bound_states_0}, which then of course also needs to vanish resulting in these nearly equal time scales.
In the next step, we want to compare the initial mixed state to different approximations of the Kadanoff-Baym equations, to see e.g., which impact the off-diagonal elements and the memory have on the thermalisation time or the life time of the states. Therefore, we look at the spectral functions of the full Kadanoff-Baym equations compared to the diagonal approximation in \cref{fig:Formation_of_bound_states_1}. The important takeaway is, that the width's $\Gamma_{n,n}$ and therefore the life times and the peak shifts of the states are quite similar although the Hartree self energy enters in a different manner. In the full Kadanoff-Baym equations, the Hartree self energy is matrix-valued, cf. \cref{10}, and therefore more terms are summed up and can contribute to the equation of motion than just the single diagonal term in the diagonal approximation in \cref{Comparison_2}.
Nevertheless, they do not seem to contribute much, indicating that the off-diagonal elements are close to zero. But in both cases the bound state has partially melted, in leading order due to the Hartree term dependent on the values of $\lambda$, $T_{\rm{bath}}$ and $\mu_{\rm{bath}}$, because there is now a nearly 70\% probability to find the ground state at a ``positive" energy, $\omega > 0$.

\begin{figure}[h]
  \centering
    \includegraphics[width=1.2\columnwidth,clip=true]{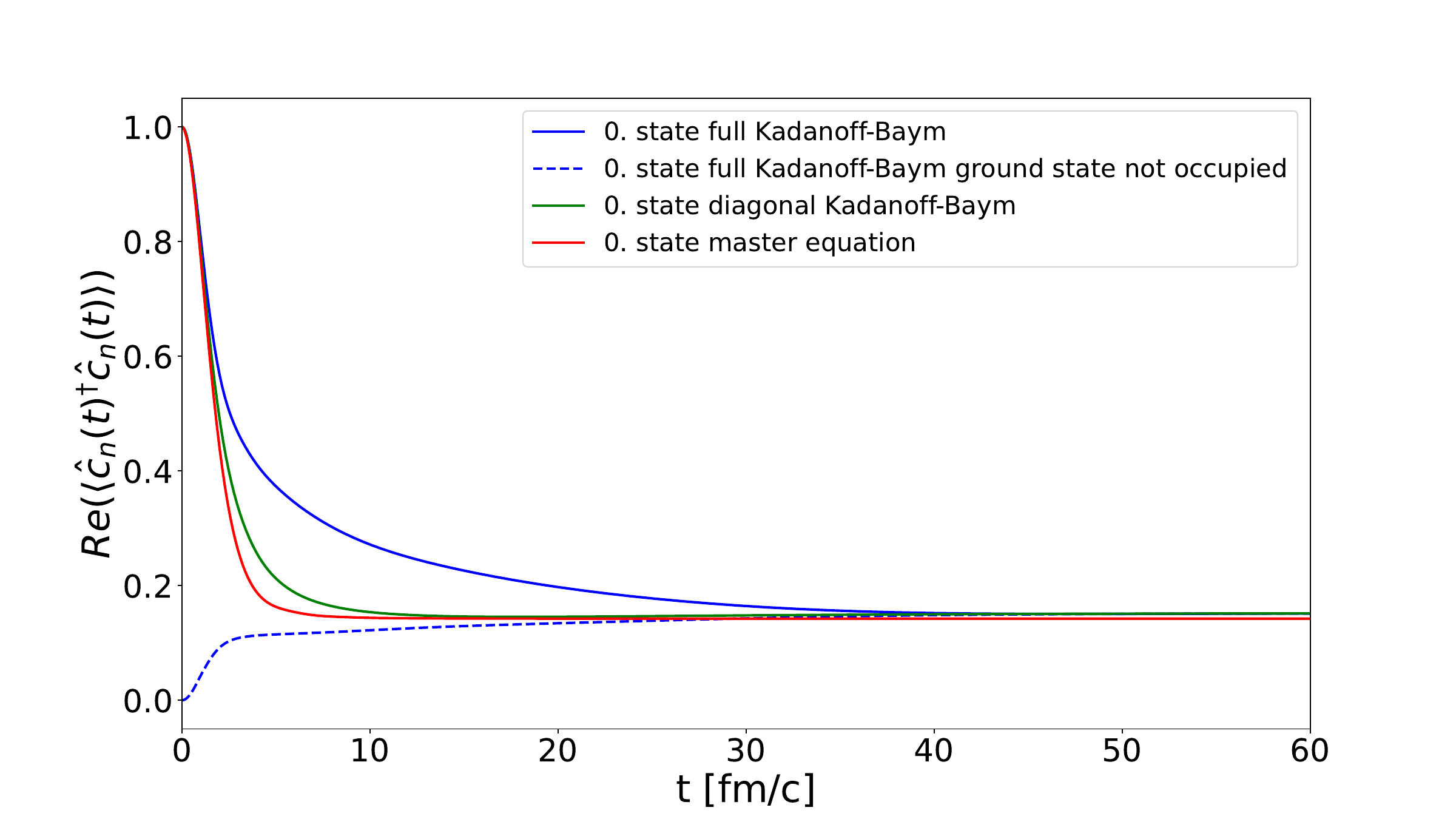}
    \caption{
     The time evolution of the bound state for all parameters identical for the full Kadanoff-Baym equation, the diagonal approximation and the plain master equation. For the full Kadanoff-Baym equation two different initial contitions were considered, namely \cref{Formation_of_bound_states_1} for the initially occupied state equivalent to the two approximations and \cref{Formation_of_bound_states_4}, were the ground state is not occupied initially.
    }
    \label{fig:Formation_of_bound_states_2}
\end{figure}

In \cref{fig:Formation_of_bound_states_2}, the central comparison of the approximations introduced in \cref{subsec:Approximations_to_the_KBE} with the full Kadanoff-Baym equation is depicted. One can observe that the relaxation of the full Kadanoff-Baym equation is by far the slowest for both initial conditions, \cref{Formation_of_bound_states_1} and when choosing
\begin{equation}
\begin{split}
c^{<}_{3,3}(0,0)=1.0, \, c^{<}_{8,8}(0,0)=0.7, \, c^{<}_{16,16}(0,0)=0.3,
\end{split}
\label{Formation_of_bound_states_4}
\end{equation}
with no initial bound state occupied, compared to the diagonal approximation and the simple master equation. 
This is due to the fact that decoherence is neglected in the quantum-kinetic master equation and in the diagonal Kadanoff-Baym equation, where one can observes a faster (nearly) exponential decay. This missing of the decoherence in the diagonal Kadanoff-Baym equation (and also in the quantum-kinetic master equation) prevents the formation of coherences, cf. bottom of \cref{fig:Formation_of_bound_states_0}, in the first $2-3 \fm/c$ for the case of initially off-diagonal elements unoccupied. These formed coherences, even if they are small ($\mathcal{O}(10^{-2})$), need to decohere during the evolution of the system, which delays the thermalization process. This phenomenon was also observed in the Lindblad formalism, cf. \cite{Rais:2025fps}, but is yet not fully understood.
Furthermore, the loss of memory in the master equation accelerates thermalization there slightly, but overall the impact of the memory seems not to be the leading reason for the delay, as the quantum-kinetic master equation and the diagonal Kadanoff-Baym equation differ only minimally compared to the full Kadanoff-Baym equation.
The omission of the Hartree shift in the quantum-kinetic master equation yields a very small variation in the final occupation numbers, which can be seen from the red line in \cref{fig:Formation_of_bound_states_2} that is slightly below the others. This issue can be addressed by using the shifted energy peaks of the spectral function in the master equation instead of the bare on-shell energies. 

\begin{figure}[]
    \hspace*{\fill}%
    \includegraphics[width=1.125\columnwidth,clip=true]{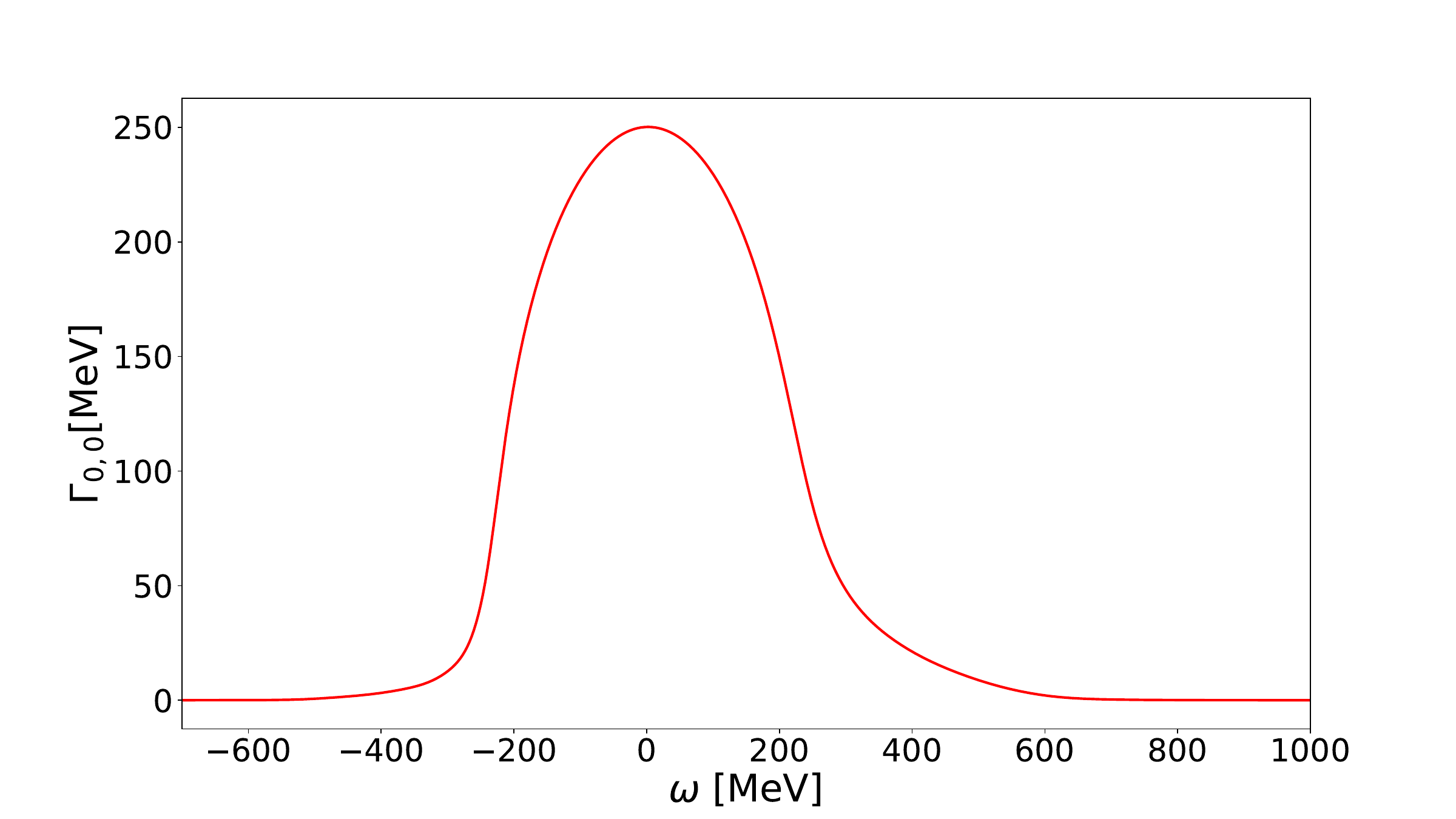}
    \hspace*{\fill}%

    \caption{
     The width $\Gamma_{0,0}(\omega,\bar{t}=52 \fm)$ as the imaginary part of the retarded self energy of the bound state.
    }
    \label{fig:Formation_of_bound_states_4}
\end{figure}

\begin{figure}[]
    \hspace*{\fill}%
    \includegraphics[width=1.125\columnwidth,clip=true]{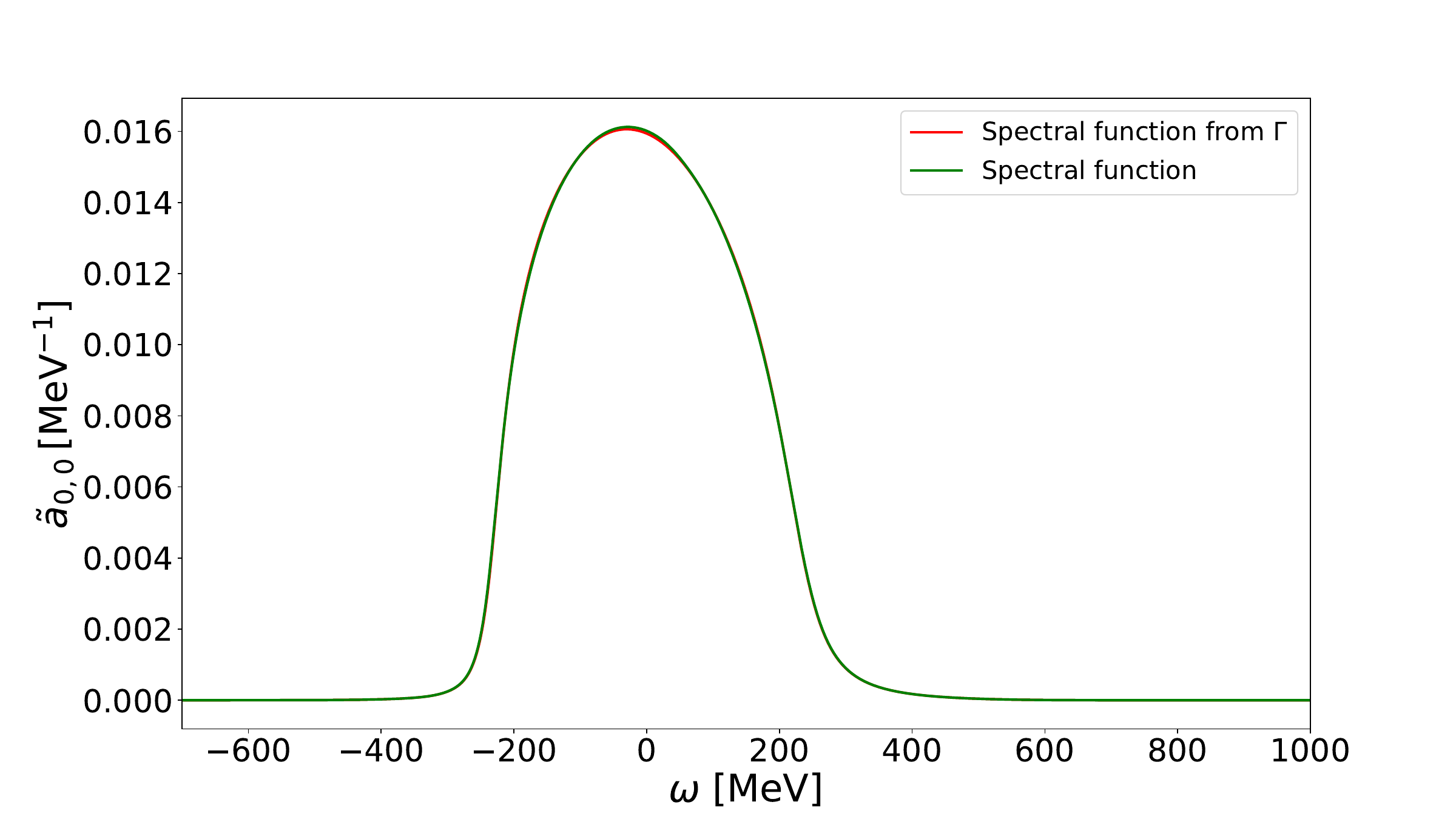}
    \hspace*{\fill}%

    \caption{
     The spectral function of the bound state compared for \cref{14} (green) and \cref{Formation_of_bound_states_0} (red).
    }
    \label{fig:Formation_of_bound_states_5}
\end{figure}

\begin{figure}[]
    \hspace*{\fill}%
    \includegraphics[width=1.125\columnwidth,clip=true]{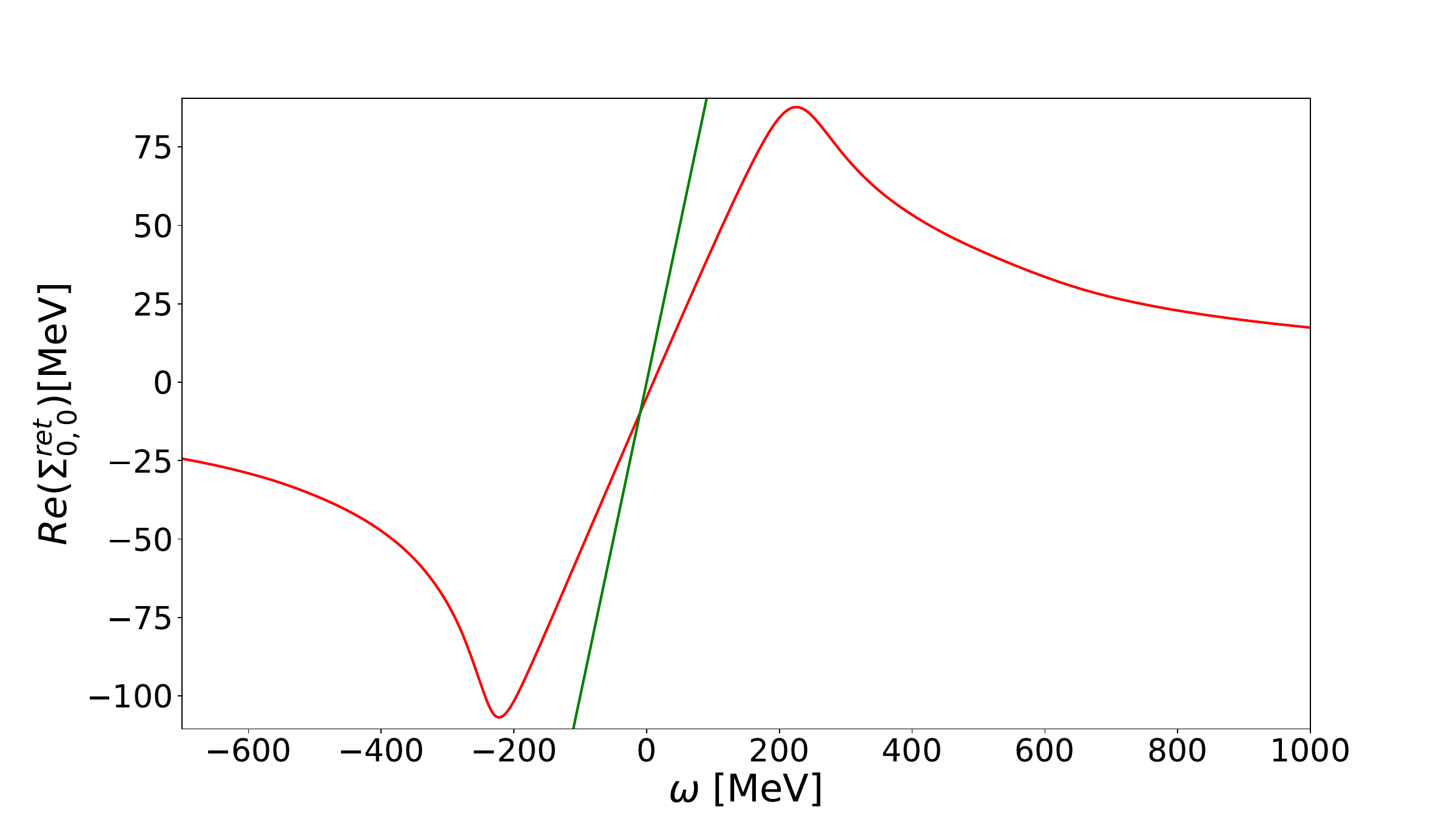}
    \hspace*{\fill}%

    \caption{
     The real part of the retarded self energy $\mathrm{Re}(\Sigma^{\mathrm{ret}}_{0,0}(\bar{t}=52 \fm,\omega)$ (red) of the bound state and $\omega - E_0$ (green).
    }
    \label{fig:Formation_of_bound_states_3}
\end{figure}

\begin{figure}[h]
\begin{center}
\includegraphics[width=1.1 \linewidth,clip=true]{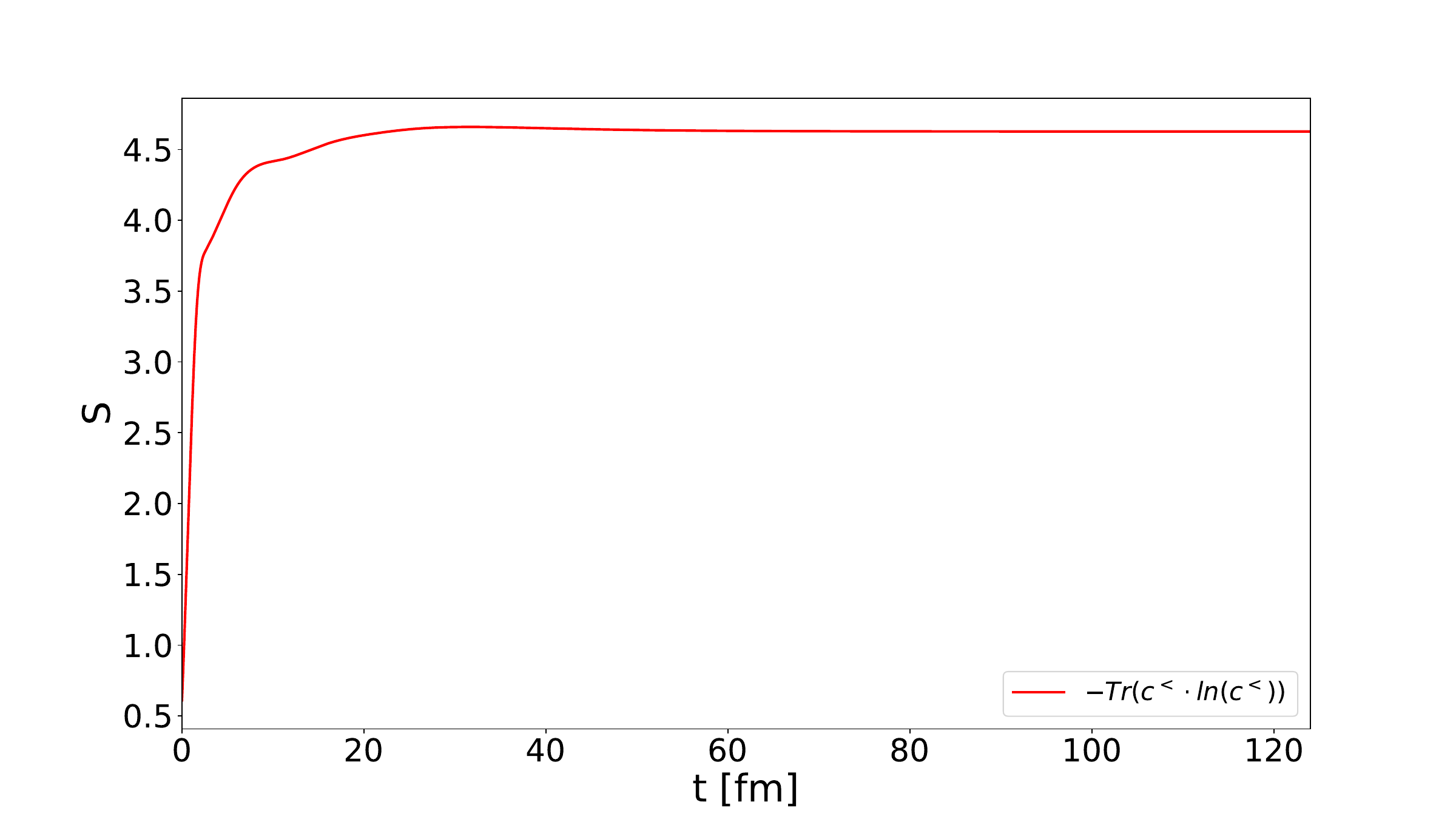}
\caption{The entropy evolution for the case of the initial mixed state in \cref{fig:Formation_of_bound_states_0}.}
\label{fig:Formation_of_bound_states_6}
\end{center}
\end{figure}

According to \cref{15.1,15.2}, we want to show exemplarily, without including the Hartree self energy here, how the collisions affect $\Gamma_{n,n}(\omega)$. In \cref{fig:Formation_of_bound_states_4} the width of the bound state is depicted for a fixed $\bar{t} = 52 \fm$. The $\Gamma$ has a maximum of $250 \MeV$, which is approximately half the width of the spectral function in \cref{fig:Formation_of_bound_states_5}. The position of the maximum of the spectral function and $\Gamma$ coincide, which is given by the minimum of the denominator of \cref{Formation_of_bound_states_0}, more precisely the zero of the first term $\omega -E_n - \mathrm{Re}(\Sigma^{\mathrm{ret}}_{n,n}(\bar{t},\omega)$. This is visualized in \cref{fig:Formation_of_bound_states_3}, where the real part of the retarded self energy is plotted together with the linear part $\omega - E_0$. The point where both functions intersect locates this maximum. 
The \cref{fig:Formation_of_bound_states_5} can also be seen as a consistency test, where \cref{14} is compared to \cref{Formation_of_bound_states_0}, showing up to some numerical uncertainties perfect agreement.
In accordance with \cref{subsection:Eigenstates_and_eigenvalues}, we will now investigate the thermodynamic properties such as entropy and energy of the (sub)system.
An illustrative example of the evolution of entropy is given in \cref{fig:Formation_of_bound_states_6}. The entropy function exhibits an increasing trend over time, and becomes constant when the system reaches equilibrium. The entropy is non-zero at the beginning, because the initial state is not a pure one, but given by \cref{Entropy_Heat_5} for the mixed initial condition \cref{Formation_of_bound_states_1}. Then the entropy rises rapidly but non-monotonously, with a peak at around $30 \fm$. It should be noted that this does not violate the second law of thermodynamics, as the analysis is conducted on an open (sub-)system and neglects the evolution of the bath itself.
Finally, we want to emphasize another fact, that the system has indeed reached (perfect) thermal equilibrium, additionally to \cref{16}, which has already been shown in \cite{2024PhLB..85138589N}. The new approach uses the Kubo-Martin-Schwinger (KMS) boundary condition \cite{doi:10.1143/JPSJ.12.570, PhysRev.115.1342,KBBook}, which is a fundamental restriction on any Green's function in (perfect) thermal equilibrium on the imaginary time axis, leading to the condition,
\begin{equation}
\begin{split}
\frac{c^{<}_{n,n}(\omega,\bar{t})}{c^{>}_{n,n}(\omega,\bar{t})}=  e^{-\beta_{\mathrm{syst}}(\omega- \mu_{\mathrm{syst}})},
 \label{Formation_of_bound_states_7}
\end{split}
\end{equation}
which can be analyzed numerically.
A short derivation of \cref{Formation_of_bound_states_7} is given in \cref{KMS}. 
The chemical potential and the temperature of the system appearing on the right-hand side can be fitted properly in a range of $\omega \in \{ \omega: c^{>}_{n,n}(\omega,\bar{t})\neq 0 \}$. 

\begin{figure}[h]
\begin{center}
\includegraphics[width=1.1 \linewidth,clip=true]{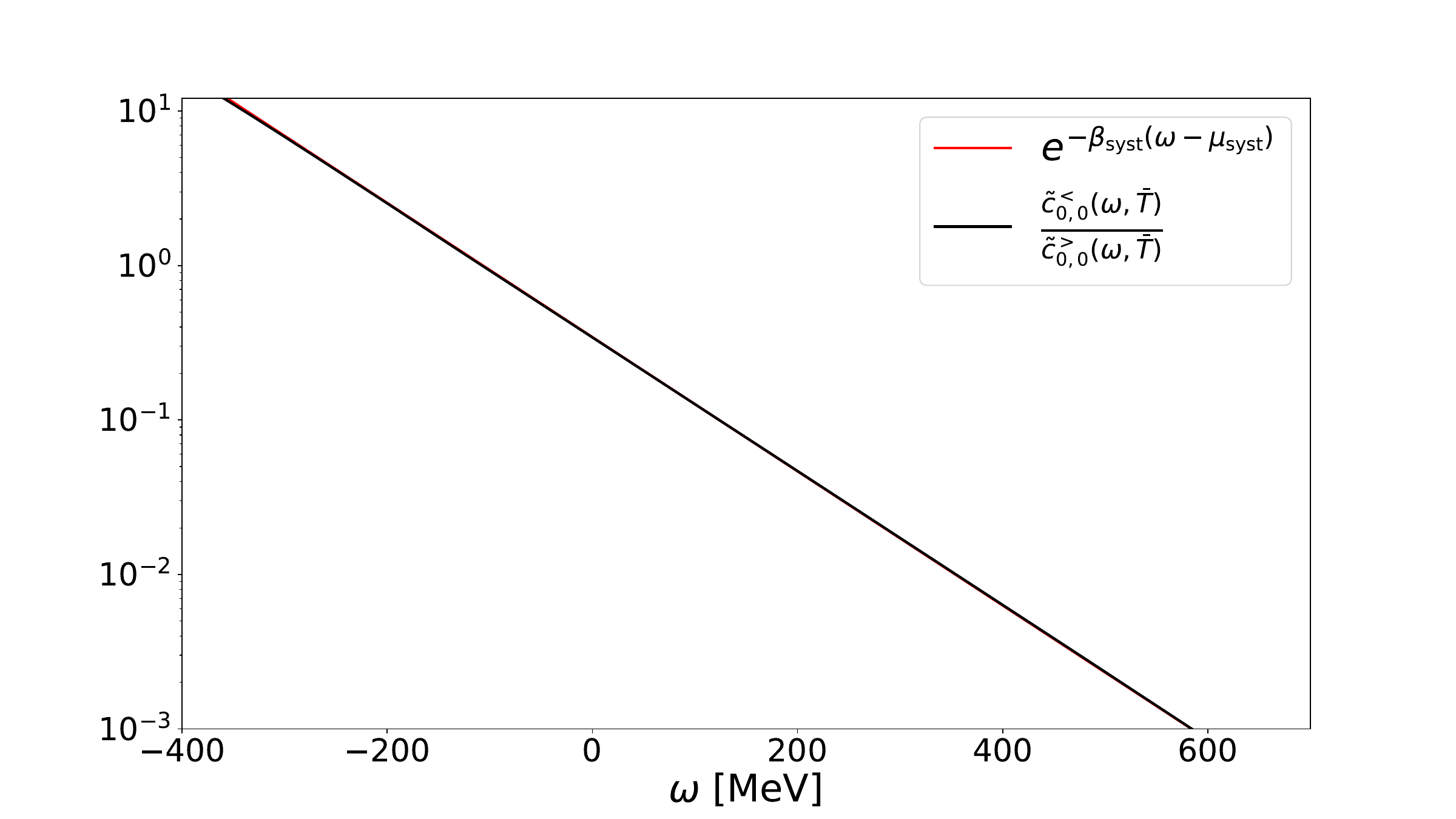}
\caption{The KMS condition of the initial pure state in \cref{fig:Formation_of_bound_states_0} used to extract the temperature $T_{\mathrm{syst}} \approx 99.993 \MeV$ ($T_{\mathrm{bath}} = 100 \MeV$) and chemical potential $\mu_{\mathrm{syst}} \approx -106.659 \MeV$ at $\bar{t}=63 \fm$.}
\label{fig:Formation_of_bound_states_8}
\end{center}
\end{figure}

We show this for the simulation done in \cref{fig:Formation_of_bound_states_0} exemplarily in \cref{fig:Formation_of_bound_states_8}, where we find a temperature of the system, that is the same as the temperature of the bath. The difference from \cite{2024PhLB..85138589N} is, that the chemical potential is now increased because there are now two particles in the system, not only one. 
A great advantage, when using \cref{Formation_of_bound_states_7} instead of \cref{16}, is that no integrals of the spectral function with the Bose/Fermi distribution need to be calculated for the fit of $T_{\mathrm{syst}}$ and $\mu_{\mathrm{syst}}$. Also the number of points in the fit is restricted to the number of basis states $S$ in \cref{16}, which is usually much less than the points in the interval $\omega \in \{ \omega: c^{>}_{n,n}(\omega,\bar{t})\neq 0 \}$ therefore, the accuracy of the fit is drastically improved through the usage of the KMS condition.

\subsection{Deuteron configuration} \label{Deuteron configuration}

In this subchapter, we will direct our attention to a more detailed examination of the system configuration tuned for deuterons. Our objective is to analyze the dependence of formation time and life time on temperature and coupling strength. Therefore, we have to set more realistic parameters, starting with the mass $\mu_S = \frac{m_N^2}{m_N + m_N}=\frac{m_N}{2} = 469 \MeV$. Furthermore, we increase the temperature moderately to $T_{\mathrm{bath}} = 120 \MeV$ and expand the system to $L = 27.5 \fm$. To achieve the deuteron binding energy of the ground state, the depths of the potential has to be changed to $V_0 = 17.8 \MeV$. The other parameters for the bath particle are kept as in \cref{Formation of bound states in the Kadanoff-Baym approach}. As initial conditions, we will compare the occupied and unoccupied ground state with only one particle in the system,
\begin{equation}
\begin{split}
c^{<}_{0,0}(0,0)=1.0, 
\end{split}
\label{Deuteron_configuration_0}
\end{equation}
and
\begin{equation}
\begin{split}
c^{<}_{8,8}(0,0)=0.7, \, c^{<}_{16,16}(0,0)=0.3,
\end{split}
\label{Deuteron_configuration_1}
\end{equation}
were everything else was set to zero.

To better understand the dependence of the formation time on the coupling constant $\lambda$, we simulated for three different values, $\lambda = 0.42$, $\lambda = 0.21$, and $\lambda = 0.105$.
\begin{figure}[h]
\begin{center}
\includegraphics[width=1.05 \linewidth,clip=true]{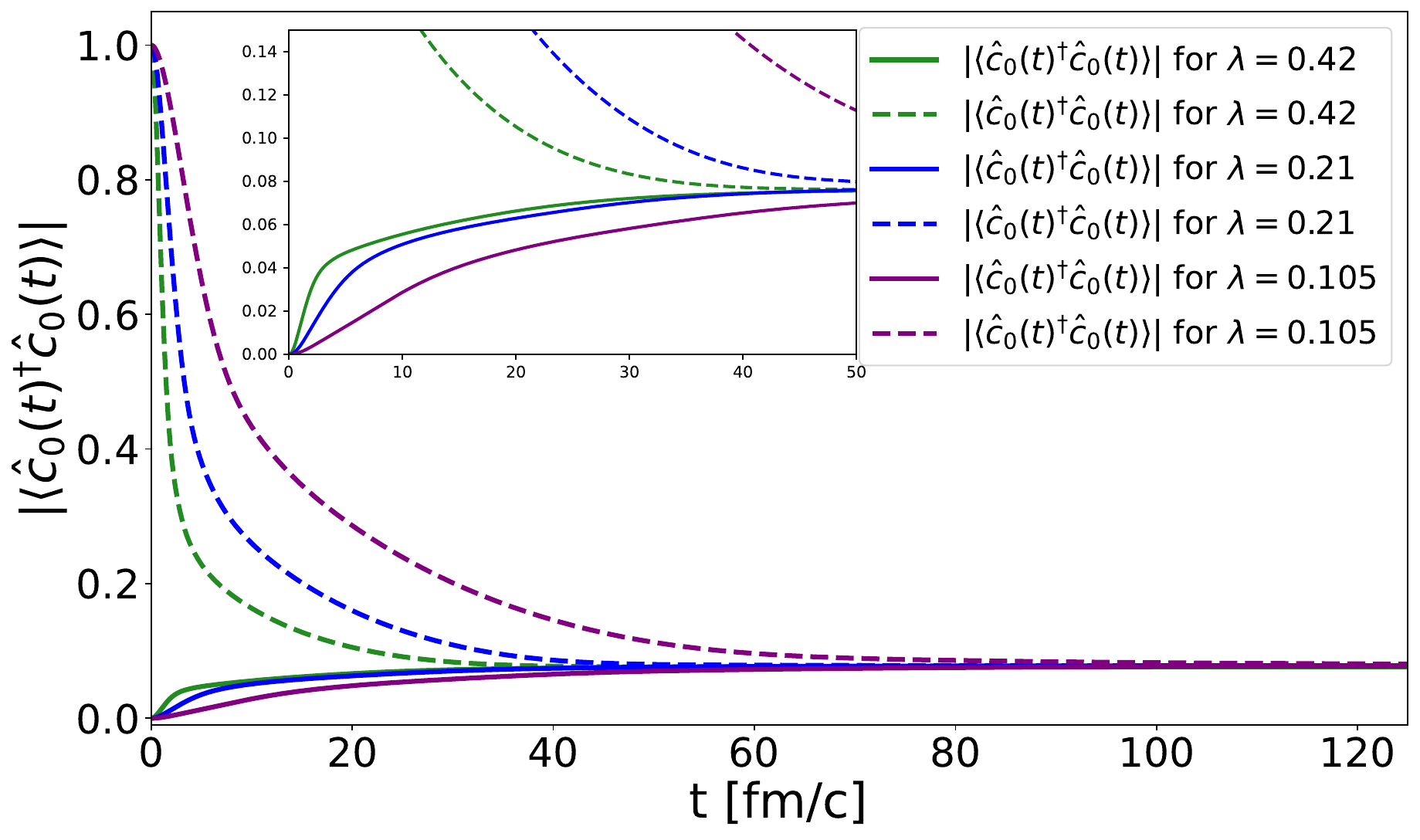}
\caption{The time evolution of the ground state for different coupling strengths and two different initial conditions, \cref{Deuteron_configuration_0,Deuteron_configuration_1}.}
\label{fig:Deuteron_configuration_0}
\end{center}
\end{figure}
In \cref{fig:Deuteron_configuration_0} the time evolution of the ground state is analyzed in dependence of the coupling strength. In addition, the clear trend of larger coupling leads to a qualitatively faster equilibration. The time scale for this to happen is not related to the coupling strength via a simple power law, e.g. $\propto \lambda^2$, as the coupling enters the dispersive self energy. Interesting to observe is that for the initial condition \cref{Formation_of_bound_states_4}, with the unoccupied ground state, a fast rise to a near the final equilibrium value takes place within the first $5-10 \fm$ (not for the very low coupling constant) followed by a quite longer time to fully reach the equilibrium value.

As a second step, we will look at the change in the spectral functions when the coupling is modified. 
\begin{figure}[h]
\begin{center}
\includegraphics[width=1.1 \linewidth,clip=true]{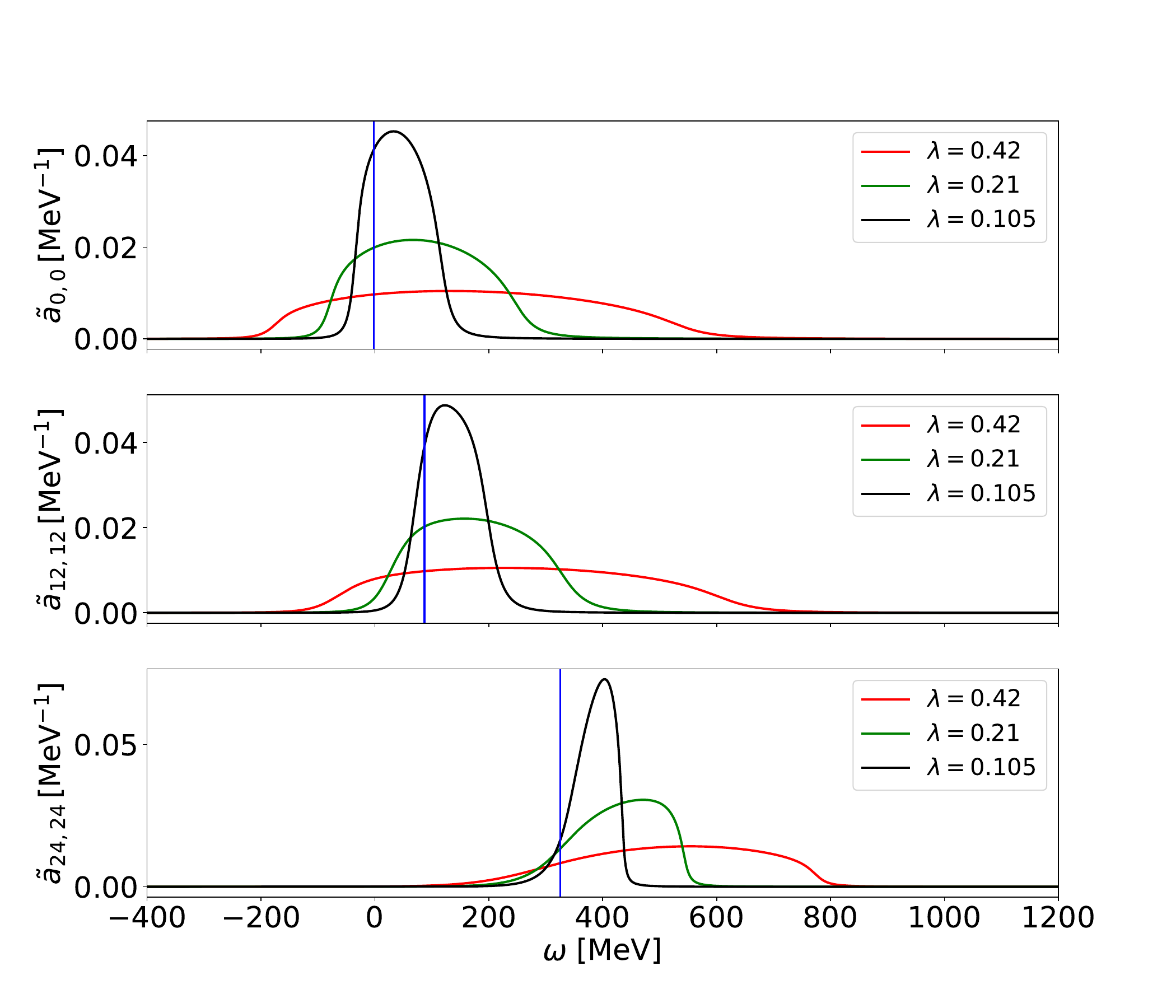}
\caption{The spectral functions of the corresponding states for different coupling strengths at $\bar{t}=63 \fm$. The vertical blue line again denotes the bare on-shell energy of the mode.}
\label{fig:Deuteron_configuration_1}
\end{center}
\end{figure}
In \cref{fig:Deuteron_configuration_1}, the spectral functions are depicted for the different coupling constants. One can clearly see how the quasi-particle peak melts when increasing the coupling strength and how the pole is shifted to higher values in $\omega$. The broadening effect is here, of course, due to the dependence $\Gamma \propto \Sigma^{\mathrm{ret}} \propto \Sigma^{>}\mp \Sigma^{>} \propto \lambda^2$, cf. \cref{15.1}. But since $\Gamma$ is actually $\Gamma(\omega)$, the full width at half maximum (FWHM) is no longer constant and, therefore, the approximation by setting the FWHM to $\Gamma$ no longer holds.
Because in all cases the spectrum is shifted to higher values of $\omega$, the ground state is only bound by a finite probability significantly less than 50\%, we want to investigate how this effects the wave functions of the system. 
In coalescence approaches \cite{PhysRevC.59.1585}, the wave function of the bound state is needed to somehow project out parts of the wave functions of the constituent particles, to obtain particle yields at the end. 
To check the validity of using the bare wave function of the bound state instead of the wave function that includes the interaction with the bath, we will compare them in the following. 

\begin{widetext}

\begin{figure}[h] 
\begin{center}
\includegraphics[width=1.0 \linewidth,clip=true]{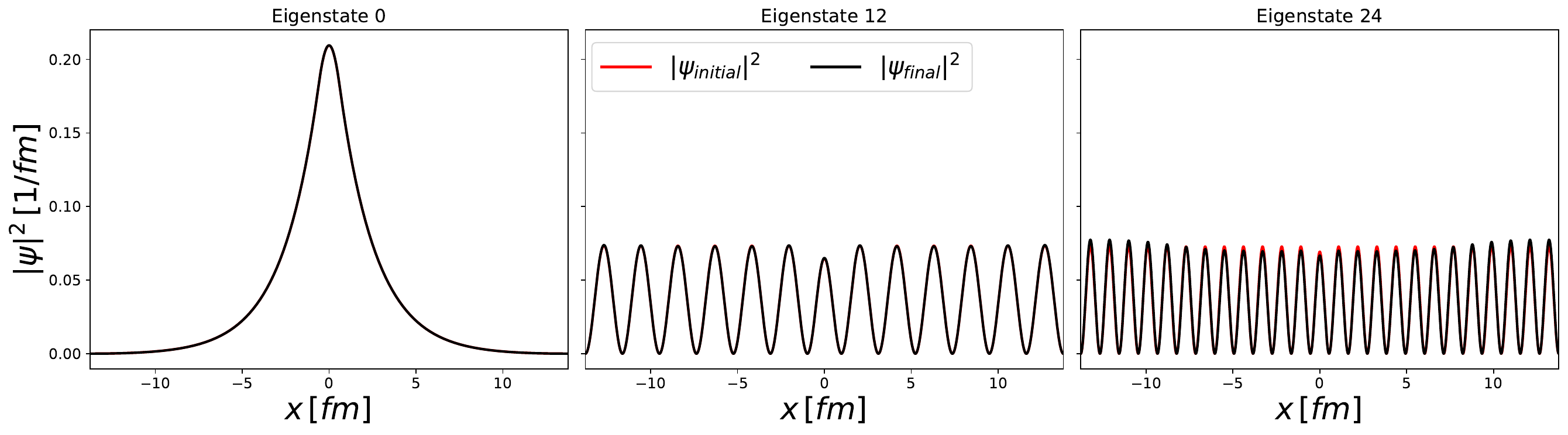}
\caption{The wave functions, $\Psi_{\mathrm{initial}}=\phi$ and $\Psi_{\mathrm{final}}=\psi(\bar{t})$ (c.f \cref{eigenfunction}), of the ground state and two higher exited states at $\bar{t}=124 \fm$ and $\lambda = 0.42$.}
\label{fig:Deuteron_configuration_2}
\end{center}
\end{figure}

\end{widetext}

In \cref{fig:Deuteron_configuration_2} the absolute square of the wave function is shown for the initial (non-interacting eigenfunctions of $h_0$) and final (interacting eigenfunctions of $\rho(t)$) wave functions. In the case of a finite set of eigenstates there seems to be only minor changes in the higher modes, but the ground state in particular does not change its shape although its energy distribution according to the spectral function is clearly shifted to positive values in $\omega$. 

\begin{figure}[h]
\begin{center}
\includegraphics[width=1.05 \linewidth,clip=true]{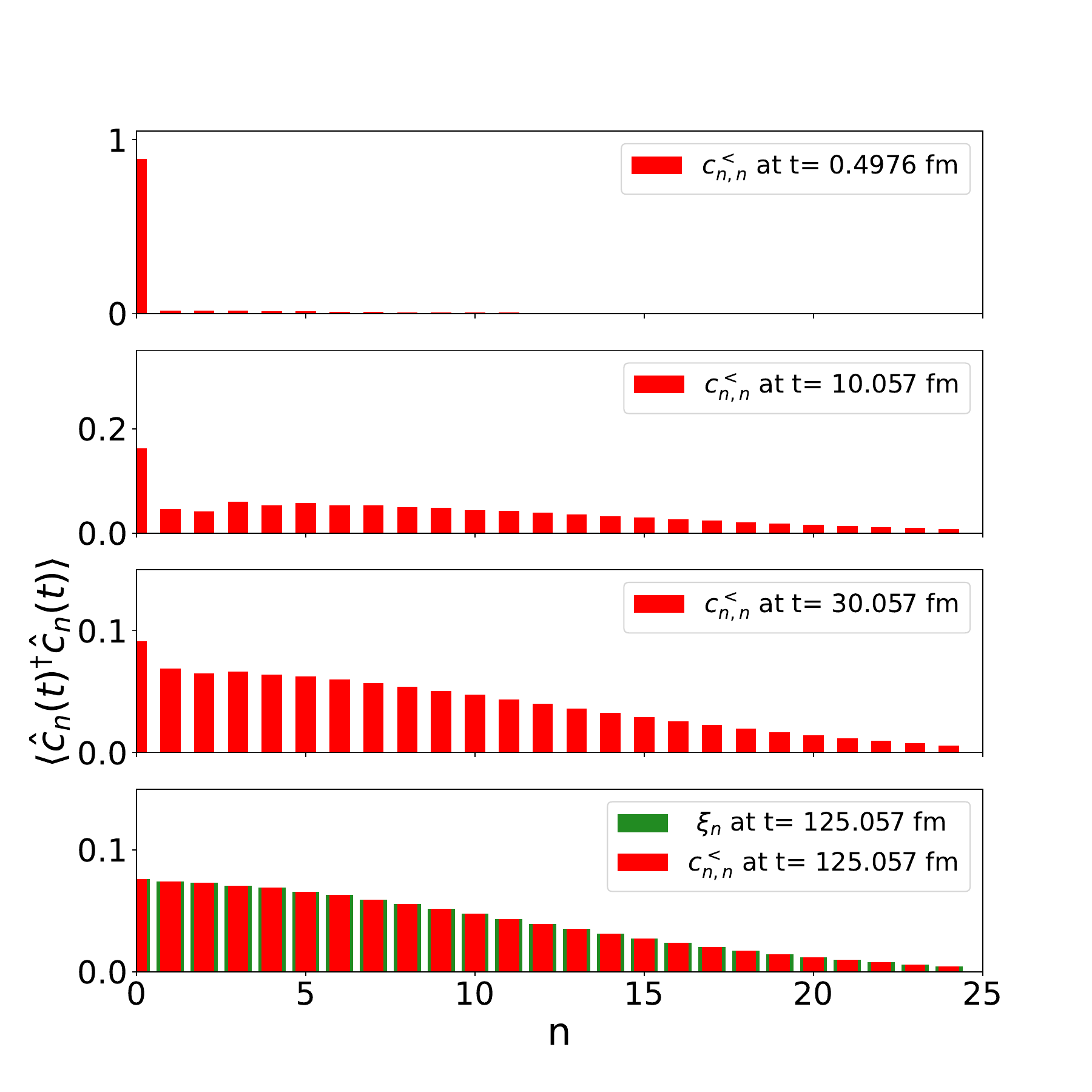}
\caption{The occupation number of the various quantum states at different times for the initial condition \cref{Deuteron_configuration_0}.}
\label{fig:Deuteron_configuration_3}
\end{center}
\end{figure}

At the end of this subsection, we want to apply the procedure of \cref{subsection:Eigenstates_and_eigenvalues}, cf. \cref{Entropy_Heat_0.8}, to show that indeed the deuteron-like system has fully thermalized. Therefore, the occupation numbers of the quantum states are shown in \cref{fig:Deuteron_configuration_3} at different times. At the final time, the eigenvalues $\xi_n$ are plotted in comparison, which shows in this case no significant deviation.

\subsection{Bound states in 3 dimensions} \label{Bound states in 3 dimensions}

The ansatz of expanding the field operator in an energy eigenbasis can in principle be applied straightforwardly in higher dimensions. In addition, the same numerical algorithm can be applied up to some small changes. This is possible due to the mapping of quantum numbers onto a single ``super" index independent of the underlying symmetry, e.g. spherical or Cartesian. For a detailed explanation, see \cref{3 dim extension}. 
The numerical difficulties arise because of the appearance of degeneracies that naturally scale with the dimensionality. This means that lower energies are achieved with the same basis size as in the one-dimensional case, which makes effects of higher excitations unattainable and restricts us to lower energetic modes.
\begin{table}[h]
\begin{center}
\begin{tabular}{|c|c|c|c|c|}
\hline
$n_{\mathrm{super}}$ & n & l & m & $E_{n,l,m} \, \mathrm{in} \, [\MeV]$\\
\hline
0 & 1 & 0 & 0 & -2.23 \\
\hline
1 & 2 & 0 & 0 & 7.85 \\
\hline
2 & 2 & 1 & -1 & 8.37 \\
\hline
3 & 2 & 1 & 0 & 8.37 \\
\hline
4 & 2 & 1 & 1 & 8.37 \\
\hline
5 & 3 & 0 & 0 & 25.27 \\
\hline
6 & 3 & 1 & -1 & 24.7 \\
\hline
7 & 3 & 1 & 0 & 24.7 \\
\hline
8 & 3 & 1 & 1 & 24.7 \\
\hline
9 & 3 & 2 & -2 & 13.78 \\
\hline
10 & 3 & 2 & -1 & 13.78 \\
\hline
11 & 3 & 2 & 0 & 13.78 \\
\hline
12 & 3 & 2 & 1 & 13.78 \\
\hline
13 & 3 & 2 & 2 & 13.78 \\
\hline
14 & 4 & 0 & 0 & 51.46 \\
\hline
\end{tabular}
\caption{The mapping of the spherical 3 dimensional symmetric system up to the s - state $\psi_{4,0,0}$. The energies of the states are related to the potential \cref{Bound_states_in_3D_3} and its choosen values. }
\label{fig:mapping of quantum numbers 3D spherical}
\end{center}
\end{table}
In \cref{fig:mapping of quantum numbers 3D spherical} an exemplary mapping of the states for a spherically symmetric 3-dimensional system is shown. Usually the energy is degenerate and depends only on the quantum numbers $n$ and $l$ or in special cases only on $n$, cf. hydrogen atom. 

One of these changes from 1 to 3 dimensions is to compute the transition amplitudes,
\begin{equation}
\begin{split}
 V_{b,n,j,k} = \int d^3\vec{r}  \int d^3\vec{r'} \phi^{*}_{b}(\vec{r}) \phi_{n}(\vec{r}) V_{\rm{int}}(|\vec{r}-\vec{r'}|) \tilde{\phi}_{j}(\vec{r'})\tilde{\phi}^{*}_{k}(\vec{r'}),
 \label{Bound_states_in_3D_0}
\end{split}
\end{equation}
where the indices $b$, $n$, $j$ and $k$ are now ``super" - indices. The calculation of the transition amplitudes necessitates the summation of a considerable number of lattice points, given by  $N_r ^2 \cdot N_\theta ^2 \cdot N_\phi ^2$, at lowest order to calculate one of $S^2 \cdot B^2$ transition amplitudes. 
Possible is a screened Coulomb potential,
\begin{equation}
\begin{split}
V_{\rm{int}}(|\vec{r}-\vec{r'}|) = \frac{\lambda}{|\vec{r}-\vec{r'}| +d},
 \label{Bound_states_in_3D_1}
\end{split}
\end{equation}
which is not considered in the following or a standard Gaussian potential,
\begin{equation}
\begin{split}
V_{\rm{int}}(|\vec{r}-\vec{r'}|) = \frac{\lambda}{\sqrt{(2 \pi)^3 }\sigma^3} e^{-|\vec{r}-\vec{r'}|^2/(2 \sigma^2)}.
\label{Bound_states_in_3D_1.1}
\end{split}
\end{equation}
Improvements to the runtime can be achieved through the assumption of a local, or so-called s-wave interaction, 
\begin{equation}
\begin{split}
V_{\rm{int}}(|\vec{r}-\vec{r'}|) =  \lambda \, \delta^{(d)}(|\vec{r}-\vec{r'}|),
 \label{Bound_states_in_3D_2}
\end{split}
\end{equation}
which results in only $N_r \cdot N_\theta \cdot N_\phi$ points and were already used in our earlier work \cite{2024PhLB..85138589N}. As an important remark, we want to point out here that in \cref{Bound_states_in_3D_1.1,Bound_states_in_3D_2} the coupling strength $\lambda$ becomes dimension-ful with $[\lambda] = l^{d-1}$.

 We now want to have a look at the bound-state formation in 3 dimensions. Similarly to the one-dimensional scenario, we apply here a ``spherical box",
\begin{equation}
\begin{split}
V_S(\vec{r}) \defeq \begin{cases}
-V_{0} &\text{if $|\vec{r}| \leq \frac{a}{2}$},\\
\,\,\,\,\, 0 &\text{if $|\vec{r}| > \frac{a}{2}$,}\\
\,\,\,\,\, \infty &\text{if $|\vec{r}| > R,$} 
\end{cases} 
 \label{Bound_states_in_3D_3}
\end{split}
\end{equation}
were we have choose $V_0 = 292.6 \MeV$, $a= 1.2 \fm$ and $R=10 \fm$ to mimic one bound state of about $-2.2 \MeV$, which is roughly the binding energy of the deuteron. We are again restricted to finite- and constant-volume calculations, which of course can not approximate the real expanding fireball in heavy-ion collisions. The (fermionic) basis sizes are chosen to be $S=30$ for the system, with wave functions according to \cref{Extension_3D_3}.

For the mass in the corresponding Schrödinger equation the reduced mass of the two nucleons is chosen to $\mu_S = 496 \MeV$. The coupling strength according to \cref{Bound_states_in_3D_2} is chosen as $\lambda = 389.3 \fm^2$ and the bath temperature remains at $100 \MeV$. 

For the (bosonic) bath particles we assume a mass of $m_B = 469 \MeV$ (increased to reach higher energies with the bath states, in order to enable more scattering processes), a chemical potential of $\mu_{\mathrm{bath}} = -10.0 \MeV$ and a basis size of $B=27$. We found, that it is unsuitable to put the bath particles in a larger spherical box around the system. The reason for this phenomenon is the orthogonality of the spherical harmonics, which results in selection rules that are so restrictive with regard to the transition amplitudes that only a small number of transitions are permitted. Consequently, equilibration and, by extension, thermalization are both impossible. 

To at least partly resolve this issue, the symmetry of the total system has to be broken. It was hypothesized that the spherical system would be placed within a rectangular bath, with the lengths of the sides, $L_x=26\fm, \, L_y=24\fm$ and $L_z=28 \fm$, which could vary if necessary. This would result in the dissolution of the degeneration energy eigenstates of the system. 

In light of the aforementioned background, the present study aims to investigate the formation of a deuteron-like state in a bath of other, heavier mesons. First, we simulate how one initially existing deuteron, $c_{0,0}^{<}(0,0) = 1$ and all other zero, can equilibrate within a bath of around $\mathcal{O}(100)$ particles, with an accuracy up to a tolerance of $\epsilon_{\rm{abs}}=2 \cdot 10^{-5}$. 
\begin{figure}[h]
\begin{center}
\includegraphics[width=1.1 \linewidth,clip=true]{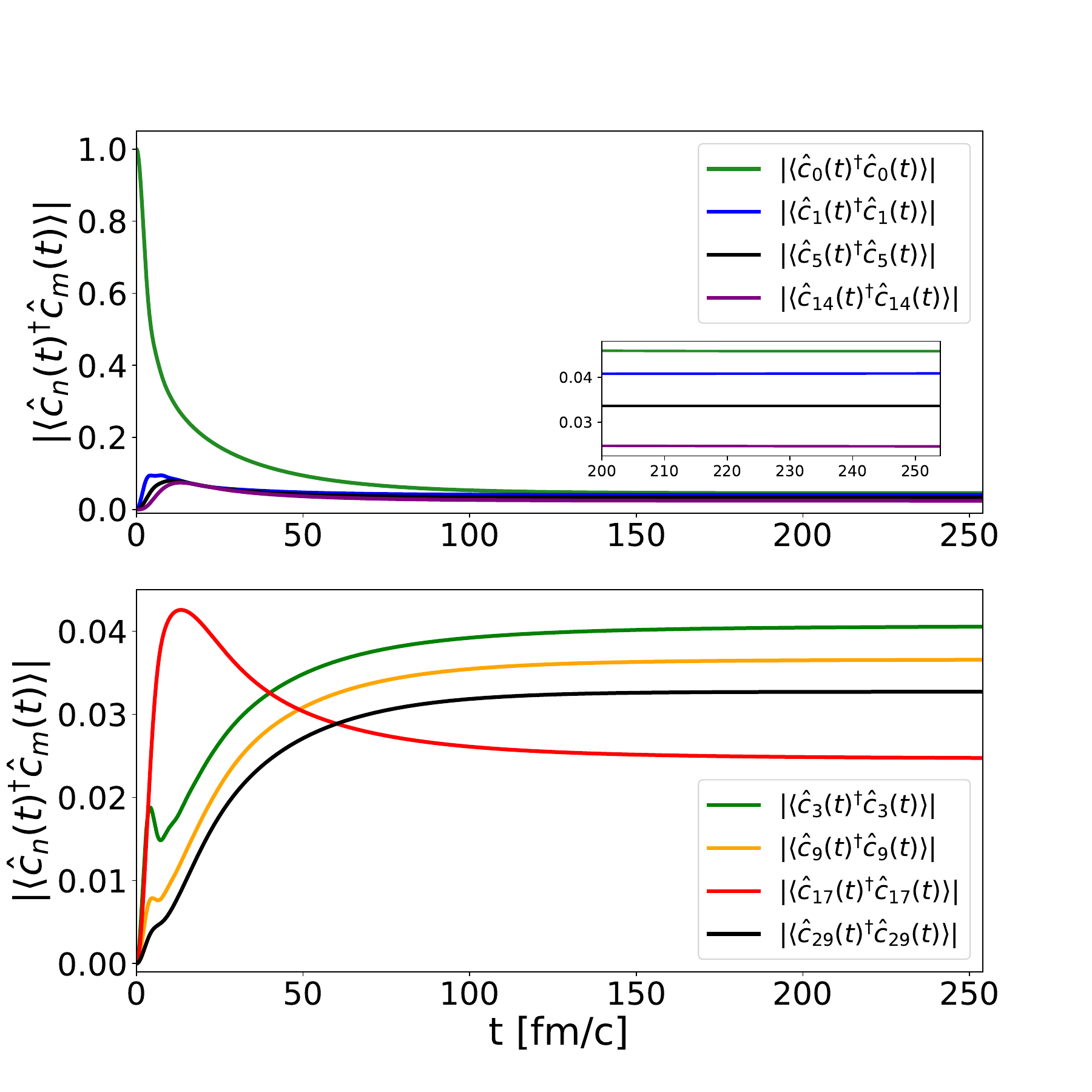}
\caption{The temporal evolution of the s-states (upper panel) and different non s-states (lower panel).}
\label{fig:Formation_of_3D_bound_states_0}
\end{center}
\end{figure}
The occupation numbers of the quantum states are suppressed in accordance with their energies, as depicted in \cref{fig:Formation_of_3D_bound_states_0}. The states develop in the direction of a thermal equilibrium, whereby the development is clearly non-exponential and shows tendencies of ``over-shooting" at early times, which was not observed in the one-dimensional case. 
The equilibration times are quite long, although a very large cross-section is employed. 

\begin{figure}[h]
\begin{center}
\includegraphics[width=1.1 \linewidth,clip=true]{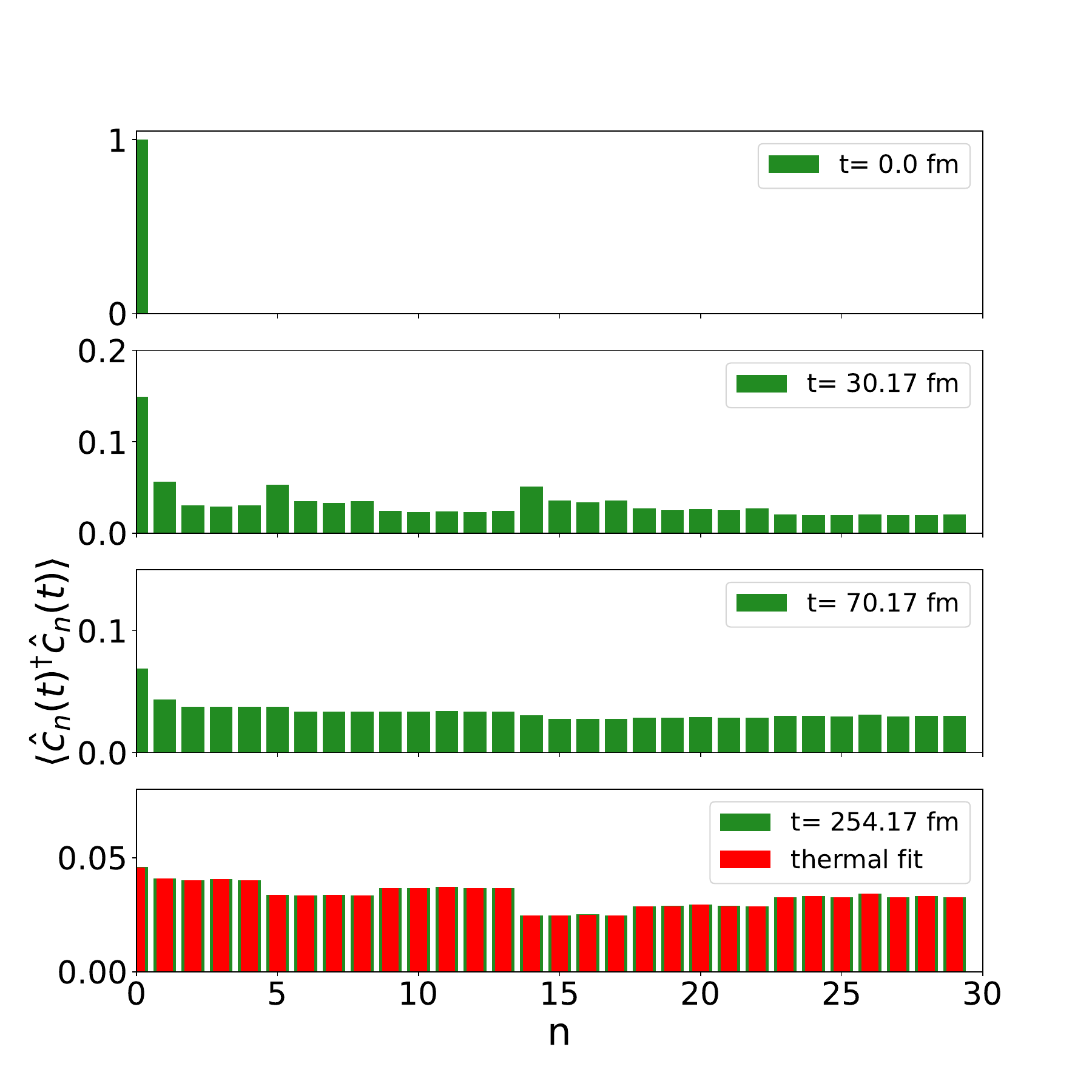}
\caption{The evolution of the occupation numbers for the initial occupied ground state in three dimensions towards its equilibrium state at temperature $T_{\mathrm{syst}}=100.82 \MeV$ and chemical potential $\mu_{\mathrm{syst}}= -264.89 \MeV$.}
\label{fig:Formation_of_3D_bound_states_0.1}
\end{center}
\end{figure}

\begin{figure}[h]
\begin{center}
\includegraphics[width=1.1 \linewidth,clip=true]{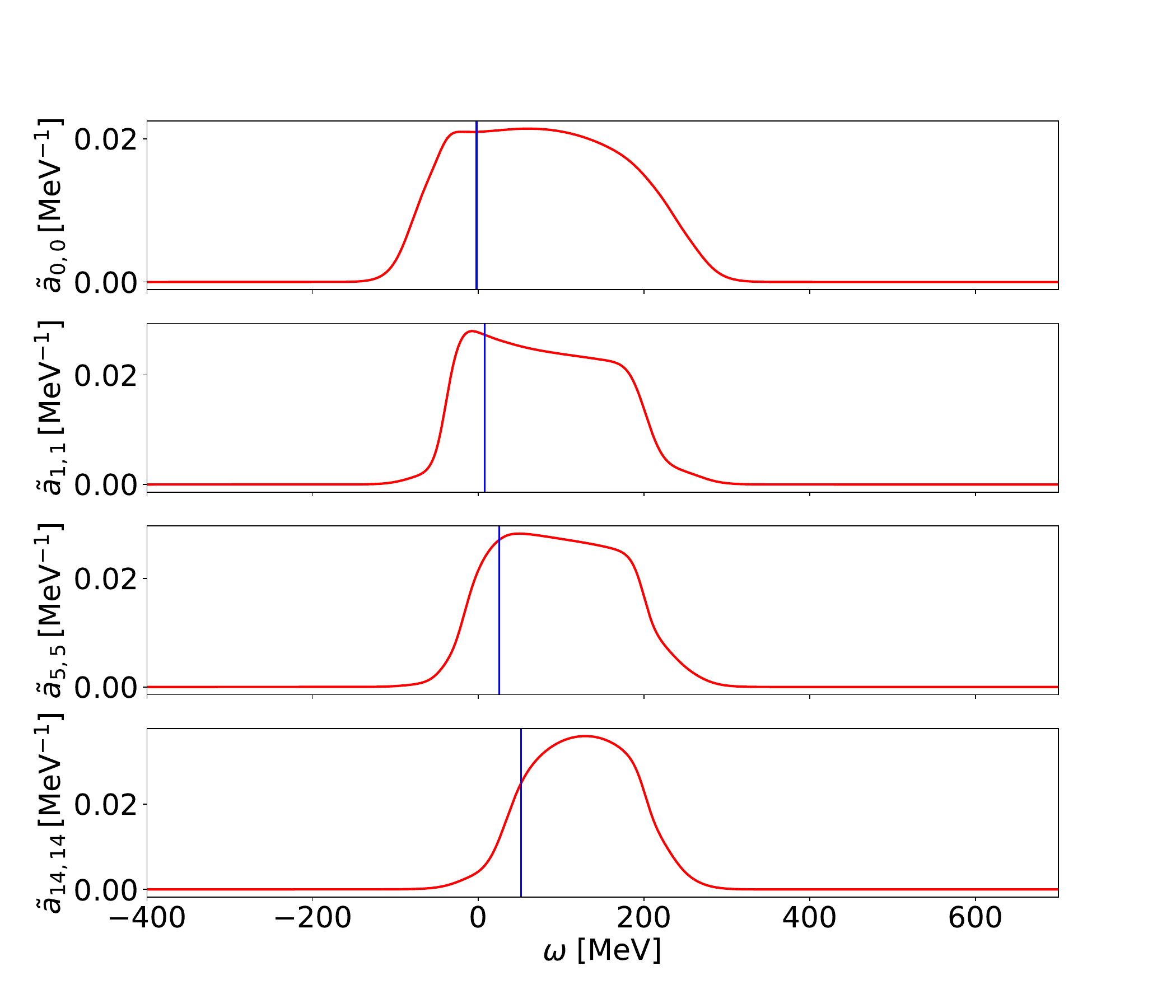}
\caption{The spectral functions $\tilde{a}_{i,i}(\omega,\bar{t})$ of the s-states (upper panel \cref{fig:Formation_of_3D_bound_states_0}) at $\bar{t} = 127 \fm$.}
\label{fig:Formation_of_3D_bound_states_1}
\end{center}
\end{figure}

\begin{figure}[h]
\begin{center}
\includegraphics[width=1.1 \linewidth,clip=true]{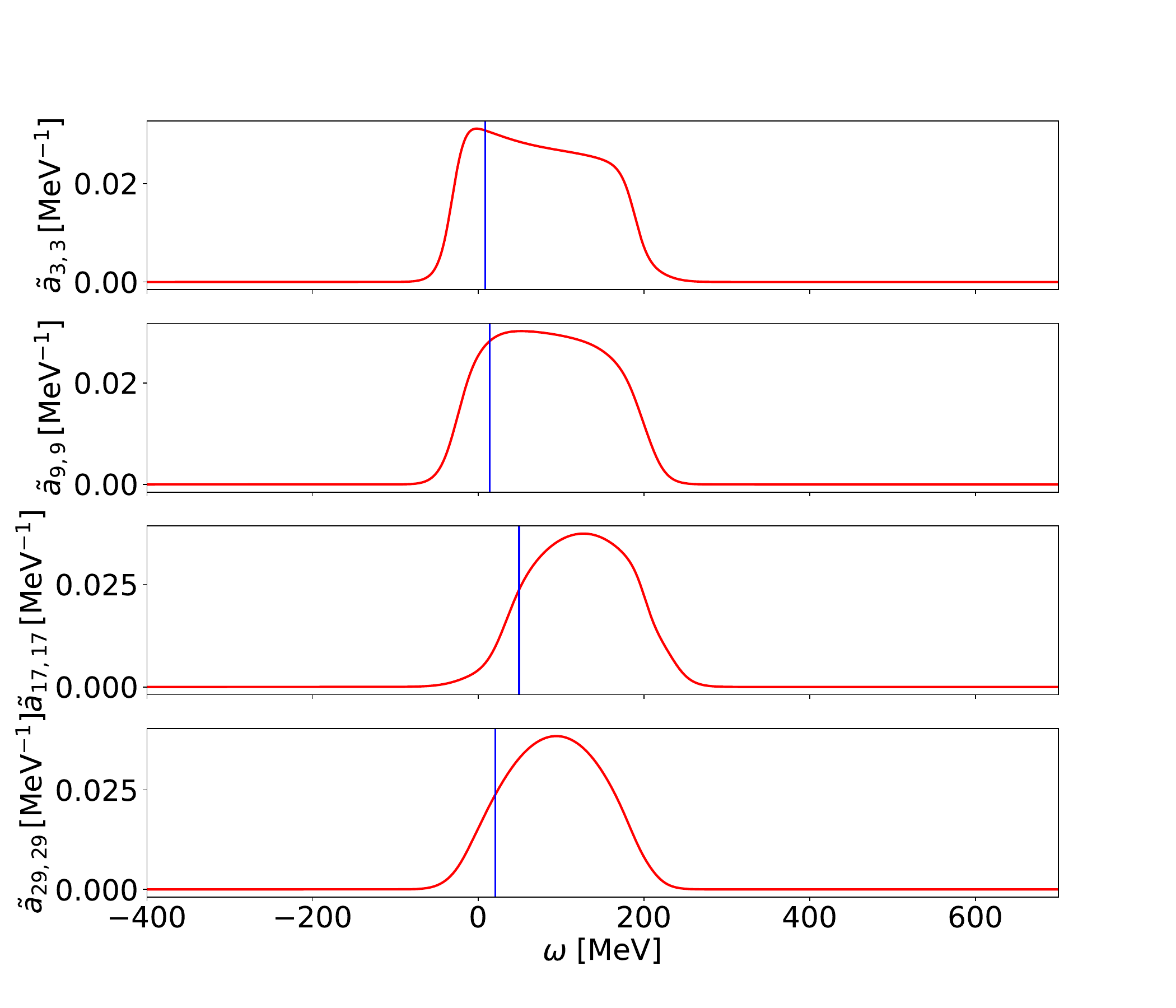}
\caption{The spectral functions $\tilde{a}_{i,i}(\omega,\bar{t})$ of the non s-states (lower panel \cref{fig:Formation_of_3D_bound_states_0}) at $\bar{t} = 127 \fm$.}
\label{fig:Formation_of_3D_bound_states_2}
\end{center}
\end{figure}

In order to verify that the final state is at the temperature of the bath, we perform the thermal fit in accordance to \cref{16}, which has also been used in \cite{2024PhLB..85138589N}. In \cref{fig:Formation_of_3D_bound_states_0.1}, the occupation numbers for the three dimensional deuteron scenario are shown at initial, intermediate, and final times. It is noteworthy that the explicit breaking of degeneracy, induced by the spherical symmetry of the external potential \cref{fig:mapping of quantum numbers 3D spherical}, can also be discerned through coupling to the heat bath. This is evident in the observation that the magnetic quantum number $m$ now exerts an influence on the occupation numbers.
In both figures, \cref{fig:Formation_of_3D_bound_states_1} and \cref{fig:Formation_of_3D_bound_states_2}, the spectral functions of various s- and non-s states are shown. The blue lines indicate again the bare on-shell energy of the corresponding energy eigenstate of the single-particle Hamiltonian $h_0$. As an overall trend, we can identify that the widths become smaller for higher states, indicating that they are slightly less coupled, but the coupling is strong enough to ensure equilibration.
A non-local interaction, cf. \cref{Bound_states_in_3D_1,Bound_states_in_3D_1.1}, could accelerate the process of thermalization without increasing the coupling parameter $\lambda$, because it enables a larger range of overlap of the wave functions and therefore a larger set of transition amplitudes are non-zero.

\subsection{Open bosonic systems with Kadanoff-Baym equations} \label{subsec: Open bosonic systems with Kadanoff-Baym equations}

The Kadanoff-Baym approach also allows to study bosonic particles, e.g. cold atoms, trapped in an external, e.g. harmonic, potential coupled to a surrounding heat bath. These kind of systems have been studied for closed systems by T. Gasenzer, J. Berges et. al. \cite{PhysRevA.72.063604} with the Kadanoff-Baym approach in one dimension, and first studies towards open systems have been conducted with the use of master equations and Kadanoff-Baym equations, but only at mean-field level in \cite{handle:20.500.11811/8961}. 

The external potential for this study is harmonic, as often used in experiments,
\begin{equation}
\begin{split}
V(r)_{S/B} \defeq \frac{1}{2}\Omega_{S/B} r^2.
\label{BEK_0}
\end{split}
\end{equation}
The seasoned reader may draw parallels with the renowned Caldeira-Leggett model \cite{CALDEIRA1983587}, wherein an oscillator and heat-bath oscillators are coupled.
It is not feasible to achieve a genuine Bose-Einstein condensation using this methodology, as such a phenomenon is only observable in the continuum limit. Unfortunately, given that the computational burden increases drastically with the total number of particles and that it is necessary to utilize increasingly smaller time steps to accurately resolve the dynamics, it is not possible to attain a system comprising more than $\mathcal{O}(100)$ particles \cite{PhysRevLett.116.225304}.

An attempt was made to simulate in different dimensions; however, it was found that even in two or three dimensions, the degeneracy factor $g_n = \binom{\mathrm{dim} + n -1}{n}$ of the different energy levels was so restrictive that thermalization in three dimensions can in principle be achieved, but the physical relevance is at least questionable, if only the three lowest energy eigenstates are considered in the system ($n_x = n_y =n_z =3 \, \rightarrow \, 27$ total eigenstates).

In one dimension, we simulate 11 bosonic (system) particles in a harmonic potential at temperature $T_{\rm{bath}} = 100 \MeV$, a chemical potential $\mu_{\mathrm{bath}} = -3 \MeV$, and with frequencies $\Omega_{\rm{S/B}}=39.46 \MeV$ and a coupling constant of $\lambda=0.2$. For this interaction a non-local Gaussian potential is used,
\begin{equation}
\begin{split}
V_{\rm{int}}(|r-r'|) =  \frac{\lambda}{\sqrt{2 \pi \sigma^2}} e^{-|r-r'|^2/(2 \sigma^2)}
\label{BEK_1}
\end{split}
\end{equation}
where $\sigma$ is set to $0.5 \fm$. The precision is set to $\epsilon_{\rm{abs}}=2 \cdot 10^{-6}$ here. This is already an improvement compared to \cite{PhysRevA.72.063604} and \cite{PhysRevLett.116.225304}, but still far from real macroscopic occupation numbers. Nevertheless, we can already observe some interesting features such as bosonic spectral functions, changes of the ground-state wave function due to the interaction with the heat bath and the energy decomposition as outlined in  \cref{subsection:Eigenstates_and_eigenvalues}. 
\begin{figure}[]
    \includegraphics[width=1.125\columnwidth,clip=true]{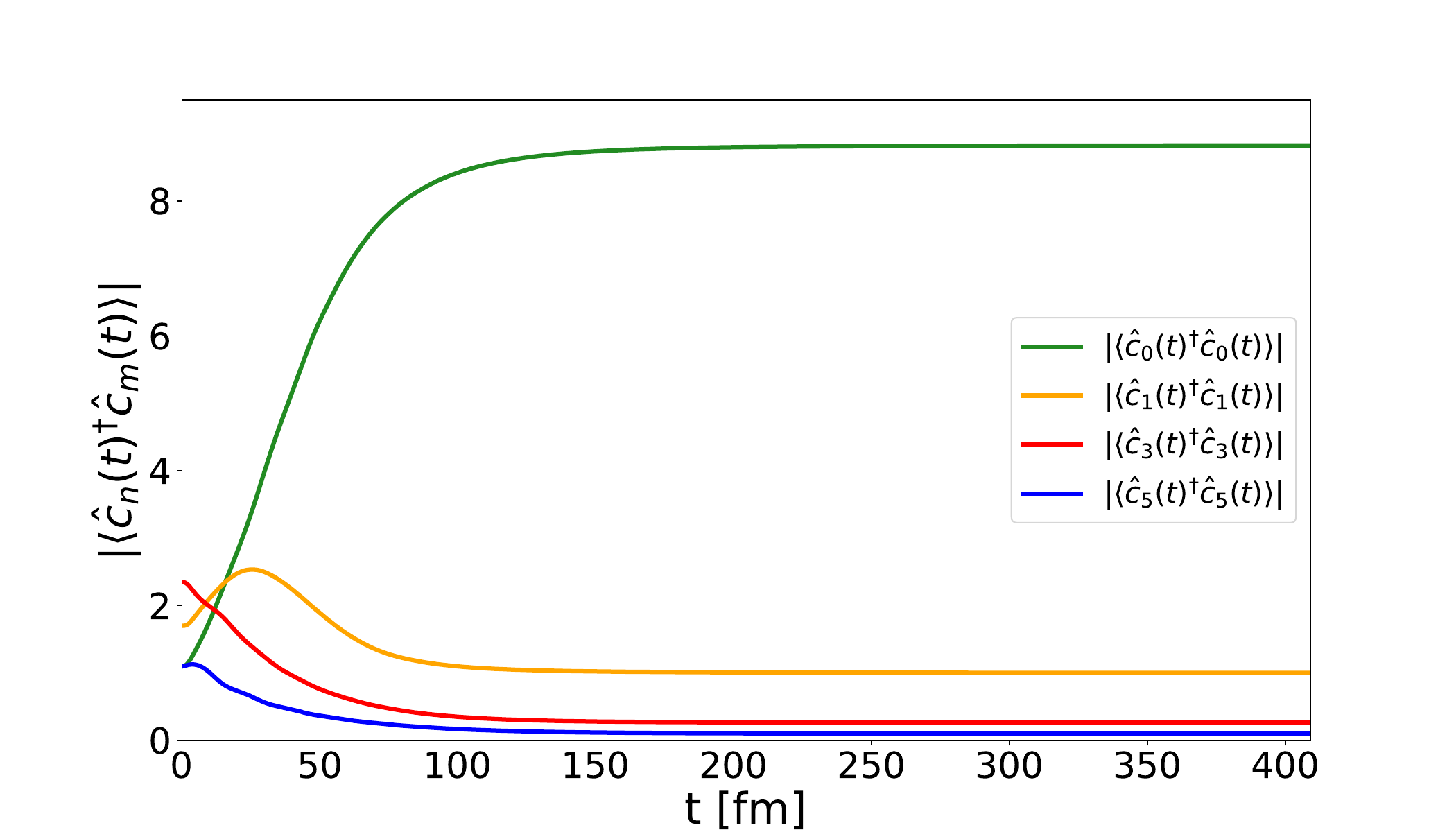}

    \caption{
     The time evolution of the ground state and higher excited states of bosonic particles in a harmonic trap.
    }
    \label{fig:Open_bosonic_system}
\end{figure}
In \cref{fig:Open_bosonic_system}, the time evolution of various quantum states is depicted. We observe a clear splitting of the states and an enhancement of the ground-state population even in this small system. In the long time limit, it is also clear, that the system has reached its equilibrium state. 

\begin{widetext}

\begin{figure}[h] 
\begin{center}
\includegraphics[width=1.0 \linewidth,clip=true]{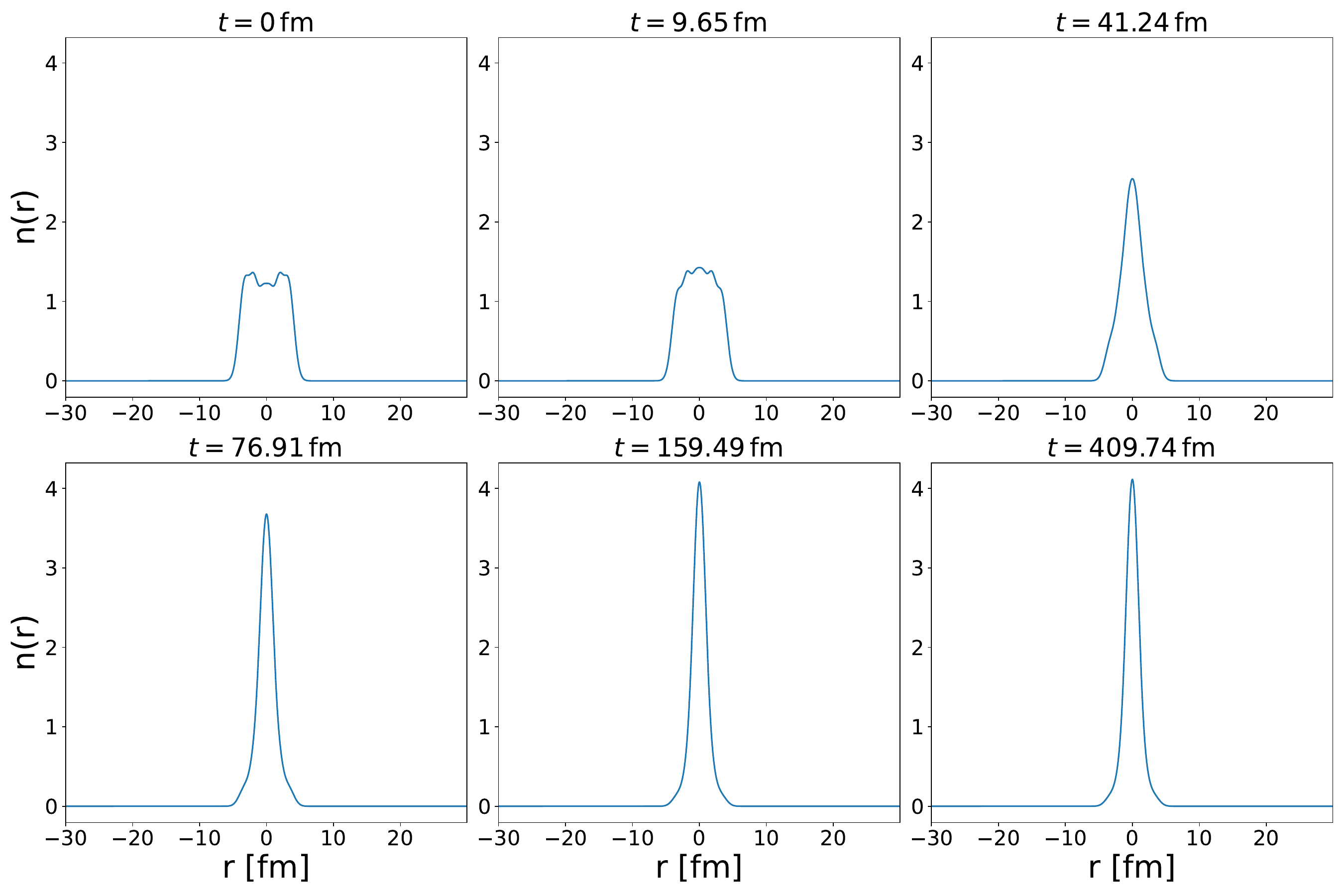}
\caption{The density $n(r,t) = i \, S^{<}(r,t,r,t)$ at different times during the evolution depicted in \cref{fig:Open_bosonic_system}.}
\label{fig:Open_bosonic_system_0}
\end{center}
\end{figure}

\end{widetext}

In \cref{fig:Open_bosonic_system_0}, the evolution of the density 

$n(r,t) =i \, S^{<}(r,t,r,t)$ is shown for the Bose gas. It is intuitive, that the enhancement of the ground-state will let the ground-state wave function, which is of Gaussian bell shape, dominate in the late time for the density.  

\begin{figure}[]
    \includegraphics[width=1.125\columnwidth,clip=true]{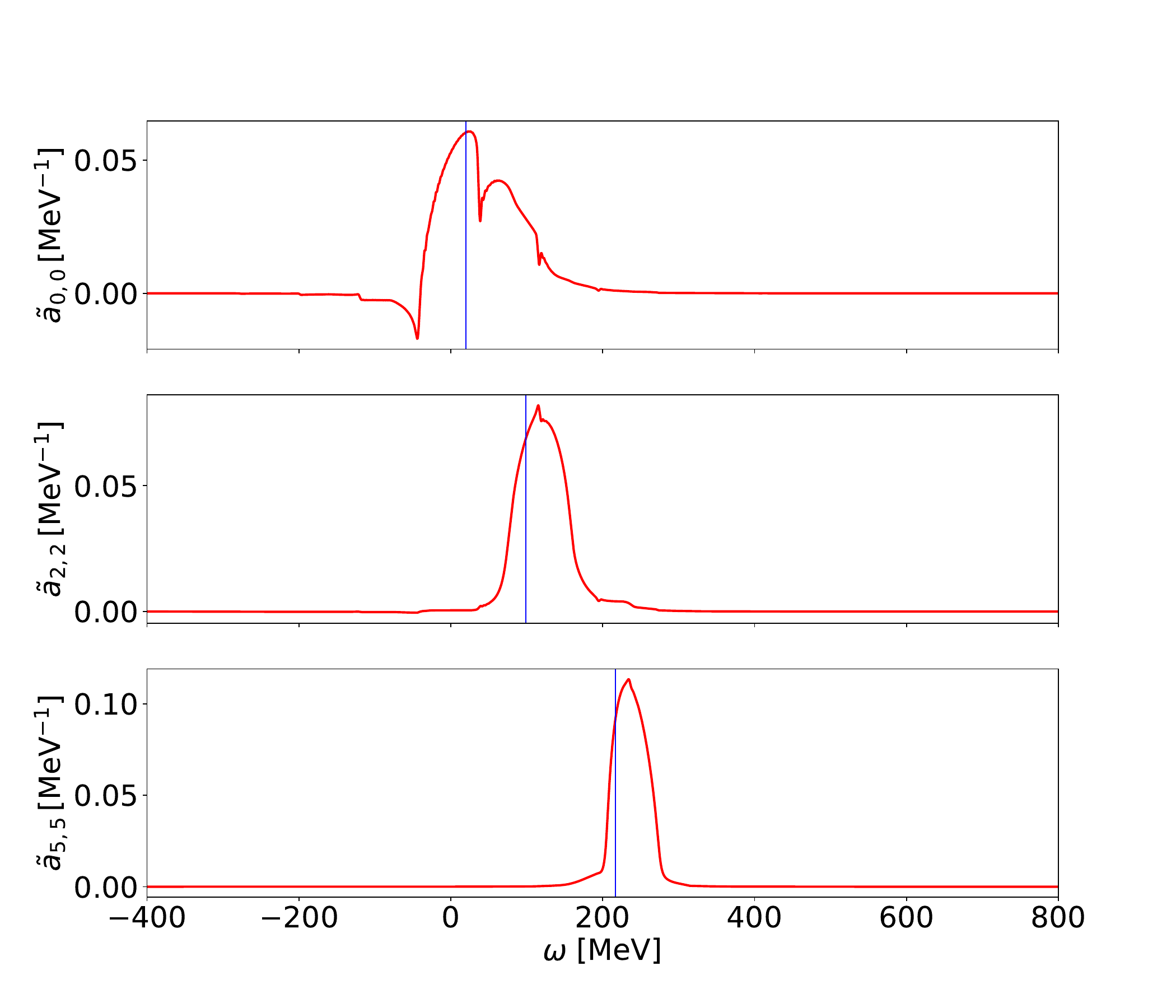}

    \caption{
     The spectral function of the ground state and some higher states of bosonic particles in a harmonic trap at $\bar{t} = 204.5$. The blue lines represent the values of the bare energy $E_n = (n+0.5)\Omega_S$.
    }
    \label{fig:Open_bosonic_system_1}
\end{figure}

However, a more intriguing question would be, how the spectral function of the ground-state in particular looks like. In \cref{fig:Open_bosonic_system_1}, the spectral functions of three exemplary quantum states are shown. It is quit remarkable, that a purely bosonic feature of the spectral function can be observed here, namely that the spectral function can become negative and somehow lose its interpretation of a stochastic distribution function, where its value represents the probability of finding a particle in this state with a given energy $\omega$. Note,however that the integral over all the spectral functions is still normalized, i.e. its integral over $\omega$ also yields unity for bosons. The zero value of the spectral function, which is also still visible for the second exited state and very clear to see for the ground state, represents the value of the chemical potential $\mu_{\rm{syst}} \approx -50 \MeV$ for this specific system. A short derivation of this can be found in \cref{Appendix:Bosonic_spectral_functions}. It should be noted that the energy peaks have been slightly shifted to higher values, thus indicating a higher effective value of $\Omega_S$ due to the interaction with the heat bath. But because of the broadening, especially the ground state contains negative contributions, which make it hard to define an effective frequency.

To further investigate the shift of the peak energy, we compare the wave function of the ground state with interaction to the bare non-interacting wave function.
\begin{figure}[h]
\begin{center}
\includegraphics[width=1.1 \linewidth,clip=true]{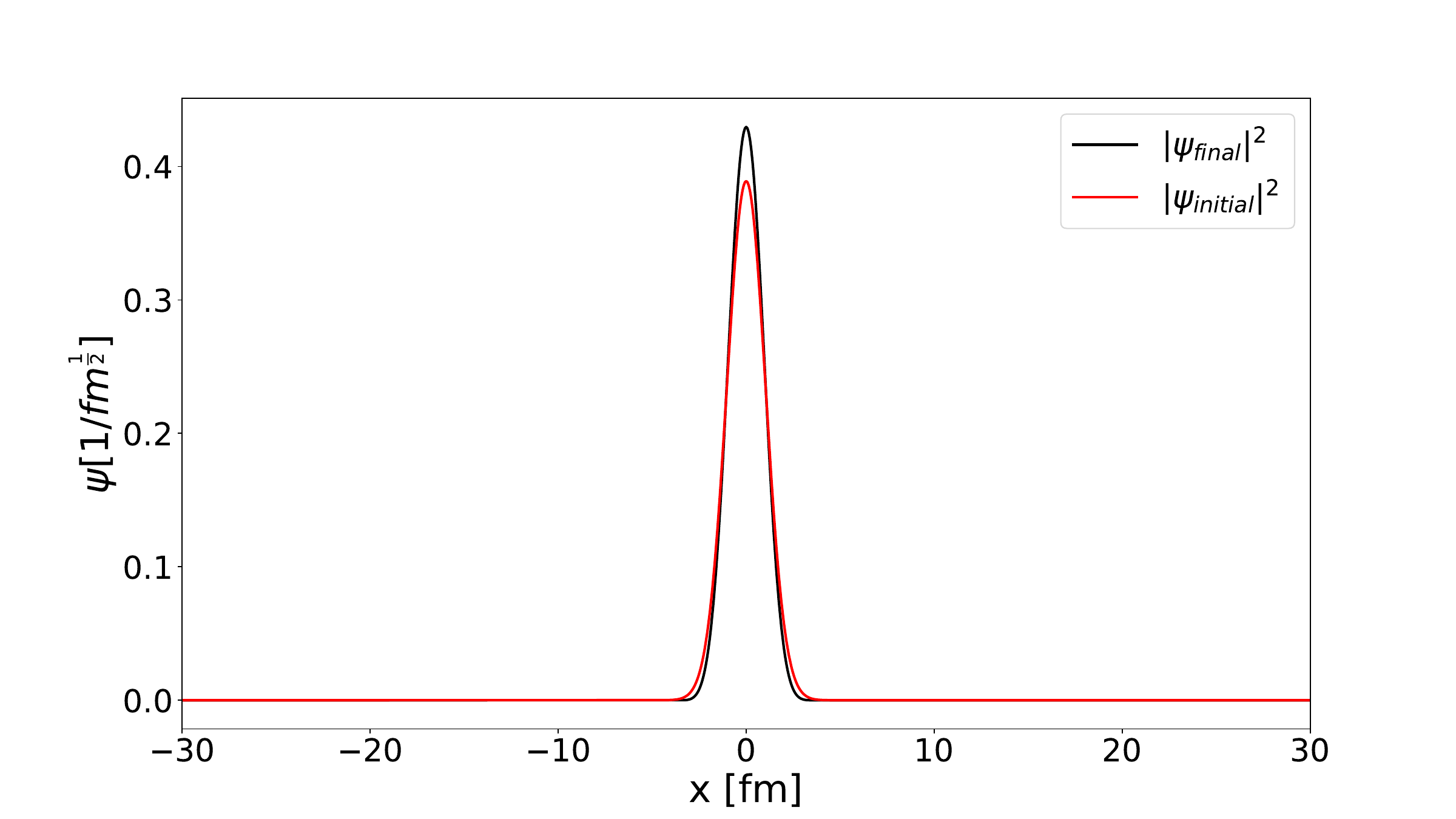}
\caption{The ground state wave function squared of the initial non-interacting and final interacting system for the case of bosons in a harmonic trap in a bath of bosonic particles also in a harmonic potential. The diagonalization was performed with the Armadillo library \cite{Sanderson_2019,sanderson2025armadilloefficientframeworknumerical}.}
\label{fig:Open_bosonic_system_2}
\end{center}
\end{figure}

We can see in \cref{fig:Open_bosonic_system_2}, that the ground state remains symmetric in position space, but becomes narrower due to artifacts from higher, even modes of the initial Hamiltonian $h_0$, which can suggest an increase in the effective ground-state energy $E^{\rm{peak}}_{0}>E_0 = \frac{\Omega_S}{2}$, because the effective width of the ground-state wave function of the oscillator scales as $L_{\rm{eff}} \propto \Omega_s^{-\frac{1}{2}}$.

\begin{figure}[h]
\begin{center}
\includegraphics[width=1.1 \linewidth,clip=true]{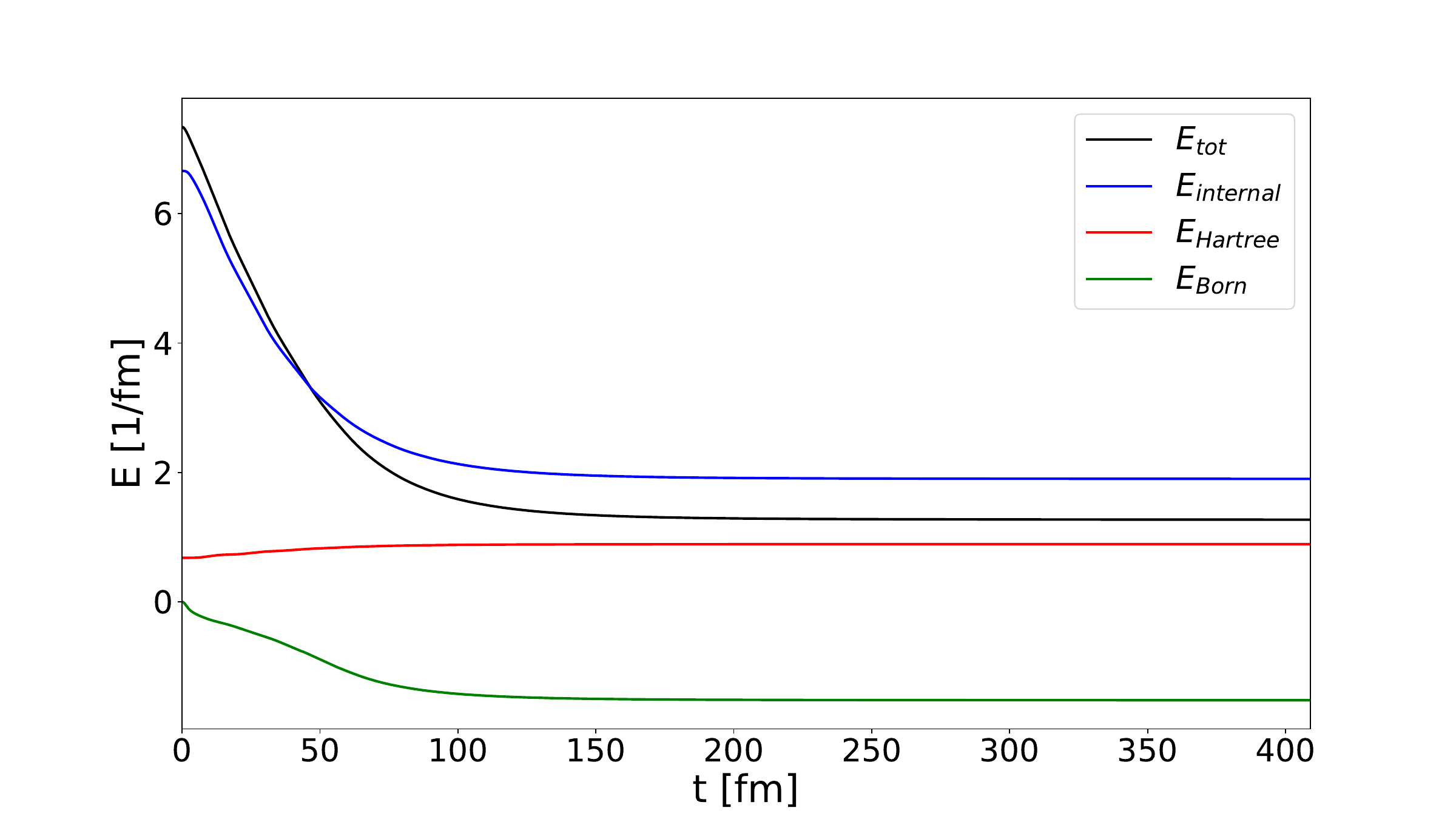}
\caption{The total energy \cref{Entropy_Heat_8} and its separate contributions for the case \cref{fig:Open_bosonic_system}.}
\label{fig:Open_bosonic_system_3}
\end{center}
\end{figure}
 
In \cref{fig:Open_bosonic_system_3}, we can see the evolution of the total energy and its constituents for an initial condition specified in \cref{fig:Open_bosonic_system}. After initialization, there is a sufficient decrease of the internal energy caused by the population of lower modes. This behavior is strongly depending on the initial condition, but will always lead to a strong in/decrease in the early times. The Hartree contribution is in this case nearly time-independent, because the bath is always in equilibrium (Born-Markov approximation). An effect which is well known from earlier calculations, e.g. \cite{Juchem_2004}, is the build-up of correlations smoothly after initialization on a much shorter time scale than (pre)-equilibration or thermalization. The overall energy balance is negative, so during the thermalization process more heat was transferred from the system to the bath than vice versa, so the system was cooled.

\section{Summary and Conclusions}

This investigation sought to demonstrate the utility of the Kadanoff-Baym equations for the analysis of open quantum systems. 

The inhomogeneous Green's functions utilized in this study have been employed in prior studies on closed systems \cite{dahlen2006propagating, dahlen2007solving, Stan_2009, PhysRevA.82.033427, PhysRevB.82.155108}. The application of open quantum systems to this field is a relatively recent development \cite{2024PhLB..85138589N}, as evidenced by the use of alternative density matrix approaches, such as the Lindblad equation \cite{Gorini:1975nb, 1976CMaPh..48..119L, DIOSI1993517,Rais:2024jio,Rais:2025fps}, which have been predominantly studied.

The question of equilibration and decoherence in an open quantum system for the inhomogeneous case in one dimension has been previously examined in \cite{2024PhLB..85138589N}, where the merits of this approach over the Lindblad equation (e.g. the appearance of spectral functions) were discussed. 
The advantages are particularly important in identifying the energy states of the interacting system and clarifying the issue of the persistence of a bound state. 
Moreover, the paper demonstrated that the approach is self-consistent and the numerics have been sufficiently tested using the requisite spectral functions, cf. \cref{fig:Formation_of_bound_states_5}.

To better understand the formation of bound states and to investigate the justification for their description by classical Boltzmann equations in heavy ion collions, we compared the full Kadanoff-Baym equations with some approximations in \cref{subsec:Approximations_to_the_KBE}.  

Our analysis revealed that the mere memory of the quantum-kinetic master equation to the diagonal Kadanoff-Baym equation results in a relatively minor adjustment to the relaxation time. However, incorporating additional correlations within the full Kadanoff-Baym equation leads to a substantial increase, surpassing threefold, in the relaxation time. In the context of decoherence, it was recently suggested in \cite{Rais:2025fps} that decoherence is an important factor in extending thermalization times.

A further investigation of the effect of correlations and decoherence (diagonal vs. full Kadanoff-Baym equations) on the spectral properties showed that in this approximation the mean field/Hartree term is very well approximated, due to its diagonal-dominant appearance, cf. \cref{Comparison_2} and \cref{10}. 

In the one-dimensional scenario, two further criteria were employed to confirm the equilibrium position: namely, the entropy and the KMS condition (fluctuation-dissipation theorem). It was also observed that the entropy in the (sub)system does not necessarily have to increase or can also decrease. This corresponds to a cooling process of the system by the coupled bath that is also explicitly shown in \cref{fig:Open_bosonic_system_3} for a bosonic system. 
It should be noted that, irrespective of the direction of the process (i.e. cooling or heating), the entropy of the system is expected to reach a state of equilibrium with the bath.

However, it has been demonstrated that entropy alone does not provide any more detailed specifications of the equilibrium state with regard to the temperature and chemical potential. In addition to the standard scheme, in which the occupation numbers (i.e. the diagonal elements of $c^{<}$) are approximated by an integral of a product of the spectral function and the respective Fermi/Bose distribution over $\omega$, an alternative method is employed. This is due to the fact that only maximal $S$ states can be used for the fit, a method that remains effective for fermions. Nevertheless, this approach is inconvenient for bosons, as the attempt to obtain the temperature and chemical potential necessitates integrating over the pole of the Bose distribution, thereby hindering the convergence of the minimization procedure. An alternative method, which subsequently emerged as a significantly more efficient and numerically stable approach, is the KMS condition. 
In this approach, one of the $S$ states can be selected independently, thereby enhancing the accuracy of the fit due to the closer alignment of points in the definition range, i.e. the parameter space, of the distribution. This can be achieved either by increasing the number of grid points in the two-time plane or by ensuring an average time-step size that is not too small. 

In \cref{Bound states in 3 dimensions}, the formal differences that arise during the transition to higher dimensions were presented and the measures for reusing or modifying the existing algorithm are explained.

The simplified deuteron problem has then been simulated using a concrete example with spherical symmetry, showing that thermalization is also achieved in three dimensions. 

 The spectral functions of the three-dimensional case were depicted in \cref{fig:Formation_of_3D_bound_states_1,fig:Formation_of_3D_bound_states_2}. Concerning the existence of the bound state in the heat bath, it has been found that it no longer needs to be completely bound, cf. \cref{fig:Formation_of_3D_bound_states_1}, as it is now distributed over a broader energy range. The shift to the positive part of the spectrum is predominantly driven by the Hartree term. 
 
Furthermore, it should be noted that the increase in dimension (due to the hardware limitations) does not provide anywhere near the number of states in three dimensions that are available in one dimension. A future increase in the number of basis states could at least provide new scattering states. This may also be relevant in the simulation of Bose-Einstein condensates, which is touched on in \cref{subsec: Open bosonic systems with Kadanoff-Baym equations}, since the dynamics of such effects can only be resolved with a sufficiently large basis and a large number of particles in the continuum limit.

However, the present calculations for trapped bosons in one dimension have surpassed the existing calculations, \cite{PhysRevA.72.063604}, regarding the particle number, and a clear tendency to separate the ground state has been recognized in \cref{fig:Open_bosonic_system,fig:Open_bosonic_system_0}. In this context, the possibility of forcing the spectral function of the bosonic ground state to negative values in a certain range $\omega <\mu_{\mathrm{syst}}$ has also been observed, cf. \cref{fig:Open_bosonic_system_1}, as is shown formally in \cref{Appendix:Bosonic_spectral_functions}. 

Finally, the change in energy levels due to the interaction with the heat bath, $H_\mathrm{SB}$, was linked to the new eigenstates of the system. In contrast to the solution of the time-independent Schrödinger equation for $h_0$, the full Hamiltonian $H_\mathrm{S} + H_\mathrm{SB}$ was diagonalized and the effect on the ground-state wave function of the harmonic oscillator is shown in \cref{fig:Open_bosonic_system_2}.

From a practical point of view, the standard kinetic master equations basically allow the description of bound states via elastic collisions with the environment  particles \cite{DANIELEWICZ1991712,PhysRevC.104.034908,Sun:2021dlz,Neidig:2021bal}. In further investigations, it is planed within the non-equilibrium Green's function framework to implement potentials, that enable the formation of multiple, stronger bound states like charmonia, e.g. $J/\Psi$ \cite{workinprogress}. As for the deuteron, only the relative motion of the $J/\Psi$ can be tracked in the Kadanoff-Baym formalism for the single-particle Green's function, which was also mentioned in \cite{Blaizot:2018oev} for the (reduced) density matrix in the Lindblad formalism. However, for a QCD-accurate treatment, analogous to the Lindblad formalism, cf. \cite{Blaizot:2017ypk,Blaizot:2018oev}, the color singlet/octet structure introduced by the non-abelian $SU(3)$ pairing of a quark and an anti-quark must be taken into account. This implies that the dynamics of two Green's functions has to be computed, $\vec{S}(1,1') = (S_s(1,1'), S_o(1,1') )^{T}$, where the subscripts denote either the singlet (s) or octet (o). These two can be expanded in a discrete, but not necessarily the same basis, and then coupled independently of each other to a bath/plasma in the same way as in $H_{\mathrm{SB}}$, cf. \cref{Hamiltonian}. In order to allow for transitions between the singlet and octet state, an additional mixing-coupling has to be introduced. This can be realized, for example, via $H_{\mathrm{int, s/o}} = \int dr \int dr' \, \hat{\psi}_s(r,t)^{\dagger} \hat{\psi}_o(r,t) \, V_{\rm{int, s/o}}(|r-r'|) \,  \hat{\phi}(r',t)^{\dagger} \hat{\phi}(r',t) + h.c.$, which can be interpreted as a bath/plasma-catalyzed transition from a singlet to an octet or vice versa. This type of interaction would not lead to a conservation of the particle number for a single component alone, but it would for the total number of singlets and octets, if the reactions are equally weighted with the same potential as indicated in $H_{\mathrm{int, s/o}}$.

\begin{acknowledgements}
The authors acknowledge for the support the European Union's Horizon 
2020 research and innovation program
under grant agreement No 824093 (STRONG-2020).
T.N. thanks Tomoi Koide for fruitful discussions.
T.N. thanks Jan Rais for supportive discussions.
T.N. is supported by Helmholtz Forschungsakademie Hessen für FAIR (HFHF).
We acknowledge support by the Deutsche Forschungsgemeinschaft (DFG) through the CRC-TR 211 ``Strong-interaction matter under extreme conditions". 
All calculations have been done on the Goethe-HLR (CSC) in Frankfurt am Main.

\end{acknowledgements}

\begin{widetext}
\appendix

\section{Numerics, parallelization and adaptive time stepping/integration}\label{sec:Numerics}

\subsection{Algorithm for the two-time propagation of the Green's functions}\label{subsec:Algorithm}

To integrate the Kadanoff-Baym equation numerically, we discretize the time arguments of $c^{\gtrless}_{n,m}(t,t')$ in (at this point) equidistant grid of units $\Delta$. 
The non-equilibrium, inhomogenous Green's function is propagated in accordance with the methodology outlined in \cite{DANIELEWICZ1984305, Juchem_2004, dahlen2006propagating, dahlen2007solving, Stan_2009, PhysRevB.82.155108, Meirinhos_2022}, building upon the earlier work of S. Köhler et al. \cite{ KOHLER1999123}. It is possible to express the functional time-evolution of the development of the Green's function in a general abstract form, 

\begin{equation}
\begin{split}
c^{>}(t+\Delta,t') &= \mathcal{F}\Bigl[c^{>}(t,t'), V_{\mathrm{eff}}\bigl[n_{B/F}, V \bigl] \Bigr], I^{>}_{1}(t,t')\bigl[c^{\gtrless}, V \bigl] \Bigr], \\
c^{<}(t,t'+\Delta) &= \mathcal{F}\Bigl[c^{<}(t,t'), V_{\mathrm{eff}}\bigl[n_{B/F}, V \bigl] \Bigr], I^{<}_{2}(t,t')\bigl[c^{\gtrless}, V \bigl] \Bigr], \\
c^{<}(t+\Delta,t+\Delta) &= \mathcal{F}\Bigl[c^{<}(t,t), V_{\mathrm{eff}}\bigl[n_{B/F}, V \bigl] \Bigr], I^{<}_{1/2}(t,t)\bigl[c^{\gtrless}, V \bigl] \Bigr]. \\
 \label{Numerics_0}
\end{split}
\end{equation}

It should be noted that, in the initial time step (where, for simplicity, $t = t' = 0$), all collision integrals $I^{\gtrless}_{1/2}$ vanish and $c^{\gtrless}(0,0)$ is defined by the initial and boundary conditions.
Integrating the first equation for $c^{>}$ in \cref{12}

\begin{equation}
\begin{split}
c^{>}_{n,m}(t+\Delta,t') &= c^{>}_{n,m}(t,t') + \int_{t}^{t+\Delta} d \bar{t} \,\frac{\partial c^{>}_{n,m}(\bar{t},t')}{\partial \bar{t}} \\
&\approx c^{>}_{n,m}(t,t') + \Delta \Bigl[ -i \sum_{i}^{S} V_{\mathrm{eff}n,i}(t) c^{>}_{i,m}(t,t') + I^{>}_{n,m 1}(t,t') \Bigr] + \mathcal{O}(\Delta^{2}) \\
&= \sum_{i}^{S} (\delta_{n,i} -i V_{\mathrm{eff}n,i}(t) \Delta + \mathcal{O}(\Delta^{2}) ) \, c^{>}_{i,m}(t,t') + \Delta \, I^{>}_{n,m 1}(t,t') + \mathcal{O}(\Delta^{2}) \\
&\approx \sum_{i}^{S} \underbrace{[e^{-i V_{\mathrm{eff}}(t) \Delta }]_{n,i}}_{\defeq U(\Delta)_{n,i}} \, c^{>}_{i,m}(t,t') + \Delta \, I^{>}_{n,m 1}(t,t') 
\label{Numerics_1}
\end{split}
\end{equation}

by using the standard Euler method reveals the potential for a $U(1)$ transformation similar to a time-evolution operator $U(t)$. Mathematically, it is just an application of the ``variation of constants" method known from inhomogeneous linear ordinary differential equations. In the following we drop the indices and use an implied matrix notation. This $U(t)$ allows for a redefinition of the Green's function via \cite{ KOHLER1999123,dahlen2007solving}

\begin{equation}
\begin{split}
c^{\gtrless}(t,t') &= U(t) C^{\gtrless}(t,t') U^{\dagger}(t')
\label{Numerics_2}
\end{split}
\end{equation}

and enables us with the help of \cref{12} to write down equations of motion for $C^{\gtrless}$. For $C^{>}$ it takes the form 

\begin{equation}
\begin{split}
\frac{\partial C^{>}(t,t')}{\partial t} &= \frac{U^{\dagger}(t)}{\partial t} c^{>}(t,t') U(t') + U^{\dagger}(t) \frac{c^{>}(t,t')}{\partial t} U(t') = U^{\dagger}(t) I^{>}_{ 1}(t,t') U(t')
\label{Numerics_3}
\end{split}
\end{equation}

where we used the properties of $U$ to cancel the first term in \cref{Numerics_1}. Using now \cref{Numerics_3,Numerics_2} in \cref{Numerics_1}, we can write \cite{dahlen2007solving}

\begin{equation}
\begin{split}
c^{>}(t+\Delta,t') &= U(t+\Delta) C^{>}(t+\Delta,t') U^{\dagger}(t') = U(t+\Delta) \Biggl[ C^{>}(t,t') + \int_{t}^{t+\Delta} d \bar{t} \, \frac{\partial C^{>}(\bar{t},t')}{\partial \bar{t}} \Biggr] U^{\dagger}(t') \\
&= U(\Delta) \Biggl[ c^{>}(t,t') + \int_{t}^{t+\Delta} d \bar{t} \, U(t-\bar{t}) I^{>}_{ 1}(\bar{t},t') \Biggr] \approx U(\Delta) \Biggl[ c^{>}(t,t') + \Bigl(\int_{t}^{t+\Delta} d \bar{t} \, U(t-\bar{t}) \Bigr) I^{>}_{ 1}(t,t') \Biggr]
\label{Numerics_4}
\end{split}
\end{equation}

where we assumed in the last step, that the collision integral is constant in this infinitesimal time interval $[t, t+\Delta]$. The remaining integral can be solved analytically

\begin{equation}
\begin{split}
&\int_{t}^{t+\Delta} d \bar{t} \, U(t-\bar{t}) = \int_{t}^{t+\Delta} d \bar{t} \, U^{\dagger}(\bar{t}-t) = \int_{0}^{\Delta} d \bar{t'} \, U^{\dagger}(\bar{t'}) = \frac{e^{i V_{\mathrm{eff}}(t) \Delta }-1}{i V_{\mathrm{eff}}(t)} \defeq \bar{K}(t,\Delta) \\
&\rightarrow \, K(t,\Delta) \defeq U(\Delta) \bar{K}(t,\Delta) = -i \Biggl[ \frac{1-e^{-i V_{\mathrm{eff}}(t) \Delta }}{V_{\mathrm{eff}}(t)} \Biggr].
\label{Numerics_5}
\end{split}
\end{equation}

A corresponding equation for $c^{<}$ can be obtained in full analogy to \cref{Numerics_4},

\begin{equation}
\begin{split}
c^{<}(t,t'+\Delta) &= U(t) C^{<}(t,t'+\Delta) U^{\dagger}(t'+\Delta) = U(t) \Biggl[ C^{<}(t,t') + \int_{t'}^{t'+\Delta} d \bar{t} \, \frac{\partial C^{<}(t,\bar{t})}{\partial \bar{t}} \Biggr] U^{\dagger}(t'+\Delta) \\
&= \Biggl[ c^{<}(t,t') \mp \int_{t'}^{t'+\Delta} d \bar{t} \, I^{<}_{2}(t,\bar{t}) U^{\dagger}(t'-\bar{t}) \Biggr] U^{\dagger}(\Delta) \approx \Biggl[ c^{<}(t,t') \mp \Bigl(\int_{t'}^{t'+\Delta} d \bar{t} \, U^{\dagger}(t'-\bar{t}) \Bigr) I^{<}_{2}(t,t') \Biggr] U^{\dagger}(\Delta) \\
&\rightarrow c^{<}(t,t'+\Delta) \approx  c^{<}(t,t') U^{\dagger}(\Delta) \mp I^{<}_{2}(t,t') K^{\dagger}(t',\Delta).
\label{Numerics_6}
\end{split}
\end{equation}

The specific equation utilized to describe the propagation of the time-diagonal must be treated in a manner that is distinct from that of the previous equations. But first we notice, that the commutator in the third equation of \cref{12} is vanishing after the unitary transformation,

\begin{equation}
\begin{split}
\frac{\partial C^{<}(t,t)}{\partial t} &= \Biggl[ \frac{U^{\dagger}(t)}{\partial t} c^{<}(t,t) U(t) + U^{\dagger}(t) \frac{c^{<}(t,t)}{\partial t} U(t) +U^{\dagger}(t) c^{<}(t,t) \frac{U(t)}{\partial t} \Biggr] = U^{\dagger}(t) \underbrace{ \Bigl( \pm I^{<}_{ 1} (t,t) \mp I^{<}_{2} (t,t) \Bigr)}_{\defeq I_{\rm{diag}}^{<}} U(t).
\label{Numerics_7}
\end{split}
\end{equation}

Again integrating over an infinitesimal time interval yields the discrete version \cite{dahlen2007solving}, 

\begin{equation}
\begin{split}
&c^{<}(t+\Delta,t+\Delta) = U(t+\Delta) C^{<}(t+\Delta,t+\Delta) U^{\dagger}(t+\Delta) = U(t+\Delta) \Biggl[ C^{<}(t,t) + \int_{t}^{t+\Delta} d \bar{t} \, \frac{\partial C^{<}(\bar{t},\bar{t})}{\partial \bar{t}} \Biggr] U^{\dagger}(t+\Delta) \\
&= U(\Delta) \Biggl[ c^{<}(t,t) + \int_{t}^{t+\Delta} d \bar{t} \, U^{\dagger}(\bar{t}-t) I_{\rm{diag}}^{<}(\bar{t},\bar{t}) U(\bar{t}-t) \Biggr] U^{\dagger}(\Delta) \approx U(\Delta) \Biggl[ c^{<}(t,t) + \int_{0}^{\Delta} d \bar{t'} \, U^{\dagger}(\bar{t'}) I_{\rm{diag}}^{<}(t,t) U(\bar{t'}) \Biggr] U^{\dagger}(\Delta).
\label{Numerics_8}
\end{split}
\end{equation}

At this juncture, it becomes evident that the collision term cannot be extracted from the integral on the assumption that it remains constant during the infinitesimal time step. This is because the equation is matrix-valued, and therefore, $I_{\rm{diag}}^{<}(\bar{t},\bar{t})$ and $U(\bar{t}-t)$ cannot be interchanged arbitrarily. This issue can be circumvented by employing Hadamard's lemma, which states

\begin{equation}
\begin{split}
e^{i V_{\mathrm{eff}} \bar{t} } I_{\rm{diag}}^{<} e^{-i V_{\mathrm{eff}} \bar{t} } = I_{\rm{diag}}^{<} + i \bar{t} \, \Bigl[V_{\mathrm{eff}}, I_{\rm{diag}}^{<} \Bigr] + \frac{(i \bar{t})^{2}}{2!} \Bigl[V_{\mathrm{eff}}, \Bigl[V_{\mathrm{eff}}, I_{\rm{diag}}^{<} \Bigr] \Bigr] + \frac{(i \bar{t})^{3}}{3!} \Bigl[V_{\mathrm{eff}}, \Bigl[V_{\mathrm{eff}}, \Bigl[V_{\mathrm{eff}}, I_{\rm{diag}}^{<} \Bigr] \Bigr] \Bigr] + \mathcal{O}(\bar{t}^{4}). \\
\label{Numerics_9}
\end{split}
\end{equation}

Numerically, this can be computed recursively via \cite{dahlen2007solving},

\begin{equation}
\begin{split}
\int_{0}^{\Delta} d \bar{t'} \, U^{\dagger}(\bar{t'}) I_{\rm{diag}}^{<}(t,t) U(\bar{t'}) = \sum_{m=0}^{\infty} T_{m} \quad ; \quad T_{m} = \frac{i \Delta}{m+1} \, \Bigl[V_{\mathrm{eff}}, T_{m-1} \Bigr] \quad ; \quad T_{0} = I_{\rm{diag}}^{<}(t,t).
\label{Numerics_10}
\end{split}
\end{equation}

As evidenced in prior research \cite{dahlen2007solving, PhysRevA.82.033427}, maintaining $m \leq m_{\rm{max}} = 3$ ensures sufficient accuracy. Inserting \cref{Numerics_10} into \cref{Numerics_8} completes the set of time stepping equations,

\begin{equation}
\begin{split}
c^{>}(t+\Delta,t') &= U(\Delta) c^{>}(t,t') + K(t,\Delta) I^{>}_{ 1}(t,t'), \\
c^{<}(t,t'+\Delta) &=  c^{<}(t,t') U^{\dagger}(\Delta) \mp I^{<}_{2}(t,t') K^{\dagger}(t',\Delta), \\
c^{<}(t+\Delta,t+\Delta) &= U(\Delta) \Biggl[ c^{<}(t,t) + \sum_{m=0}^{m_{\rm{max}}} T_{m} \Biggr] U^{\dagger}(\Delta).
\label{Numerics_11}
\end{split}
\end{equation}

It should be mentioned here, that one recovers the standard Euler method, when expanding the exponentials in $U(\Delta)$ and $K(t,\Delta)$ and setting $m_{\rm{max}}=0$ respectively.
In the numerical calculation, the following procedure is applied in sequence \cite{KOHLER1999123,dahlen2007solving, PhysRevA.82.033427}:

\begin{enumerate}
  \item The Green's functions $c^{\gtrless}$ are initialized at time argument $t_0 = 0$.
  \item Compute $\Sigma_{H}$ and together with $h_0$ then $V_{\mathrm{eff}}$ at time $t$.
  \item Compute the step operators $U(\Delta)$ and $K(t,\Delta)$.
  \item Compute $\Sigma^{\gtrless}$, which are needed for the collision integrals $I_{1/2}^{\gtrless}$.
  \item Evaluate the collision integrals at times $(t,t')$ needed in \cref{Numerics_11}.
  \item Propagate the Green's functions $c^{\gtrless}$ according to \cref{Numerics_11}.
  \item Compute $\Sigma_{H}$ and together with $h_0$ then $V_{\mathrm{eff}}$ at time $t+\Delta$.
  \item Compute the step operators $U(\Delta)$ and $K(t+\Delta,\Delta)$.
  \item Compute $\Sigma^{\gtrless}$, which are needed for the collision integrals $I_{1/2}^{\gtrless}$.
  \item Evaluate the collision integrals at times $t+\Delta$ and $t'+\Delta$ needed in (\ref{Numerics_11}).
  \item The collision integrals and the effective potentials are modified according to the Trapezoidal rule: \\ $V_{\mathrm{eff}} \rightarrow \Bigl[ V_{\mathrm{eff}}(t) + V_{\mathrm{eff}}(t+\Delta) \Bigr] /2 $; $I_{1/2}^{\gtrless} \rightarrow \Bigl[ I_{1/2}^{\gtrless}(t, \cdot) + I_{1/2}^{\gtrless}(t+\Delta,\cdot) \Bigr] /2$
  \item Propagate the Green's functions $c^{\gtrless}$ according to \cref{Numerics_11}, set $t=t+\Delta$, and return to 2.
  \label{Numerics_procedure}
\end{enumerate}

The predictor-corrector method, how such methods as described above are called, is known as ``Heun's method".

\subsection{Parallelisation of the predictor-corrector Algorithm}

In order to facilitate parallelization, it is essential to conduct a detailed examination of the structure of the Kadanoff-Baym equations. 
In light of the fact that the Green's functions and the self energies in accordance with \cref{7} are skew-Hermitian, and the Green's functions in addition fulfill boundary conditions (depending on whether fermionic or bosonic), it is now necessary to ascertain which Green's functions or self energies are required to determine the collision integrals.
Therefore, let us revisit \cref{7} in more depth. 

We start with $I_{1}^{>}$ at the point in the two-time plane $(t,t')$ and suppose we are at time step $n_t = t/\Delta$ at arbitrary time $t' \in [0,t-\Delta] \rightarrow t > t'$ . The Green's functions are required as,

\begin{equation}
\begin{split}
&\forall \, \bar{t} \in [0,t] \quad  c^{>}(\bar{t},t') \defeq \begin{cases}
c^{>}(t',\bar{t})^{*} &\text{if $\bar{t} \le t'$},\\
1 \mp c^{<}(t',t') &\text{if $\bar{t} = t'$},\\
c^{>}(\bar{t},t') &\text{if $\bar{t} > t'$}
\end{cases}   \\
&\forall \, \bar{t} \in [0,t'] \quad c^{<}(\bar{t},t').
\label{Numerics_12}
\end{split}
\end{equation}

The self energies, and thus the Green's functions within them, enter via

\begin{equation}
\begin{split}
&\forall \, \bar{t} \in [0,t] \quad  \Sigma^{>}(t, \bar{t}),\\
&\forall \, \bar{t} \in [0,t] \quad  \Sigma^{<}(t, \bar{t}) = \Sigma^{<}(\bar{t},t)^{*}.
\label{Numerics_13}
\end{split}
\end{equation}

For $I_{2}^{<}$, analogous considerations can be made at the point $(t',t)$. Here the Green's functions and self energies are required as, 

\begin{equation}
\begin{split}
&\forall \, \bar{t} \in [0,t] \quad  c^{<}(t, \bar{t}) \defeq \begin{cases}
c^{<}(\bar{t},t)^{*} &\text{if $\bar{t} \le t$},\\
c^{<}(t, \bar{t}) &\text{if $\bar{t} = t$}
\end{cases}   \\
&\forall \, \bar{t} \in [0,t'] \quad c^{>}(t,\bar{t}), \\
&\forall \, \bar{t} \in [0,t] \quad  \Sigma^{>}(\bar{t},t)= \Sigma^{>}(t, \bar{t})^{*},\\
&\forall \, \bar{t} \in [0,t] \quad  \Sigma^{<}(\bar{t},t).
\label{Numerics_14}
\end{split}
\end{equation}

$I_{1}^{<}$ is only needed for the diagonal time step when $t=t'$, 

\begin{equation}
\begin{split}
&\forall \, \bar{t} \in [0,t] : \\
&c^{<}(\bar{t},t),\\
&c^{>}(\bar{t},t)\defeq \begin{cases}
c^{>}(t,\bar{t})^{*} &\text{if $\bar{t} \le t$},\\
1 \mp c^{<}(t,t) &\text{if $\bar{t} = t$},\\
\end{cases} \\
&\Sigma^{>}(\bar{t},t)= \Sigma^{>}(t, \bar{t})^{*},\\
&\Sigma^{<}(\bar{t},t).
\label{Numerics_15}
\end{split}
\end{equation}

For a fixed time $t'$, the collision integrals generally involve the following Green's functions in the two-time plane. Firstly, the Green's functions, which enter the self energies and are situated on the border of the current state of the calculation, with at least one time argument designated as $t$. And secondly, those which depend on earlier time arguments and are located on lines parallel to the time axes which are intersecting at point $(t', t')$. In accordance with the diagonal time step \cref{Numerics_8}, the two described areas are found to be equal \cite{BalzerPHD}.

In a real calculation, the whole non-equilibrium Green's function ($c^{\gtrless} \in \mathbb{C}$) must be stored to compute the collision integrals. 
The computational burden associated with solving these equations increases significantly with the number of time steps. This is due to the necessity of evaluating the time-integrals on the one hand and, on the other hand, in the $i$-th time step $(2 \cdot i +1)$ propagations for $S^{2}$ matrix elements have to be evaluated. 

Consequently, the quantity of data accumulated in each time step increases steadily. This is because a number of $(2 \cdot i +1) \cdot S^{2}$ correlation functions must be stored when performing a time step. Assuming double precision, the total memory demand is given by roughly 

\begin{equation}
\begin{split}
16 \cdot N_{t}^{2} \cdot S^{2}
\label{Numerics_15.1}
\end{split}
\end{equation}

bytes \cite{BalzerPHD, PhysRevA.82.033427} (temporally allocated self energies and collision integrals not counted), where $N_{t}$ is the total number of time steps in the calculation, which can quickly exceed the $\mathcal{O}(100)$ gigabyte scale. In order to describe physical systems at realistic temperatures and stronger couplings, we are constrained to employ small $\Delta$ and/or large $S$, which results in increased computation times and reduced simulation time ranges.
An excessively large $\Delta$ will inevitably result in the failure of convergence and unstable propagation, including the emergence of divergent collision integrals and the contravention of particle number conservation. Consequently, the simulation of a thermalization process is generally a challenging problem that requires fine tuning when setting the dimension of the eigenbasis $S$ or the step size $\Delta$, where in the future we will explore the possibility of making it adaptive. 

In addition to the physical requirements of the physical system, the numerical propagation of \cref{Numerics_11} is similarly limited by the available computer hardware and performance. 

The calculations with distributed memory were done on the Goethe-HLR (CSC) in Frankfurt am Main. On the newer ``general1" node, comprising two Intel Xeon Gold 6148 (Skylake) CPUs, with 192 or 768 gigabyte RAM and 40(80) cores (hyperthreading) per node, were accessible. However, to ensure optimal utilization of the CSC cluster, it is imperative to implement reliable high-performance computing methodologies that incorporate message-passing paradigms (to accommodate the distributed-memory architecture and highly parallel computing). This is crucial to achieve results within the prescribed computation time limit (21 days) of the computer cluster.

It is feasible to augment the efficiency of the aforementioned propagation algorithm by parallelizing all significant loops in OpenMP (open multi-processing). However, this approach would inherently constrain our calculations to smaller basis sets, $S,B$, and grid sizes, $N_t$, due to the restricted RAM and the limited number of CPU cores. This approach, however, still refers to the inhomogeneous Green's function as a shared-memory object, and thus fails to address the primary issue in propagation, namely the enormous amount of dynamic memory required to account for the Kadanoff-Baym equations' non-Markovian structure.

Throughout this section, we will show how existing parallel algorithms for large-scale computing with distributed memory can be improved and we will also establish a time stepping process that will be effective in accomplishing these large-scale computing tasks in the first place. For the purpose of circumventing the necessity to frequently repeat and spend a considerable amount of time accessing non-local memory, it is crucial to plan out a well-adapted distribution of memory and to utilize the specific structure of the collision integrals examined in the previous section \cite{BalzerPHD, PhysRevA.82.033427}. The resulting algorithm is designed to minimize communication between computer nodes.

The following discussion on distributed memory computing is based on the Message Passing Interface (MPI) \cite{10.5555/330577}. It is important to note that MPI employs a distinct approach in comparison to OpenMP. In MPI, multiple instances of the same program are initiated, with the number $\# ranks$ of instances determining the total number of MPI ranks. The process of parallelization is obtained by solving independent but yet synchronized subtasks on each rank. This is accomplished through the exchange of information via point-to-point and/or collective communications. The conventional method for identifying a particular MPI process, which involves designating the initial MPI rank as zero, is analogous to the convention of loops in C/C++ also starting at zero and continuing up to $\# ranks -1$. 

From the perspective of computational efficiency, it is advantageous in the context of parallel algorithms involving numerous MPI processes to have direct access for each MPI process to the Green's functions necessary to calculate specific collision integrals $ I_{1/2}^{\gtrless}$. This eliminates the need for Green's functions to be communicated between different MPI ranks, allowing for the rapid evaluation of collision integrals.
A scheme for implementing precomputed self energies is achieved through a decomposition of the discretized two-time plane into columns and rows of distinct perpendicular blocks \cite{BalzerPHD, PhysRevA.82.033427}. The distribution of memory is then allocated in a manner that assigns the inhomogeneous Green's function in different domains to specific MPI ranks, contingent on the number of MPI processes available \cite{BalzerPHD, PhysRevA.82.033427}.  

The optimal configuration involves storing the Green's function at point $(t_i,t'_j)$ in the memory of two MPI processes, designated $p_{a}$ and $p_{b}$. In the discretized two-time plane, any point can be expressed as a function of the (constant) grid spacing, $t_i=i \cdot \Delta$, $t'_j=j \cdot \Delta$, where $i,j \in \mathbb{N}_0$. The values for $p_{a}, p_{b} \in [0, \# ranks-1]$ can be computed in a straightforward manner

\begin{equation}
\begin{split}
&p_{a} = (t_i / \Delta) \, \% \, \# ranks = i \, \% \, \# ranks  ,\\
&p_{b} = (t_j / \Delta) \, \% \, \# ranks = j \, \% \, \# ranks ,
\label{Numerics_16}
\end{split}
\end{equation}

where $\%$ denotes the modulo operator. Of course this is not an optimal handling of the memory, when the two MPI processes are identical $p_{a}=p_{b}$. This will happen if, for example the corresponding grid coordinates $(i,j)$ are a $\# ranks$ multiple of the same rank. As an example we consider two MPI ranks, then all Green's functions at $(t_i,t'_j)=(2\cdot i \Delta, 2 \cdot j \Delta); \, i,j \in \mathbb{N}_0$ will be stored twice in rank $p_0$. In general, we have 

\begin{equation}
\begin{split}
&\sum_{a=0}^{\# ranks -1} \,\, \sum_{n=0}^{p_{a}+ \# ranks \cdot n< N_t} \, 2 \cdot N_t \cdot 16 \cdot S^2 \,  [byte]
\label{Numerics_17}
\end{split}
\end{equation}

reserved. If one divides \cref{Numerics_17} by the (shared) total amount of memory required \cref{Numerics_15.1}, 

\begin{equation}
\begin{split}
 \sum_{a=0}^{\# ranks -1} \,\, \sum_{n=0}^{p_{a}+ \# ranks \cdot n< N_t} \, \frac{2}{N_t} = 2,
\label{Numerics_18}
\end{split}
\end{equation}

the Green's function is stored twice. The memory stored per MPI rank instead decreases $\propto \frac{1}{\# ranks}$ \cite{BalzerPHD, PhysRevA.82.033427}, thus including more and more ranks will reduce the memory, which has to be stored in each rank, and enables in principle large $N_t$.  

If the memory is distributed in an appropriate manner as previously described, it allows for a straightforward parallel treatment of the collision integrals. In light of the preceding discussion, it is evident that Green's functions entering the collision integral directly are accessible, whereas those entering indirectly via the self energies are not. This is in accordance with the findings presented in \cref{Numerics_12,Numerics_15}. Fortunately, the structure of the self energies \cref{9} is compatible with the distribution of memory, thereby allowing them to be computed for given time arguments on a single MPI rank. To achieve optimal efficiency, a strategy that involves the parallel computation of $\Sigma^{\gtrless}$ on each individual rank  is recommended, followed by the subsequent dissemination of those values to the other ranks through collective MPI communication. Moreover, the step operators $U(\Delta)$ and $K(t, \Delta)$, must be calculated at each time step. In self-interacting systems, the MPI rank responsible for storing the diagonal time Green's function at time $(t,t)$ is uniquely capable of computing the Hartree term, but in the case of system-bath interaction, any rank could theoretically assume the task, obviating the need for additional communication, which would be a time-saving measure. Once the step operators and self energies have been distributed to all MPI ranks, the time steps for the Green's function can be calculated in both the $t$ and $t'$ directions. To achieve this, the collision integrals, $I_{1/2}^{\gtrless}$, are calculated at the specified points in time for the respective MPI ranks. Subsequently, the respective values of $c^{>}(t+\Delta,t')$ and $c^{<}(t',t+\Delta)$ can be calculated one at a time in sequence. At this juncture, the hybrid MPI-OpenMP parallelization, which has recently been incorporated into our algorithm, becomes operational. This enables the calculation of the distinct matrix elements, $c^{\gtrless}_{n,m} \, \forall n,m \in [0,S-1]$, for each MPI rank according to \cref{Numerics_11} using OpenMP. As previously stated, propagation to $c^{<}_{n,m}(t+\Delta,t+\Delta)$ on the time diagonal, which requires a distinctive approach according to the final equation in \cref{Numerics_11}, is carried out by the same MPI rank, propagating both $c^{>}(t+\Delta,t)$ and $c^{<}(t,t+\Delta)$, as for these steps $c^{\gtrless}(t,t)$ must be available at that rank. Once the corrector steps have been performed and the newly calculated Green's function has been correctly distributed to its storage locations, the diagonal elements are sent to new MPI ranks. In the following table, we will provide a detailed overview of the extended, parallelized algorithm (based on (\cite{BalzerPHD, PhysRevA.82.033427}).

\begin{enumerate}
  \item The Green's functions $c^{\gtrless}$ are initialized at every MPI rank $ n \in [0,\# rank-1]$ at time argument $t_0 = 0$.
  \item On each MPI rank the wave functions $\phi_m$ and the corresponding eigenvalues $E_m$ of $h_0$ are read in.
  \item The ``transition amplitudes" $V_{b,n,j,k}$ are calculated via \cref{10} using the full hybrid MPI-OpenMP parallelization and are broadcasted in sequence by each rank (via MPI\_Bcast) to all other ranks. In the case, that the interaction is with a heat bath only, the Hartree self energies can be computed in a similar manner.
  \item The actual calculation starts at time $t_i$ ($i=0$ initially) on each MPI rank. Memory for the self energies and collision integrals is allocated locally on each MPI rank.
  \item If the system is (also) self-interacting, it is necessary to diagonalize the effective potential $V_{\mathrm{eff}}$ on the ``main" MPI rank $i\%(\# ranks)$ and utilize the eigenvectors and eigenvalues with the armadillo library \cite{Sanderson_2019,sanderson2025armadilloefficientframeworknumerical} to compute the step operators $U(\Delta)$ and $K(t_i,\Delta)$. Subsequent to this, a broadcast is employed on the rank $i\%(\# ranks)$ to all other MPI ranks.
  \item The self energies $\Sigma^{>}(t_i,t_j)$ and $\Sigma^{<}(t_j,t_i)$ are calculated $\forall j \leq i$ on the corresponding MPI ranks $j\%(\# ranks)$ using the intrinsic OpenMP parallelization, which is a novel approach. If $i > (\# ranks)$ at least one MPI rank has to compute more than one pair of self energies $\Sigma^{\gtrless}$. Finally, at the end a broadcast of the self energies $\Sigma^{\gtrless}$ to each other MPI rank is performed. On each MPI rank, the received self energies have to be stored at the right positions.
  \item After a MPI\_Barrier statement, that forces synchronization of all ranks, till all needed self energies are received, the collision integrals $I_{1}^{>}(t_i,t_j)$ and $I_{2}^{<}(t_j,t_i)$ are calculated $\forall j \leq i$ on the corresponding MPI ranks $j\%(\# ranks)$ and afterwards the Green's function $c^{>}(t_{i+1},t_{j})$ and $c^{<}(t_{j},t_{i+1})$ are obtained via \cref{Numerics_11}. On MPI rank $i\%(\# ranks)$ especially, the Green's function $c^{<}(t_{i+1},t_{i+1})$ is evaluated. Again the loop over $\gtrless$ and the matrix elements of the Green's functions is parallelized intrinsically via OpenMP. At the end, each MPI rank (all if $i>(\# ranks)$) that participated in the propagation process of the Green's functions has to store them in their local memory and send them (via MPI\_Isend) to the next ``main" MPI rank $(i+1)\%(\# ranks)$. From this we can conclude, that the new main MPI rank does not have to send anything to itself. To be able to perform the corrector steps in the following, the collision integrals $I_{1/2}^{<}(t_i,t_i)$ in the time diagonal have to be sent as well to the new main rank.
  \item After the new main MPI rank has received (by MPI\_Recv) the collison integrals and Green's functions, additional memory for the new self energies and collision integrals is allocated.
  \item The effective potential $V_{\mathrm{eff}}$ is now diagonalized with the armadillo library \cite{Sanderson_2019,sanderson2025armadilloefficientframeworknumerical} on the new main MPI rank, which is the $(i+1)\%(\# ranks)$th rank. Consequently, the step operators $U(\Delta)$ and $K(t_{i+1},\Delta)$ are calculated on this rank and broadcast to all other MPI ranks.
  \item The self energies, $\Sigma^{>}(t_{i+1},t_j)$ and $\Sigma^{<}(t_j,t_{i+1})$, are calculated for all $j \leq i+1$ in order to accommodate the additional load of one MPI rank, if not all MPI ranks were already utilized in the fifth step. Finally, on each MPI rank, a broadcast of the self energies, denoted by $\Sigma^{\gtrless}$, is performed to the other MPI ranks, denoted by $l \neq j \% (\# ranks)$. On each MPI rank, the received self energies must be stored.
  \item On each MPI rank the collision integrals $I_{1}^{>}(t_{i+1},t_j)$ and $I_{2}^{<}(t_j,t_{i+1})$ are calculated $\forall j \leq i+1$ on the corresponding MPI ranks and afterwards (ensured by MPI\_Barrier) the Green's functions $c^{>}(t_{i+1},t_{j})$ and $c^{<}(t_{j},t_{i+1})$ are obtained via equation (\ref{Numerics_11}) but with the Trapezoidal rule: \\ $V_{\mathrm{eff}} \rightarrow \Bigl[ V_{\mathrm{eff}}(t) + V_{\mathrm{eff}}(t+\Delta) \Bigr] /2 $; $I_{1/2}^{\gtrless} \rightarrow \Bigl[ I_{1/2}^{\gtrless}(t, \cdot) + I_{1/2}^{\gtrless}(t+\Delta,\cdot) \Bigr] /2$.\\
  On MPI rank $(i+1)\%(\# ranks)$ the Green's function $c^{<}(t_{i+1},t_{i+1})$ is corrected. The Green's functions are stored in the local memory and transmitted (via MPI\_Isend) to the new main MPI rank $(i+1)\%(\# ranks)$. Once every communication is received, the propagation step corresponding to the $i$-th iteration is completed, and the process begins again at step 4 with $i\rightarrow i+1$.
\end{enumerate} \label{Numerics_procedure2}

Should one desire to extend the propagation of the Green's functions beyond the computing time limit of the cluster, each MPI rank stores its data at the conclusion of the calculation. However, the quantity of data to be stored on the scratch system becomes exceedingly large for advanced time evolution. This significantly impedes the algorithmic process and, in particular, necessitates the consumption of considerable resources on reinitialization.

\subsection{Extension to 3 dimensions}\label{3 dim extension}

The practical aspect of the numerical approach presented for solving the Kadanoff-Baym equations is that it is independent of the dimensionality of the space, which is a significant advantage. The most straightforward approach is to focus on the dimension $d = 3$, which will be relevant later. The only additional fundamental condition is that the system under consideration can evolve into discrete states \cite{Keldysh:1964ud, dahlen2006propagating, dahlen2007solving, Stan_2009,PhysRevA.82.033427}. The entire spatial dependence, for instance, of the wave functions or the interaction, is encapsulated in the transition amplitudes $V_{m,a,k,j}$ and is integrated out before the actual calculation begins. 

Although more complex from a numerical point of view, the method for computing these quantities is unambiguous. However, the addition of higher dimensions introduces a new challenge: the degeneracy of energy eigenstates. To achieve a comparable energy cut-off to one dimension, one must consider a larger number of states. In three dimensions, the Green's functions can be expanded using either Cartesian, 

\begin{equation}
\begin{split}
S^{>}(1,1') &= -i \sum_{n_x,n_y,n_z} \sum_{n'_x,n'_y,n'_z} \langle \hat{c}_{n_x,n_y,n_z}(t) \hat{c}_{n'_x,n'_y,n'_z}(t')^{\dagger} \rangle \phi_{n_x,n_y,n_z}(\vec{r}) \phi^{*}_{n'_x,n'_y,n'_z}(\vec{r'}), \\
S^{<}(1,1') &= \mp i \sum_{n_x,n_y,n_z} \sum_{n'_x,n'_y,n'_z} \langle \hat{c}_{n'_x,n'_y,n'_z}(t')^{\dagger} \hat{c}_{n_x,n_y,n_z}(t) \rangle \phi_{n_x,n_y,n_z}(\vec{r}) \phi^{*}_{n'_x,n'_y,n'_z}(\vec{r'}),
\label{Extension_3D_1}
\end{split}
\end{equation}

or spherical coordinates,

\begin{equation}
\begin{split}
S^{>}(1,1') &= -i \sum_{n_1,l_1,m_1} \sum_{n'_1,l'_1,m'_1} \langle \hat{c}_{n_1,l_1,m_1}(t) \hat{c}_{n'_1,l'_1,m'_1}(t')^{\dagger} \rangle \phi_{n_1,l_1,m_1}(\vec{r}) \phi^{*}_{n'_1,l'_1,m'_1}(\vec{r'}), \\
S^{<}(1,1') &= \mp i \sum_{n_1,l_1,m_1} \sum_{n'_1,l'_1,m'_1} \langle \hat{c}_{n'_1,l'_1,m'_1}(t')^{\dagger} \hat{c}_{n_1,l_1,m_1}(t) \rangle \phi_{n_1,l_1,m_1}(\vec{r}) \phi^{*}_{n'_1,l'_1,m'_1}(\vec{r'}).
\label{Extension_3D_2}
\end{split}
\end{equation}

In the case of Cartesian coordinates the quantum numbers are as usual $n_{x/y/z} \in [0, n_{\rm{max}}]$ and for spherical coordinates the quantum numbers are $n \in \mathbb{N}^{+}$, $l \in [0, n-1]$ and $m \in [-l,l]$. The eigenfunctions are given in this case via a product ansatz,

\begin{equation}
\begin{split}
\phi_{n_x,n_y,n_z}(\vec{r}) &= \phi_{n_x}(x) \, \phi_{n_y}(y) \, \phi_{n_z}(z) ,\\
\phi_{n,l,m}(\vec{r}) &= \mathrm{R}_{n,l}(r) \rm{Y}_{l,m}(\theta,\varphi).
\label{Extension_3D_3}
\end{split}
\end{equation}

The degeneracy enters naturally here, because in spherical coordinates the energies usually do not depend on $m$ and in Cartesian coordinates, when certain symmetries lead to degeneracies of the spectrum in different directions.

In both cases it is possible to find a bijective ``super" - index $n_s(n_x,n_y,n_z)$ or $n_s(n,l,m)$, which labels the states in a certain order, such that the same algorithm can be applied just with new computed transition amplitudes.

\subsection{Adaptive time stepping and integration }\label{subsec:Adaptivetimestepping}

Depending on the initial conditions, stiffness, or large coupling constants, small time-step sizes are often necessary to correctly resolve the dynamics and avoid instabilities. However, in order to observe long-time effects, such as thermalization, it is necessary to be able to enlarge the time-step size until sufficient accuracy is reached to have access to these time scales. Otherwise, an excessive number of time steps would be required, resulting in a computation time that scales at least cubically with the maximal number of time steps, $N_t$. This would cause the time per step to increase, thereby slowing down the overall calculation. 

One potential solution to this issue is to implement an adaptive multi-step method, e.g. the variable Adams method, as suggested in \cite{Meirinhos_2022,Blommel:2024zsr}.
In the present case, we propose the use of the adaptive ``Heun'' method as the simplest multi-step method. This is because the developed algorithm is already in this form, so there is no need for further changes in communication or parallelization.

The important point in all adaptive methods is the crucial estimation of the local error. Therefore, the consistency of the single-step method used must be adapted in the predictor step. A single-step method (e.g., simple Euler method) for solving an initial value problem

\begin{equation}
\begin{split}
\dot{y}(t) = \mathcal{F}\Bigl( t, y(t), \int y(t) \Bigr); \quad y(t_0) = y_0
\label{Numerics_19}
\end{split}
\end{equation}

is said to be of consistency order $k$, if for a prediction $y_{i+1}$ the single-step method complies 

\begin{equation}
\begin{split}
\lim_{\Delta_i \rightarrow 0} \,\, \sup_{t_{i} \in [t_{0} , t_{f}]} \frac{|y_{i+1} - y(t_{i}+\Delta_i)|}{\Delta_{i}^{k+1}} < \infty
\label{Numerics_20}
\end{split}
\end{equation}

for a given step size $\Delta_{i}$. This implies that there exist constants $C$ and $\Delta_0$ for which 

\begin{equation}
\begin{split}
\sup_{t_{i} \in [t_{0} , t_{f}]} |y_{i+1} - y(t_{i}+\Delta_{i})| \leq C \Delta_{i}^{k+1} 
\label{Numerics_21}
\end{split}
\end{equation}

holds $\forall \Delta_i \, \in (0, \Delta_0)$. It is evident that the simple Euler method is of consistency order 1, while the subsequent application of the "Heun" method, with consistency order 2, can be demonstrated by examining the error of the corresponding integration method, in this case, the simple Riemann blocks or trapezoidal rule.

It should be noted that in real-world applications, the precise value of a function at a given point in time is not always known. In such cases, it is common to compare the result obtained with a more accurate method, denoted by $\hat{y}_{i+1}$, which has a higher consistency value, $k' > k$.

\begin{equation}
\begin{split}
|y_{i+1} - y(t_{i}+\Delta_{i})| = |y_{i+1} - \hat{y}(t_{i}+\Delta_{i})| + \mathcal{O}(\Delta_{i}^{k'+1}).
\label{Numerics_22}
\end{split}
\end{equation}

To achieve a given tolerance $\epsilon_{\rm{abs}}$ with a optimal step size $\hat{\Delta}_{i}$, we demand 

\begin{equation}
\begin{split}
\epsilon_{\rm{abs}} = C \hat{\Delta}_{i}^{k+1}.
\label{Numerics_23}
\end{split}
\end{equation}

Inserting now \cref{Numerics_23,Numerics_22}) in \cref{Numerics_21} yields 

\begin{equation}
\begin{split}
\hat{\Delta}_{i} = \Delta_{i} \sqrt[k+1]{\frac{\epsilon_{\rm{abs}}}{|y_{i+1} - \hat{y} (t_{i}+\Delta_{i})|}} = \Delta_{i} \, \rm{eq_{i}}
\label{Numerics_24}
\end{split}
\end{equation}

for the new step size, where $\rm{eq_{i}}$ denotes the error quotient at this time step. Given the high cost of evaluating this error, in practice a safety factor of $0.9$ is often included. Also limiting factors for the maximal/minimal scaling are useful to avoid repeated adjusting o
f the step size via

\begin{equation}
\begin{split}
\hat{\Delta}_{i} =  \Delta_{i} \cdot \rm{min}( \rm{max_{scal}}, max( \rm{min_{scaling}}, \rm{safety\_factor} \cdot \rm{eq_{i}}))
\label{Numerics_25}
\end{split}
\end{equation}

and the correction of the step size is only performed, when $\rm{eq_{i}}<1$, and the same time step is repeated till $\rm{eq_{i}} \geq 1$. If this is achieved, the subsequent time step commences with the step size that has satisfied the requisite accuracy of the preceding step. 

When applying this procedure to the Kadanoff-Baym equations, one encounters a matrix-valued (integro)-differential equation, which necessitates the consideration of an appropriate error estimation. A suitable orientation for this error estimation is the diagonal time step $|c^{<}(t_{i+1},t_{i+1}) - \hat{c}^{<}(t_{i+1},t_{i+1})|$. In particular, the maximum norm is employed,

\begin{equation}
\begin{split}
|c^{<}(t_{i+1},t_{i+1}) -\hat{c}^{<}(t_{i+1},t_{i+1})| = \max_{n,m \in [0, S-1]} \, |c_{n,m}^{<}(t_{i+1},t_{i+1}) -\hat{c}_{n,m}^{<}(t_{i+1},t_{i+1})|.
\label{Numerics_26}
\end{split}
\end{equation}

In the previously described parallelized algorithm, the step size correction is initiated at step 11. on the new main MPI rank. However, in the event that a new step size is required to achieve the desired level of accuracy, a step back to 5. is necessary.

The entire mechanism of step-size control is based on the assumption that the right-hand side of \cref{Numerics_19} is known in an exact analytical form. However, this is not the case for integro-differential equations, as the integrals are approximated numerically via, for instance, the compound trapezoidal rule, which inherently introduces errors. Provided that the step size remains constant throughout the calculation, the addition of errors is not problematic as long as the order of accuracy for the value of the integral is at least of the order of consistency of the multi-step method employed. To illustrate, we demonstrate that Heun's method remains consistent when employing the compound trapezoidal rule for integral evaluation. In this context, the subsequent time step prediction is provided by

\begin{equation}
\begin{split}
y_{i+1} &= y_{i} + \frac{\Delta}{2} \Biggl[ \mathcal{F}\Bigl(t_i, y(t_i), \int y(t) + \mathcal{O}(\Delta^2) \Bigr) + \mathcal{F}\Bigl(t_{i+1}, y(t_{i+1}), \int y(t) + \mathcal{O}(\Delta^2)\Bigr) \Biggr] + \mathcal{O}(\Delta^3) \\
&= y_{i} + \frac{\Delta}{2} \Biggl[ \mathcal{F}\Bigl(t_i, y(t_i), \int y(t) \Bigr) + \mathcal{O}(\Delta^2) + \mathcal{F}\Bigl(t_{i+1}, y(t_{i+1}), \int y(t) \Bigr) + \mathcal{O}(\Delta^2) \Biggr] + \mathcal{O}(\Delta^3) \\
&= y_{i} + \frac{\Delta}{2} \Biggl[ \mathcal{F}\Bigl(t_i, y(t_i), \int y(t) \Bigr) + \mathcal{F}\Bigl(t_{i+1}, y(t_{i+1}), \int y(t) \Bigr) \Biggr] + \mathcal{O}(\Delta^3).
\label{Numerics_27}
\end{split}
\end{equation}

Thus, the error remains to be of order $\mathcal{O}(\Delta^3)$. But if the step size is adaptive, this is no longer the case, because the error of the compound trapezoidal rule scales with order $\mathcal{O}(\Delta_{\rm{max}}^2)$, where $\Delta_{\rm{max}} = \max_{j \in [0, N_t -1]} \, \Delta_j$. This approach may result in time steps that are less consistent than those obtained through the application of Heun's method, thus affecting the overall consistency of the solution obtained, 

\begin{equation}
\begin{split}
y_{i+1} &= y_{i} + \frac{\Delta_i}{2} \Biggl[ \mathcal{F}\Bigl(t_i, y(t_i), \int y(t) \Bigr) + \mathcal{F}\Bigl(t_{i+1}, y(t_{i+1}), \int y(t) \Bigr) \Biggr] + \mathcal{O}(\Delta_{\rm{max}}^2 \, \Delta_i).
\label{Numerics_28}
\end{split}
\end{equation}

The integration on a non-uniform grid can be enhanced by employing higher-order polynomials for interpolation, thereby achieving a higher order of accuracy. The two fundamental approaches under consideration are the Newton polynomials \cite{Meirinhos_2022} and the Lagrange polynomials, which will be the subject of a further discussion later, but also other approaches are possible \cite{Blommel:2024zsr}. The integrals we aim to solve are of the form

\begin{equation}
\begin{split}
\int_{t_{0}}^{t_i} d\bar{t} \, K(t,\bar{t},t') = \sum_{n=1}^{i} \int_{t_{n-1}}^{t_n} d\bar{t} \, K(t,\bar{t},t'),
\label{Numerics_29}
\end{split}
\end{equation}

where $K(t,\bar{t},t')$ is a product of Green's functions $c^{\gtrless}$ and self energies $\Sigma^{\gtrless}$ respectively. To compute these integrals on a non-uniform grid, the function $K$ is approximated by a polynomial of order $p$ (we will neglect the external times $t,t'$ in the following)

\begin{equation}
\begin{split}
P(\bar{t}) &= \sum_{j=0}^{p} a_j \cdot \phi_j(\bar{t}) \\
K(\bar{t}) &= P(\bar{t}) + E_p(\bar{t}),
\label{Numerics_30}
\end{split}
\end{equation}

where $\phi_j$ denotes a set of basis polynomials, e.g standard, Newton, or Lagrange basis and $E_p$ the error of the interpolation at the time point. In order to proceed, the coefficients $a_j$ must be calculated using the grid points $t_j$ and the corresponding function values $K_j$. This results in a set of linear equations, which can be written in matrix form as follows,

\begin{equation}
\begin{split}
\sum_{j}^{p} \phi_j(t_l) a_j = K_l.
\label{Numerics_31}
\end{split}
\end{equation}

$\phi_j(t_l)$ denotes the so-called ``Vandermonde" matrix, which is regular, as far as all grid points are distinct. The numerical value to be solved for in linear equations of this type scales as $\mathcal{O}(p^3)$, so it is common to attempt to identify a basis set where the Vandermonde matrix has a simple form, such as triangular (Newton) or diagonal (Lagrange). However, direct approaches, as described in \cite{Blommel:2024zsr}, which employ fast LU decomposition, are also used. 
The focus of this discussion will be on the Lagrange polynomials, which will prove to be numerically more efficient for calculational purposes. 
A disadvantage of the Lagrange polynomials in comparison to the Newton polynomials is that they cannot be enlarged, when adding new data points and want to increase the degree of the polynomial. In this case, one has to calculate all coefficients $a_j$ again. The effort to solve the system of linear equations is undoubtedly the simplest in the case of Lagrange basis, because the Vandermonde matrix is diagonal and no ``divided differences" have to be calculated. However, the construction of the Lagrange basis is more complicated than the Newton basis, resulting in a numerical effort of the order of $\mathcal{O}(p^2)$ in both cases. This can be seen directly from the definition of the basis polynomials, 

\begin{equation}
\begin{split}
n_j(\bar{t}) &= \prod_{n=0}^{j-1} (\bar{t}-t_n), \\
l_j(\bar{t}) &= \prod_{n=0; n\neq j}^{p} \frac{\bar{t}-t_n}{t_j-t_n}.
\label{Numerics_32}
\end{split}
\end{equation}

The Newton polynomials are said to be more stable, especially when the grid points are no longer uniformly distributed and the degree of the polynomial is higher ($>4$), see Runge's phenomenon. This will limit us in the order of the polynomial interpolation, but we will leverage the significant advantage of the Lagrange polynomials, namely that not only the associated basis functions, $l_j$, are independent of the interpolation values, $K_j$, implies that different sets of interpolation values, $K_j$, with identical interpolation points, $\bar{t}_j$, can be interpolated expediently once the basis functions, $l_j$, have been established, but also the coefficients of the polynomial are directly given by the interpolating function values $a_j=K_j$.

This is precisely the case for the Kadanoff-Baym equations, where the Green's functions $c_{n,m}^{\gtrless}$ are all situated on the same grid in the two-time plane. Consequently, at a given time in this plane, the Lagrange basis polynomials are evaluated only once for all $S^2$ Green's functions, which renders this approach more advantageous as the number of included eigenstates $S$ increases. All integrals are then reduced to integrals over the Lagrange basis polynomials, which are known analytically. A problem, which cannot be circumvented in all direct quadrature methods, is the fact, that for every $n$ in \cref{Numerics_29} this procedure has to be done. The aforementioned approach thus appears to be an even more efficacious methodology when one's objective is to ascertain long-term behavioral tendencies. Then \cref{Numerics_29} takes the following form, dropping the outer time indices,

\begin{equation}
\begin{split}
\int_{t_{0}}^{t_i} d\bar{t} \, K(\bar{t}) &= \sum_{n=1}^{i} \int_{t_{n-1}}^{t_n} d\bar{t} \, K(\bar{t}) \approx \sum_{n=1}^{i} \int_{t_{n-1}}^{t_n} d\bar{t} \, P_n(\bar{t}) = \sum_{n=1}^{i} \int_{t_{n-1}}^{t_n} d\bar{t} \, \sum_{j=0}^{p} K^{(n)}_j \cdot l^{(n)}_j(\bar{t}) \\
&= \int_{t_{0}}^{t_p} d\bar{t} \, \sum_{j=0}^{p} K_j \cdot \prod_{f=0;f \neq j}^{p} \frac{\bar{t}-t_f}{t_j-t_f} + \sum_{n=p+1}^{i} \int_{t_{n-1}}^{t_n} d\bar{t} \, \sum_{j=n-p}^{n} K_j \cdot \underbrace{\prod_{f=n-p;f \neq j}^{n} \frac{\bar{t}-t_f}{t_j-t_f}}_{=l^{(n)}_j(\bar{t})} \\
&= \sum_{j=0}^{p} K_j \cdot \int_{t_{0}}^{t_p} d\bar{t} \,  l^{(p)}_j(\bar{t})  + \sum_{n=p+1}^{i} \sum_{j=n-p}^{n} K_j \cdot \int_{t_{n-1}}^{t_n} d\bar{t} \, l^{(n)}_j(\bar{t}).
\label{Numerics_33}
\end{split}
\end{equation}

Now, we want to know what we have gained in terms of accuracy at the end. Therefore, we want to have a closer look at the error $E_p(t)$ in \cref{Numerics_30} and also the resulting error of the integrals we calculate. We will do this here for the case of the Newton polynomial, but for other polynomials, it will follow similarly. From standard textbook knowledge we know from Taylor expansions and mean value theorem, that the error can be estimated by

\begin{equation}
\begin{split}
E_p(t) = K(t) - P(t) = \frac{K^{(p+1)}(\xi)}{(p+1)!} \prod_{i=0}^p (t-t_i),
\label{Numerics_35.1}
\end{split}
\end{equation}

where $\xi \in [t_0, t_p]$. Here we can also see, that for equidistant time steps the error can be estimated upwards by 

\begin{equation}
\begin{split}
E_p(t) \leq \frac{K^{(p+1)}(\xi)}{(p+1)!} ((p-1) \cdot \Delta)^{p}.
\label{Numerics_35.11}
\end{split}
\end{equation}

But how does this now effect the error of the integral? We remember from \cref{Numerics_29}, that we only have to do this for the small divided intervals $[t_{n-1}, t_n]$. Integrating \cref{Numerics_35.1} over one time step yields,

\begin{equation}
\begin{split}
\int_{t_{n-1}}^{t_n} d\bar{t} \, E_p(\bar{t}) = \frac{K^{(p+1)}(\xi)}{(p+1)!} \int_{t_{n-1}}^{t_n} d\bar{t} \, \prod_{i=0}^p (\bar{t}-t_{n-i}) \leq \frac{1}{(p+1)!} \max_{t_{n-p} < \xi < t_n}|K^{(p+1)}(\xi)|  \int_{t_{n-1}}^{t_n} d\bar{t} \, |\prod_{i=0}^p (\bar{t}-t_{n-i})|,
\label{Numerics_35.2}
\end{split}
\end{equation}

and after applying \cref{Numerics_33}, we end up with the error of the whole integration given by

\begin{equation}
\begin{split}
\int_{t_{0}}^{t_i} d\bar{t} \, E_p(\bar{t}) &= \int_{t_{0}}^{t_p} d\bar{t} \, E_p(\bar{t}) + \sum_{n=p+1}^{i} \int_{t_{n-1}}^{t_n} d\bar{t} \, E_p(\bar{t}) \leq \frac{1}{(p+1)!} \max_{t_{0} < \xi < t_p}|K^{(p+1)}(\xi)|  \int_{t_{0}}^{t_p} d\bar{t} \, |\prod_{i=0}^p (\bar{t}-t_{p-i})| \\
&+\sum_{n=p+1}^{i} \frac{1}{(p+1)!} \max_{t_{n-p} < \xi < t_n}|K^{(p+1)}(\xi)|  \int_{t_{n-1}}^{t_n} d\bar{t} \, |\prod_{i=0}^p (\bar{t}-t_{n-i})|.
\label{Numerics_35.3}
\end{split}
\end{equation}

The final step is to select the optimal degree of the polynomial, $p_{\rm{opt}}$, to minimize the error. We can see in \cref{Numerics_35.3}, that larger $p$ do not have to lead to smaller errors. This may be due to oscillatory behavior, cf. Runges phenomenon, which implies large higher derivatives in the interval of interest. And if $p$ this is chosen too small, the approximation of the integral may be too inaccurate, as evidenced by the compound trapezoidal rule, so up to now it is not clear how to select and appropriate order.

In order to identify $p_{\rm{opt}}$, a straightforward test case scenario is employed, as outlined in \cite{Blommel:2024zsr}, where a test function is integrated over the time interval utilizing a range of interpolation polynomials at varying degrees. An appropriate choice of this test function could be the collision term, that enters the diagonal time step in the two-time plane,

\begin{equation}
\begin{split}
C^{n}_{\rm{test}} &= \int_{t}^{t_{0}} d\bar{t} \, \sum_{i}^{S} \biggl[ \Sigma^{>}_{n,i}(t,\bar{t}) \, c^{<}_{i,n}(\bar{t},t) - \Sigma^{<}_{n,i}(t,\bar{t}) \, c^{>}_{i,n}(\bar{t},t) - c^{>}_{n,i}(t,\bar{t}) \, \Sigma^{<}_{i,n}(\bar{t},t) + c^{<}_{n,i}(t,\bar{t}) \, \Sigma^{>}_{i,n}(\bar{t},t) \biggr].
\label{Numerics_34}
\end{split}
\end{equation}

The optimal degree is then obtained by \cite{Blommel:2024zsr}

\begin{equation}
\begin{split}
p_{\rm{opt}} &= \min_{p \in [2,p_{\rm{max}}]} \, \Biggl[ \, \sum_{n=0}^{S-1} |C^{n}_{\rm{test}_{p}} - C^{n}_{\rm{test}_{p-1}}| \, \Biggr].
\label{Numerics_35}
\end{split}
\end{equation}

The value $|C^{n}_{\rm{test}_{p_{\rm{opt}}}} - C^{n}_{\rm{test}_{p_{\rm{opt}}-1}}|$ can then be used as an error estimation in the algorithm, which can be used to adjust the next timestep size. This is done in our case by an upper boundary of this integration error and when the current error overshoots this value, the time-step size cannot be increased in this step.

It should be noted that this selection does not require a significant computational effort in comparison to evaluating $2(2 \cdot j +1) \cdot S^{2}$ in the $j$-th time step. This mechanism is repeated at each step-size adaptation. As the selection of the optimal order is conducted on the main MPI rank, a broadcast of $p_{\rm{opt}}$ to all other ranks is inserted after its computation (before Step 7. in \cref{Numerics_procedure2}).

\subsection{On higher-order time-stepping with the variable Adams method }

As an improvement of the predictor-corrector method described in \cref{subsec:Algorithm} applied in earlier works \cite{ KOHLER1999123,DANIELEWICZ1984305, Juchem_2004, dahlen2006propagating, dahlen2007solving, Stan_2009, PhysRevB.82.155108}, in recent works \cite{Meirinhos_2022,Blommel:2024zsr} a more accurate approximation of the integrals in the last steps of \cref{Numerics_4,Numerics_6,Numerics_8} has been achieved by using higher-order polynomials for interpolation similar to the evaluation of the memory integrals in the previous \cref{subsec:Adaptivetimestepping}.

This results in the so-called (variable) Adams methods depending on the order of the polynomial used.
However, the implementation is straightforward, when using the routines for the memory integrals. 
We show this as an example for \cref{Numerics_4}, which can of course be generalized to the other time steps, \cref{Numerics_6,Numerics_8}. For the explicit/predictor step, we use the existing values of $I_1^{>}$ and $U$ for $\bar{t} \leq t_i$ to build a polynomial, which is then used for extrapolation in the integration interval $[t_i,t_i +\Delta_i]$,

\begin{equation}
\begin{split}
c^{>}(t_i +\Delta_i,t') &= U(\Delta_i) \Biggl[ c^{>}(t_i,t') + \int_{t_i}^{t_i+\Delta_i} d \bar{t} \, U(t_i-\bar{t}) I^{>}_{1}(\bar{t},t') \Biggr] \\
&= U(\Delta_i) \Biggl[ c^{>}(t_i,t') + \int_{t_i}^{t_{i+1}=t_i+ \Delta_i} d\bar{t} \, \sum_{j=0}^{p} \prod_{f=0;f \neq j}^{p} \frac{\bar{t}-t_{i-f}}{t_{i-j}-t_{i-f}} \cdot U(t_i-t_{i-j}) I^{>}_{1}(t_{i-j},t') \Biggr].
\label{Numerics_36}
\end{split}
\end{equation}

In the implicit/corrector step, all calculated values up to $t_{i+1}$ are used to construct the interpolation polynomial for the integration interval $[t_i,t_{i+1}]$. The formula can be obtained from the explicit/predictor step by shifting the indices under the integral from $i \rightarrow i+1$,

\begin{equation}
\begin{split}
c^{>}(t_i +\Delta_i,t') &= U(\Delta_i) \Biggl[ c^{>}(t_i,t') + \int_{t_i}^{t_i+\Delta_i} d \bar{t} \, U(t_i-\bar{t}) I^{>}_{1}(\bar{t},t') \Biggr] \\
&= U(\Delta_i) \Biggl[ c^{>}(t_i,t') + \int_{t_i}^{t_{i+1}=t_i+ \Delta_i} d\bar{t} \, \sum_{j=0}^{p} \prod_{f=0;f \neq j}^{p} \frac{\bar{t}-t_{i+1-f}}{t_{i+1-j}-t_{i+1-f}} \cdot U(t_{i}-t_{i+1-j}) I^{>}_{1}(t_{i+1-j},t') \Biggr].
\label{Numerics_37}
\end{split}
\end{equation}

The important and more advanced part is to manage the memory correctly. For the first two equations in (\ref{Numerics_11}) no significant changes regarding MPI have to be done, because these propagation steps are always on the same MPI rank, just the memory integrals $I^{\gtrless}_{1/2}$ and the evolution operator $U(t)$ have to be stored for the last $p+1$ time-steps (including the current one). 
In the algorithm, this is realized by implementing the arrays for the collision integrals twice, namely for the explicit and the implicit time step. Before the (explicit) memory integrals $I_{1/2}^{\gtrless}$ are calculated in step 7. of the parallelized algorithm, the array of the implicit collision integral is copied into the explicit one. Then the value of the (explicit) collision integral is calculated and overwrites the last entry coming from the array of the implicit collision integral. After performing the explicit time step, the entries of the explicit array are copied into the implicit one by a shift of 1 entry, allowing to let the first entry free, where the value of the (implicit) collision integral can be placed after its calculation in step 11. of the parallelized algorithm.

For the time diagonal, this is more difficult. In principle the steps are the same as described before, but after performing the explicit time step, the explicit array cannot be copied into the implicit one directly because the MPI rank, which is performing the implicit time-step changes. All shifted entries have to be sent, compare step 7. of the parallelized algorithm.

At the end of this rather abstract numerical section, we have to answer the question, if this amount of work has been totally worth it and can be applied without fearing any instabilities, which may arise due to error resummation or oscillatory behavior of the interpolating polynomials.
This is unfortunately another topic in mathematics, which has developed under the name ``(linear) control theory" since the 60s of the last century \cite{DGLbook,linmultistepstability}. In short: There is no guaranteed mechanism for e.g. order selection even for ordinary differential equations, but just ``(heuristic) dynamic error models" including the history of time steps in the calculation of the local error \cite{linmultistepstability}. So, what should they be for ordinary integro-differential equations?
The most important observation during our calculations was that the changes in time step size should be smooth ($\Delta_i/\Delta_{i-1} \approx 1$), which can be achieved by a limitation of the change in step size \cite{linmultistepstability}, and the order of the time stepping should be smaller than the order of integration when calculating the memory integrals. An all-purpose weapon, it is by no means, and also high orders like $p>5$ have turned out to be highly unstable, contradicting the experiences of the authors in \cite{Meirinhos_2022}.

\section{Derivation of the master equation}\label{sec:Master}

The following section presents a concise derivation of the Quantum-Kinetic-Master (QKM) equation, starting directly from the Kadanoff-Baym equations in the two-time and eigenstate representation as utilized within this work. A brief review of this derivation is provided for the sake of two key objectives: first, to highlight the connection between the full Kadanoff-Baym equations and its approximated form, and second, to elucidate the assumptions that enter the QKM equation. 

The QKM equation, which describes the temporal evolution of distribution functions for quantum states, is analogous to the Boltzmann equation, which describes quasi-particle behavior. 
 
We can use now the equation for the time-diagonal \cref{12} to derive an equation of motion for the distribution function of the quantum states, 

\begin{equation}
\begin{split}
- & \frac{\partial}{\partial t} c^{<}_{b,a}(t,t) - i [E,c^{<}(t,t)]_{b,a} = \int_{t_0}^{t}d\bar{t} \, \sum_{i}^{S} \Bigl( \Sigma^{>}_{b,i}(t,\bar{t}) \, c^{<}_{i,a}(\bar{t},t) - \Sigma^{<}_{b,i}(t,\bar{t}) \, c^{>}_{i,a}(\bar{t},t) + c^{<}_{b,i}(t,\bar{t}) \, \Sigma^{>}_{i,a}(\bar{t},t) - c^{>}_{b,i}(t,\bar{t}) \, \Sigma^{<}_{i,a}(\bar{t},t) \Bigr).
\label{QKME_1}
\end{split}
\end{equation}

Usually the on-shell energy $E_a$ can depend on time, when self interacting Hartree terms are taken into account $E_a \rightarrow E_a + \Sigma_{Ha,a}(t)$. To eliminate the commutator, the phase shifts of the Green's function (compare \cref{8}) are taken explicitly, similar to the $U(1)$ gauge transformation applied in definition \cref{Numerics_2},

\begin{equation}
\begin{split}
c^{\gtrless}_{b,a}(t,t') \rightarrow e^{-i E_b t} c^{\gtrless}_{b,a}(t,t') e^{iE_a t'} .
\label{QKME_2}
\end{split}
\end{equation}

This has the advantage that the commutator in \cref{QKME_1} is eliminated, and the time evolution of the quantum state distribution function is determined solely by the collision terms, self energies, and Green's functions within. Inserting \cref{QKME_2} in \cref{QKME_1} yields for the right hand side terms,

\begin{equation}
\begin{split}
\int_{t_0}^{t}d\bar{t} \, \sum_{i}^{S} \Sigma^{>}_{b,i}(t,\bar{t}) \, c^{<}_{i,a}(\bar{t},t) &= \int_{t_0}^{t}d\bar{t} \sum_{m,n,i}^{S} \sum_{j,k}^{B} \, (1\pm n_{B/F}(\epsilon_{j})) \, n_{B/F}(\epsilon_{k}) \, V_{b,n,j,k} \, c^{>}_{n,m}(t,\bar{t}) \, V_{m,i,k,j} \, c^{<}_{i,a}(\bar{t},t) \, e^{-i (\epsilon_{j} -E_a -\epsilon_{k} +E_n)t} \\
& \quad \quad  \quad \quad  \quad \quad \quad \quad  e^{i (\epsilon_{j} -E_i -\epsilon_{k} +E_m) \bar{t}} , \\
\int_{t_0}^{t}d\bar{t} \, \sum_{i}^{S} \Sigma^{<}_{b,i}(t,\bar{t}) \, c^{>}_{i,a}(\bar{t},t) &= \int_{t_0}^{t}d\bar{t} \sum_{m,n,i}^{S} \sum_{j,k}^{B} \, (1 \pm n_{B/F}(\epsilon_{j})) \, n_{B/F}(\epsilon_{k}) \, V_{b,n,j,k} \, c^{<}_{n,m}(t,\bar{t}) \, V_{m,i,k,j} \, c^{>}_{i,a}(\bar{t},t) \, e^{i (\epsilon_{j} -E_a -\epsilon_{k} +E_n)t} \\
& \quad \quad  \quad \quad  \quad \quad \quad \quad  e^{-i (\epsilon_{j} -E_m -\epsilon_{k} +E_i) \bar{t}} , \\
\int_{t_0}^{t}d\bar{t} \, \sum_{i}^{S} c^{<}_{b,i}(t,\bar{t}) \,  \Sigma^{>}_{i,a}(\bar{t},t) &= \int_{t_0}^{t}d\bar{t} \sum_{m,n,i}^{S} \sum_{j,k}^{B} \, (1 \pm n_{B/F}(\epsilon_{j})) \, n_{B/F}(\epsilon_{k}) \, V_{i,n,j,k} \, c^{>}_{n,m}(t,\bar{t}) \, V_{m,a,k,j} \, c^{<}_{b,i}(\bar{t},t) \, e^{i (\epsilon_{j} -E_b -\epsilon_{k} +E_m)t} \\
& \quad \quad  \quad \quad  \quad \quad \quad \quad  e^{-i (\epsilon_{j} -E_n -\epsilon_{k} +E_i) \bar{t}} , \\
\int_{t_0}^{t}d\bar{t} \, \sum_{i}^{S} c^{>}_{b,i}(t,\bar{t}) \,  \Sigma^{<}_{i,a}(\bar{t},t) &= \int_{t_0}^{t}d\bar{t} \sum_{m,n,i}^{S} \sum_{j,k}^{B} \, (1 \pm n_{B/F}(\epsilon_{j})) \, n_{B/F}(\epsilon_{k}) \, V_{i,n,j,k} \, c^{<}_{n,m}(t,\bar{t}) \, V_{m,a,k,j} \, c^{>}_{b,i}(\bar{t},t) \, e^{-i (\epsilon_{j} -E_m -\epsilon_{k} +E_b)t} \\
& \quad \quad  \quad \quad  \quad \quad \quad \quad  e^{i (\epsilon_{j} -E_n -\epsilon_{k} +E_i) \bar{t}} .
\label{QKME_3}
\end{split}
\end{equation}

As we are interested in developing a pure quantum-kinetic equation that is comparable to a Boltzmann equation for discrete quantum states, we have made several approximations, as outlined in \cite{Juchem_2004}. All quantum correlations between these quantum states will be neglected, resulting in the matrix-valued coefficients becoming vector-valued. To ensure the equations are Markovian, the vector-valued occupation numbers of the quantum states are reduced to their maximal-time argument. This involves neglecting all spectral information and retaining only the statistical information, as discussed in \cite{PhysRevC.49.1693,PhysRevC.51.3232,Juchem_2004}. This results in the following proposed replacement for the equation,

\begin{equation}
\begin{split}
c^{\gtrless}_{b,a}(t,t') \rightarrow \delta_{b,a} \, c^{\gtrless}_{b,a}(t_{\rm{max}},t_{\rm{max}}) = c^{\gtrless}_{b}(t_{\rm{max}}),
\label{QKME_4}
\end{split}
\end{equation}

where the maximal time $t_{\rm{max}} = \rm{max}(t,t')$ has been introduced \cite{Juchem_2004}. The time integration inherent to the collision integrals will now be rendered inconsequential, as all Green's functions, which previously depended on twice the local (maximal) time and the integration variable, can be extracted from the time integration. Inserting \cref{QKME_4} in \cref{QKME_3} and using $V_{b,n,k,j} = V_{n,b,j,k}^{*}$ yields,

\begin{equation}
\begin{split}
\Sigma^{>}_{b}(t) \, c^{<}_{b}(t) &= \sum_{n}^{S} \sum_{j,k}^{B} \, (1 \pm n_{B/F}(\epsilon_{j})) \, n_{B/F}(\epsilon_{k}) \, |V_{b,n,k,j}|^2 \, c^{>}_{n}(t,\bar{t}) \, c^{<}_{b}(t) \, \int_{t_0}^{t}d\bar{t} \, e^{-i(t-\bar{t})(\epsilon_{j} -E_b -\epsilon_{k} +E_n)}, \\
\Sigma^{<}_{b}(t) \, c^{>}_{b}(t) &= \sum_{n}^{S} \sum_{j,k}^{B} \, (1 \pm n_{B/F}(\epsilon_{j})) \, n_{B/F}(\epsilon_{k}) \, |V_{b,n,k,j}|^2 \, c^{<}_{n}(t) \, c^{>}_{b}(t) \, \int_{t_0}^{t}d\bar{t} \,
e^{i(t-\bar{t}) (\epsilon_{j} -E_n -\epsilon_{k} +E_b)}, \\
c^{<}_{b}(t) \,  \Sigma^{>}_{b}(t) &=  \sum_{n}^{S} \sum_{j,k}^{B} \, (1 \pm n_{B/F}(\epsilon_{j})) \, n_{B/F}(\epsilon_{k}) \, |V_{b,n,k,j}|^2 \, c^{>}_{n}(t) \, c^{<}_{b}(t) \, \int_{t_0}^{t}d\bar{t} \, e^{i(t-\bar{t}) (\epsilon_{j} -E_b -\epsilon_{k} +E_n)}, \\
c^{>}_{b}(t) \,  \Sigma^{<}_{b}(t) &=  \sum_{n}^{S} \sum_{j,k}^{B} \, (1 \pm n_{B/F}(\epsilon_{j})) \, n_{B/F}(\epsilon_{k}) \, |V_{b,n,k,j}|^2 \, c^{<}_{n}(t) \, c^{>}_{b}(t) \, \int_{t_0}^{t}d\bar{t} \, e^{-i(t-\bar{t}) (\epsilon_{j} -E_n -\epsilon_{k} +E_b)}.
\label{QKME_5}
\end{split}
\end{equation}

The first and the third and the second and the fourth term in \cref{QKME_5}, which are added/subtracted in \cref{QKME_1}, can be combined, which results in a cosine under the integral,

\begin{equation}
\begin{split}
\Sigma^{>}_{b}(t) \, c^{<}_{b}(t) + c^{<}_{b}(t) \,  \Sigma^{>}_{b}(t) &= \sum_{n}^{S} \sum_{j,k}^{B} \, (1 \pm n_{B/F}(\epsilon_{j})) \, n_{B/F}(\epsilon_{k}) \, |V_{b,n,k,j}|^2 \, c^{>}_{n}(t,\bar{t}) \, c^{<}_{b}(t) \, \int_{t_0}^{t}d\bar{t} \, 2 \, \mathrm{cos}\biggl((t-\bar{t})(\epsilon_{j} -E_b -\epsilon_{k} +E_n)\biggr), \\
\Sigma^{<}_{b}(t) \, c^{>}_{b}(t) + c^{>}_{b}(t) \,  \Sigma^{<}_{b}(t) &= \sum_{n}^{S} \sum_{j,k}^{B} \, (1 \pm n_{B/F}(\epsilon_{j})) \, n_{B/F}(\epsilon_{k}) \, |V_{b,n,k,j}|^2 \, c^{<}_{n}(t) \, c^{>}_{b}(t) \, \int_{t_0}^{t}d\bar{t} \, 2 \, \mathrm{cos}\biggl((t-\bar{t}) (\epsilon_{j} -E_n -\epsilon_{k} +E_b)\biggr).
\label{QKME_6}
\end{split}
\end{equation}

For the distribution function the equal time (anti) commutation relation can be used, $c^{>}_{n}(t) = 1 \pm c^{<}_{n}(t)$, resulting in the Bose enhancement factor or Pauli blocking factor known from the usual Boltzmann equation. The terms in \cref{QKME_1} describe ($2 \leftrightarrow 2$) scattering processes, as is well-established in the literature. Consequently, we find the typical gain and loss terms in \cref{QKME_6} here, where particles in the quantum state $b$ can be scattered out into quantum states $n$ or vice versa. 

As for the full Green's function the resulting evolution equation is depending on the initial value (occupation) of the various quantum states at time $t_0$. One would immediately ask oneself, why there is no energy-conserving delta-function as in the common Boltzmann equation? We will briefly address later, why a completely conserving approach is useless, when the energy levels are discrete, but for now we will see, what happens, when we calculate the integrals in \cref{QKME_6}, 

\begin{equation}
\begin{split}
&2 \, \int_{t_0}^{t}d\bar{t} \, \mathrm{cos}\biggl((t-\bar{t}) (\epsilon_{j} -E_n -\epsilon_{k} +E_b)\biggr) = -2 \int_{t-t_0}^{0}d\hat{t} \, \mathrm{cos}\biggl(\hat{t} (\epsilon_{j} -E_n -\epsilon_{k} +E_b)\biggr) \\
= \, &2 \int_{0}^{t-t_0}d\hat{t} \, \mathrm{cos}\biggl(\hat{t} (\epsilon_{j} -E_n -\epsilon_{k} +E_b)\biggr) = \frac{2}{(\epsilon_{j} -E_n -\epsilon_{k} +E_b)} \, \mathrm{sin}\biggl((t-t_0) (\epsilon_{j} -E_n -\epsilon_{k} +E_b)\biggr).
\label{QKME_7}
\end{split}
\end{equation}

In the limit as $t-t_0 \rightarrow \infty$ for fixed $t_0$, the sinc function converges to an energy-conserving delta function, as anticipated from the Boltzmann equation,

\begin{equation}
\begin{split}
\lim_{t-t_0 \rightarrow \infty} \frac{2}{(\epsilon_{j} -E_n -\epsilon_{k} +E_b)} \rm{sin}\biggl((t-t_0) (\epsilon_{j} -E_n -\epsilon_{k} +E_b)\biggr) = 2 \pi \delta(\epsilon_{j} -E_n -\epsilon_{k} +E_b).
\label{QKME_8}
\end{split}
\end{equation}

However, distributions are only mathematically meaningful below an integral, so in the limit of continuous energy levels. As we are not doing this, this requires some physical interpretation in analogy to \cite{PhysRevA.55.2902,PhysRevA.56.575}. 

In the case of discrete states, the distance between the levels, denoted as $\Delta E$, depends on the properties of the enclosing potential. Consequently, each discrete state $i$ represents a specific energy interval $[E_i - \frac{\Delta E}{2}, E_i + \frac{\Delta E}{2}]$.  It is likewise feasible to discretize a continuous system in this manner, which is invariably undertaken when the objective is to simulate them accurately. The occupation numbers of the quantum states can then be interpreted as an average of the continuous distribution function, $n(\omega)$, over the specific energy interval, as evidenced in \cite{PhysRevA.56.575},

\begin{equation}
\begin{split}
c_{i}^{<} = \int_{E_i - \frac{\Delta E}{2}}^{E_i + \frac{\Delta E}{2}} \frac{d\mathrm{\omega}}{2\pi} \frac{n(\omega)}{\Delta E}.
\label{QKME_9}
\end{split}
\end{equation}

Accordingly, the delta function as defined in \cref{QKME_8} should be understood to represent \cite{PhysRevA.56.575}

\begin{equation}
\begin{split}
\delta(\epsilon_{j} -E_n -\epsilon_{k} +E_b) = \frac{\delta_{\epsilon_{j} -E_n -\epsilon_{k} +E_b, 0}}{\Delta E},
\label{QKME_10}
\end{split}
\end{equation}

where $\delta_{\epsilon_{j} -E_n -\epsilon_{k} +E_b, 0}$ is the Kronecker delta. The equation has now become numerically manageable. However, the Kronecker delta is only physically meaningful if the energy gaps, represented by the symbol $\Delta E$, are constant and do not depend on the state. Nevertheless, this is only the case for the harmonic oscillator or the discretization of continuous systems, where the degree of discretization can be freely selected. In general, this is not the case. In such instances, the forced conservation of energy would preclude the occurrence of significant scattering processes, with only ``trivial" scattering being possible. 

To circumvent this issue and avoid the use of a diverging delta function, we restrict the sinc function to certain widths. Thus the final QKM equation is now obtained as,

\begin{equation}
\begin{split}
\frac{\partial}{\partial t} c^{<}_{b}(t) = & \sum_{n}^{S} \sum_{j,k}^B \Bigl( (1 \pm n_{B/F}(\epsilon_{j})) \, n_{B/F}(\epsilon_{k}) \, c^{>}_{b}(t) \, c^{<}_{n}(t) - (1 \pm n_{B/F}(\epsilon_{k})) \, n_{B/F}(\epsilon_{j}) c^{>}_{n}(t) \, c^{<}_{b}(t) \Bigr) \\ 
& |V_{b,n,k,j}|^{2} \frac{2}{(\epsilon_{j} -E_n -\epsilon_{k} +E_b)} \rm{sin}\biggl((\tilde{t}-t_0) (\epsilon_{j} -E_n -\epsilon_{k} +E_b)\biggr), \\
& \tilde{t} \defeq \begin{cases}
t_{\rm{max}} &\text{if $t > t_{\rm{max}}$,}\\
t &\text{if $t \le t_{\rm{max}}$}. 
\end{cases},
\label{QKME_11}
\end{split}
\end{equation}

where $t_{\rm{max}}$ ensures, that the sinc does not diverge for large times. 
Another limitation of the expression (\ref{QKME_7}) is that it is identically zero when $t=t_0=0$. This issue can be addressed by shifting the reference point to an earlier time than the start of the actual calculation \cite{Juchem_2004}.

It is now necessary to provide a brief recapitulation of the approximations that were employed in the derivation of the QKM equation. Two of these approximations are directly present in \cref{QKME_2}. The first is the negligibility correlations in contrast to the Kadanoff-Baym equations. The second is the Markovianity of the QKM equation. Another significant approximation is the on-shell (including Hartree terms) ansatz for the dispersion relation, which is only valid when the collisional broadening of the state is much smaller than the energy gap $\Delta E$. Otherwise, the quasi-particle interpretation is meaningless because the properties of a particle with a certain energy cannot be described by a single state of the single-particle Hamiltonian \cite{PhysRevA.56.575}.

\section{KMS Condition}\label{KMS}
This property, which is fundamental to all Green's functions (not only two-point functions), goes back to R. Kubo, P. C. Martin and J. Schwinger \cite{doi:10.1143/JPSJ.12.570, PhysRev.115.1342}.
We want to give a short derivation which is based on \cite{KBBook}.
First the $c^{<}$ Green's function can be rearranged using the cyclic property of the trace,

\begin{equation}
\begin{split}
 c^{<}_{n,m}(t,t') &= \langle \hat{c}_{m}(t')^{\dagger} \hat{c}_{n}(t) \rangle =  \frac{\mathrm{Tr}\Bigl( e^{-\beta(\hat{H}-\mu \hat{N})} \hat{c}_{m}(t')^{\dagger} \hat{c}_{n}(t) \Bigr) }{\mathrm{Tr}\Bigl( e^{-\beta(\hat{H}-\mu \hat{N})}\Bigr)}
= \frac{\mathrm{Tr}\Bigl(  \hat{c}_{n}(t) e^{-\beta(\hat{H}-\mu \hat{N})} \hat{c}_{m}(t')^{\dagger} \Bigr) }{\mathrm{Tr}\Bigl( e^{-\beta(\hat{H}-\mu \hat{N})}\Bigr)} .
\end{split}
\end{equation}

 To obtain a thermal average again, a suitable one is inserted

\begin{equation}
\begin{split}
 \frac{\mathrm{Tr}\Bigl(  \hat{c}_{n}(t) e^{-\beta(\hat{H}-\mu \hat{N})} \hat{c}_{m}(t')^{\dagger} \Bigr) }{\mathrm{Tr}\Bigl( e^{-\beta(\hat{H}-\mu \hat{N})}\Bigr)} =  \frac{\mathrm{Tr}\Bigl( e^{-\beta(\hat{H}-\mu \hat{N})} e^{\beta(\hat{H}-\mu \hat{N})} \hat{c}_{n}(t) e^{-\beta(\hat{H}-\mu \hat{N})} \hat{c}_{m}(t')^{\dagger} \Bigr) }{\mathrm{Tr}\Bigl( e^{-\beta(\hat{H}-\mu \hat{N})}\Bigr)} = \langle e^{\beta(\hat{H}-\mu \hat{N})} \hat{c}_{n}(t) e^{-\beta(\hat{H}-\mu \hat{N})} \hat{c}_{m}(t')^{\dagger} \rangle.
 \label{KMS_Condition_0}
\end{split}
\end{equation}

To compress further, we want to recapitulate the solution of the Heisenberg equation of motion for an operator, which is given as 

\begin{equation}
\begin{split}
\hat{c}_{n}(t) = e^{i\hat{H}t} \hat{c}_{n}(0) e^{-i\hat{H}t} \rightarrow \hat{c}_{n}(t-i \beta) = e^{\beta\hat{H}} \hat{c}_{n}(t) e^{-\beta\hat{H}}.
\label{KMS_Condition_1}
\end{split}
\end{equation}

For the particle number operator $\hat{N}=\sum_{i} \hat{c}_{i}^{\dagger} \hat{c}_{i}$, the commutator with the annihilation operator can be computed by using the standard product rules and the corresponding fundamental (anti) commutation relations, $\Bigl[\hat{c}_{i}, \hat{c}_{j}^{\dagger} \Bigr] = \delta_{i,j}$ for bosons and $\Bigl \{\hat{c}_{i}, \hat{c}_{j}^{\dagger} \Bigr \} = \delta_{i,j}$ for fermions. For bosons one obtains

\begin{equation}
\begin{split}
\Bigl[\hat{c}_{n}, \hat{N} \Bigr] = \sum_{i} \Bigl[\hat{c}_{n}, \hat{c}_{i}^{\dagger} \hat{c}_{i} \Bigr] = \sum_{i} \Bigl[\hat{c}_{n}, \hat{c}_{i}^{\dagger} \Bigr] \hat{c}_{i} + \hat{c}_{i}^{\dagger} \, \Bigl[\hat{c}_{n}, \hat{c}_{i} \Bigr] = \hat{c}_{n}
\end{split}
\end{equation}

and similarly for fermions

\begin{equation}
\begin{split}
\Bigl[\hat{c}_{n}, \hat{N} \Bigr] = \sum_{i} \Bigl[\hat{c}_{n}, \hat{c}_{i}^{\dagger} \hat{c}_{i} \Bigr] = \sum_{i} \Bigl\{\hat{c}_{n}, \hat{c}_{i}^{\dagger} \Bigr\} \hat{c}_{i} - \hat{c}_{i}^{\dagger} \, \Bigl\{\hat{c}_{n}, \hat{c}_{i} \Bigr\} = \hat{c}_{n}.
\end{split}
\end{equation}

So in both cases we have 

\begin{equation}
\begin{split}
\hat{c}_{n} \hat{N} = \Bigl( \hat{N} + 1 \Bigr) \hat{c}_{n} \rightarrow e^{-\beta \mu \hat{N}} \hat{c}_{n} e^{\beta \mu \hat{N}} = e^{\beta \mu} \hat{c}_{n}.
\label{KMS_Condition_2}
\end{split}
\end{equation}

Inserting \cref{KMS_Condition_1,KMS_Condition_2} in \cref{KMS_Condition_0} one is left with 

\begin{equation}
\begin{split}
c^{<}_{n,m}(t,t') = \langle e^{\beta(\hat{H}-\mu \hat{N})} \hat{c}_{n}(t) e^{-\beta(\hat{H}-\mu \hat{N})} \hat{c}_{m}(t')^{\dagger} \rangle = e^{\beta \mu} \, c^{>}_{n,m}(t-i \beta,t') .
 \label{KMS_Condition_3}
\end{split}
\end{equation}

Applying a Fourier transformation in the relative time $\Delta t$ on both sides yields

\begin{equation}
\begin{split}
c^{<}_{n,m}(\omega,\bar{t}) = e^{\beta \mu} \, \int d\Delta{t} \, e^{i \omega \Delta{t}} c^{>}_{n,m}(\Delta t -i \beta,\bar{t}) = e^{\beta \mu} \, \int d \Delta\bar{{t}} \, e^{i \omega( \Delta{\bar{t}} +i \beta)} c^{>}_{n,m}(\Delta \bar{t},\bar{t}) =  e^{-\beta(\omega- \mu)} \, c^{>}_{n,m}(\omega,\bar{t}) ,
 \label{KMS_Condition_4}
\end{split}
\end{equation}

which is the KMS condition, that should be fulfilled at any time $\bar{t}$ when the system is in equilibrium.

\section{Spectral functions}\label{Appendix:Bosonic_spectral_functions}

It is known since the invention of the Källen-Lehmann representation, that the spectral function can be written as a sum over eigenstates of the (complete interacting) Hamiltonian.

\begin{equation}
( \hat{H} - \mu \hat{N} ) \ket{n} = \underbrace{( E_n - \mu \, N_n)}_{\defeq \tilde{E}_n} \ket{n}
\label{Bosonic_spectralfunction_0}
\end{equation}

We can first derive an expression for the retarded Green’s function in thermal equilibrium ($\bar{t} \rightarrow \infty$) and later use the relation $\tilde{a}_{n,n}(\omega)=-2 \, \mathrm{Im}(c_{n,n}^{\mathrm{ret}}(\omega))$. We define the retarded Green's function as 

\begin{equation}
c_{n,n}^{\mathrm{ret}}(\Delta t) = -i \, \Theta(\Delta t) \Bigl \langle \Bigl[ \hat{c}_{n}(t), \hat{c}_{n}^{\dagger}(t') \Bigr]_{\mp} \Bigr \rangle.
\end{equation}

Now we use the definition of the thermal expectation value and write the trace explicitly in the states of the Hamiltonian and define as a shortcut for the partition function $Z=\rm{Tr}(e^{-\beta ( \hat{H} - \mu \hat{N} )})$.

\begin{equation}
\begin{split}
c_{n,n}^{\mathrm{ret}}(\Delta t) &= -i \, \frac{\Theta(\Delta t)}{Z} \sum_{a,b} e^{-\beta \tilde{E}_a} \Bigl[ e^{i(E_a-E_b)(t-t')} \bra{a}\hat{c}_{n}\ket{b} \bra{b} \hat{c}_{n}^{\dagger} \ket{a} \mp e^{-i(E_a-E_b)(t-t')} \bra{a}\hat{c}^{\dagger}_{n}\ket{b} \bra{b} \hat{c}_{n} \ket{a} \Bigr] \\
&= -i \, \frac{\Theta(\Delta t)}{Z} \sum_{a,b} \Bigl[ e^{-\beta \tilde{E}_a} \mp e^{-\beta \tilde{E}_b} \Bigr] e^{i(E_a-E_b)(t-t')} |\bra{a}\hat{c}_{n}\ket{b}|^{2}  
\end{split}
\end{equation}

Now one can Fourier transform into frequency space, where the relevant integral is

\begin{equation}
\begin{split}
\int d \Delta t e^{i \omega \Delta t} \Theta(\Delta t) e^{i(E_a-E_b)(\Delta t)} = \lim_{\epsilon \rightarrow 0^{+}} \int_{0}^{\infty} d \Delta t e^{i (\omega +i \epsilon) \Delta t} e^{i(E_a-E_b)(\Delta t)} = \lim_{\epsilon \rightarrow 0^{+}} \frac{i}{\omega + E_a - E_b + i \epsilon}
\end{split}
\end{equation}

So we have

\begin{equation}
\begin{split}
c_{n,n}^{\mathrm{ret}}(\omega) &= \frac{1}{Z} \sum_{a,b} \frac{\Bigl[ e^{-\beta \tilde{E}_a} \mp e^{-\beta \tilde{E}_b} \Bigr]}{\omega + E_a - E_b + i 0^{+}} |\bra{a}\hat{c}_{n}\ket{b}|^{2} 
\label{Bosonic_spectralfunction_2}
\end{split}
\end{equation}

To obtain the final result, we use the Sokhotski–Plemelj theorem, which states

\begin{equation}
\begin{split}
\lim_{\epsilon \rightarrow 0^{+}} \frac{1}{x \pm i \epsilon} = \mp i \pi \delta(x) + \mathcal{P}  \Bigl( \frac{1}{x} \Bigr),
\label{Bosonic_spectralfunction_1}
\end{split}
\end{equation}

where $\mathcal{P}$ denotes the Cauchy principal value. Inserting \cref{Bosonic_spectralfunction_1} into \cref{Bosonic_spectralfunction_2} yields

\begin{equation}
\begin{split}
c_{n,n}^{\mathrm{ret}}(\omega) &= \frac{1}{Z} \sum_{a,b} \Bigl[ e^{-\beta \tilde{E}_a} \mp e^{-\beta \tilde{E}_b} \Bigr] \Bigl[ -i \pi \delta(\omega + E_a - E_b) + \mathcal{P}  \Bigl( \frac{1}{\omega + E_a - E_b} \Bigr) \Bigr] \, |\bra{a}\hat{c}_{n}\ket{b}|^{2} .  
\label{Bosonic_spectralfunction_3}
\end{split}
\end{equation}

The spectral function follows as

\begin{equation}
\begin{split}
\tilde{a}_{n,n}(\omega)= -2 \, \mathrm{Im}(c_{n,n}^{\mathrm{ret}}(\omega)) &= \frac{2 \pi}{Z} \sum_{a,b} \Bigl[ e^{-\beta \tilde{E}_a} \mp e^{-\beta \tilde{E}_b} \Bigr] \delta(\omega + E_a - E_b) \, |\bra{a}\hat{c}_{n}\ket{b}|^{2} \\
&= \frac{2 \pi}{Z} \sum_{a,b} \, \Bigl[ 1 \mp e^{-\beta (\tilde{E}_b-\tilde{E}_a)} \Bigr] \,e^{-\beta \tilde{E}_a} \delta(\omega + E_a - E_b) \, |\bra{a}\hat{c}_{n}\ket{b}|^{2}.
\label{Bosonic_spectralfunction_4}
\end{split}
\end{equation}

As in our case $\hat{c}_{n}$ is a one-particle annihilation operator, the matrix element can only contribute if $N_b=N_a + 1$, so we can rewrite the energy differences in the exponent to

\begin{equation}
\begin{split}
\tilde{E}_b-\tilde{E}_a = E_{b} - \mu N_{b} - ( E_a - \mu N_a) = E_{b} - E_a - \mu,
\label{Bosonic_spectralfunction_5}
\end{split}
\end{equation}

where \cref{Bosonic_spectralfunction_0} was used. This finally yields after applying the delta function

\begin{equation}
\begin{split}
\tilde{a}_{n,n}(\omega)= \frac{2 \pi}{Z} \Bigl[ 1 \mp e^{-\beta (\omega -\mu)} \Bigr] \sum_{a,b} \,e^{-\beta \tilde{E}_a} \delta(\omega + E_a - E_b) \, |\bra{a}\hat{c}_{n}\ket{b}|^{2}.
\label{Bosonic_spectralfunction_6}
\end{split}
\end{equation}

This was already shown in \cite{Fetter} and shows indeed, that the bosonic spectral function can be negative, if $\omega - \mu < 0$ .

\end{widetext}


\bibliography{references}%

\end{document}